# The Impact of the General Data Protection Regulation (GDPR) on Online Usage Behavior

*Klaus M. Miller[1], Bernd Skiera, Julia Schmitt*

Klaus M. Miller, Assistant Professor of Marketing, HEC Paris, Rue de la Libération 1, 78350 Jouy-en-Josas, France, Hi!PARIS Chairholder, Center in Data Analytics & AI for Science, Technology, Business & Society, Phone +33-1-39-67-70-88, millerk@hec.fr.

Bernd Skiera, Professor of Electronic Commerce, Department of Marketing, Faculty of Economics and Business, Goethe University Frankfurt, Theodor-W.-Adorno-Platz 4, 60323 Frankfurt, Germany, Phone +49-69-798-34649, skiera@wiwi.uni-frankfurt.de.

Julia Schmitt, Department of Marketing, Faculty of Economics and Business, Goethe University Frankfurt, Theodor-W.-Adorno-Platz 4, 60323 Frankfurt, Germany, Phone +49-69-798-34563, Email: schmitt@wiwi.uni-frankfurt.de.

Funding:
This project has received funding from the European Research Council (ERC) under the European Union's Horizon 2020 research and innovation program (grant agreement No. 833714). Miller gratefully acknowledges support from the Hi! PARIS Center on Data Analytics and Artificial Intelligence for Science, Business, and Society and ANR/France 2030 program (ANR-23-IACL-0005).

Declarations of interest: none.

[1] Corresponding author.

# The Impact of the General Data Protection Regulation (GDPR) on Online Usage Behavior

## *ABSTRACT*

Privacy regulations aim to safeguard consumers, but can have unintended consequences on how users interact with websites. This article estimates the causal effect of the EU's General Data Protection Regulation (GDPR) on online usage behavior and decomposes it into usage frequency (unique visitors) and usage intensity (visits per unique visitor). Using a generalized synthetic control estimator across trillions of visits to 6,387 websites in 24 industries and 13 countries—11 months before and 19 months after enforcement—it compares observations subject to the GDPR (EU users or EU websites) with unaffected observations (non-EU users on non-EU websites). Weekly visits decline by 4.88% within 3 months and by 10.02% after 18 months, with the decline increasingly driven by usage frequency: by 18 months, unique visitors decline by 6.61%, whereas visits per unique visitor decline by only 0.59%. The average conceals offsetting effects consistent with a reallocation of online activity: about one quarter of websites gain significantly, and among websites losing users, the remaining users engage more intensively, whereas intensity falls where user numbers grow. Losses concentrate among hedonic and smaller websites and continue to deepen during the first wave of data-protection enforcement actions rather than being concentrated around the GDPR compliance deadline. Taken together, these patterns are more consistent with restricted data-driven user acquisition and privacy salience, than with consent friction and degraded retention as persistent explanations. Long-horizon, decompositional evaluation thus reveals GDPR's differential effect on usage frequency/intensity and heterogeneous effects not distinguished by the aggregate traffic effects reported in prior work.

*Keywords*: Economics of Privacy, Public Policy, Consumer Protection, Synthetic Control, Difference in Differences, Quasi Experiments

*JEL Codes:* K23, L86, D12, M37, O33

Internet users are increasingly concerned about their privacy online. Policymakers worldwide have enacted privacy laws to address these concerns. The European Union (EU)'s General Data Protection Regulation (GDPR), in force since 2018, is one of the most stringent privacy laws.

However, regulations such as the GDPR can also trigger unintended effects. In particular, the compliance requirements they impose could alter websites' functionality and reshape how users interact with them. These changes could negatively affect web traffic and user engagement, resulting in revenue losses for firms (Wang et al. 2024). Such losses could stem directly from users (e.g., changes to functionality might diminish users' likelihood of purchasing products on an e-commerce site) or from lost ad revenue (e.g., resulting from diminished web traffic). Such losses could also have broader societal effects as companies compensate—for example, by scaling back services, introducing fees, or eliminating jobs.

Indeed, the effects of a privacy regulation are shaped by multiple, potentially conflicting, and even unpredictable factors (Acquisti 2024)—and thus might evolve over time and vary across different websites and user bases. Crucially, the same total decline can arise from fewer users or from less usage per remaining user, and which effect dominates points to the mechanism at work and to the appropriate managerial response, as we discuss in the following section.

Against these requirements, prior evidence remains partial. Several studies have evaluated unintended effects of privacy regulations (see, e.g., Dubé et al. 2025 for a broad overview and Johnson 2023 for an overview of effects of the GDPR), including how these effects evolve (Peukert et al. 2022; Johnson et al. 2023; Lefrere et al. 2026). Yet the large-sample evidence concerns firms' compliance behavior rather than usage, and the studies that do measure user-facing usage rest on narrower samples (Goldberg et al. 2024; Lefrere et al. 2026), limiting the breadth of websites and industries over which usage effects can be traced. Above all, prior studies rarely separate whether lost traffic reflects fewer users or less usage per user, effects with very different managerial remedies.

Here, we exploit large-scale data and adopt a rigorous modeling approach to unpack the effects of the GDPR over time, devoting particular attention to website heterogeneity in characteristics that proxy the underlying mechanisms (websites' reliance on data-driven advertising and tracking, content type, data sensitivity, size, and regulatory exposure). Our analyses focus on several outcomes that represent the amount of user activity on a website (e.g., the number of weekly visits or page impressions), which we collectively refer to as "usage behavior". Moreover, we decompose our measures of usage behavior into two facets: usage frequency (e.g., the number of unique visitors) and usage intensity (e.g., usage per unique visitor). This decomposition is motivated by the fact that the two facets may respond differently to regulation, as our results confirm, with implications for how to adapt to it. Prior studies do not formally decompose total website usage behavior into usage frequency and usage intensity (see Table W1 in Web Appendix A). We use this decomposition to examine which usage facet accounts for the aggregate effect and whether the two facets reinforce or offset one another across websites.

More specifically, our analysis relies on a dataset capturing usage behavior on 6,387 unique websites spanning 24 industries, comprising the most popular websites in 13 developed countries (11 EU countries, Switzerland, and the United States). The data cover the period from July 2017 through December 2019—i.e., about 11 months before and 19 months after the GDPR's enforcement (hereafter referred to as "GDPR") on May 25th, 2018—enabling us to construct a before-and-after analysis. Within our dataset, some website–user interactions are subject to the GDPR (i.e., those involving EU websites or EU users), while others are not (i.e., those involving non-EU websites or non-EU users). The latter serves as a control group. We use a generalized synthetic control (GSC) estimator (Xu 2017) to isolate the effect of GDPR on our metrics of interest. Robustness tests support our identifying assumptions, yielding three sets of findings:

First, the GDPR reduced the average website's weekly visits by 4.88% within 3 months, deepening to 10.02% after 18 months, with weekly page impressions declining by similar amounts (3.12% and 9.28%). Decomposing this effect, usage frequency deteriorates over time (unique visitors: -0.77% at 3 months, -6.61% at 18 months), whereas usage intensity recovers (visits per unique visitor: -1.62% at 3 months, -0.59% at 18 months). Thus, the aggregate decline is increasingly a usage-frequency phenomenon.

Second, the average conceals offsetting effects consistent with reallocation rather than uniform decline: by 18 months, 54.87% of websites showed significantly negative effects on visits, yet 23.96% showed significant gains, and the share of websites without a detectable effect more than halved. Among websites that lost unique visitors, the remaining users increased their usage (+4.77% visits per unique visitor at 18 months). In contrast, intensity fell among user-gaining websites (-9.09%), an offsetting pattern that any single aggregate effect conceals.

Third, these patterns provide diagnostic evidence on the mechanisms developed below: the decline increasingly runs through usage frequency rather than intensity, concentrates among hedonic websites and smaller websites (the smallest size decile loses 21.49% of weekly visits at 18 months versus 9.04% for the largest, consistent with increased market concentration), and continues to deepen during the first wave of enforcement actions rather than being concentrated around the May 2018 compliance deadline; flat bounce rates and the absence of a usage break at the cookie consent banner rollout provide little evidence that consent friction or degraded retention are the primary persistent drivers.

These findings provide a focused extension of prior evidence on the GDPR's effect on online usage. Prior studies show that privacy regulation can reduce aggregate usage; we show that this average masks distinct responses along two facets. First, we decompose usage into frequency and intensity and show that the two can move in opposite directions across websites, an offsetting pattern consistent with a reallocation of online activity that prior aggregate effects do not distinguish. Second, our broader sample—6,387 websites across 24 industries and 13 countries—and 19-month post-GDPR horizon reveal how these facets evolve and identify both winners and losers of the GDPR. Third, we link this heterogeneity to theoretically motivated mechanisms and boundary conditions, showing that differences across websites operate primarily through usage frequency and indicating where the usage costs of privacy regulation concentrate. Together, the evidence is more consistent with restricted data-driven user acquisition and privacy salience than with consent friction or degraded retention as persistent explanations. Table W1 in Web Appendix A positions our design, sample, outcomes, and observation period relative to the closest prior studies (Goldberg et al. 2024; Lefrere et al. 2026).

## THEORY AND EXPECTED EFFECTS OF THE GDPR ON ONLINE USAGE BEHAVIOR

The General Data Protection Regulation (GDPR) became enforceable in May 2018 and conditions the processing of personal data on a lawful basis, most prominently users' informed consent. The regulation can affect online usage through both supply- and demand-side channels.

On the supply side, websites changed how they collected and used personal data. They delayed cookie placement until consent had been obtained, reduced the number of third parties receiving user data, introduced consent-management interfaces, and revised their privacy policies (Libert et al. 2018; Degeling et al. 2019; Peukert et al. 2022; Johnson et al. 2023; Lefrere et al. 2026; Miller et al. 2026). These changes may reduce websites' ability to acquire and retain users through data-intensive advertising, targeting, attribution, and personalization.

On the demand side, the GDPR changed the conditions under which users encounter and evaluate data collection. Consent interfaces make data processing salient when entering a website and require users to decide whether to accept, reject, or configure data use (Utz et al. 2019; Warberg et al. 2023).

We organize these competing effects using the privacy calculus (Dinev and Hart 2006). Users weigh the value they derive from a website against the perceived costs of sharing personal data with others. The GDPR may increase perceived costs by making data collection more salient and by adding effort to website entry. It may also increase perceived benefits by giving users greater control and confidence in how their data are processed (Xu et al. 2011; Tucker 2013; Ma et al. 2026). Because websites differ in their value to users, their reliance on personal data, and the regulatory environments in which they operate, the balance between these forces may vary systematically across websites.

### *Relationship between Usage Frequency, Usage Intensity, and Total Usage*

The GDPR may affect total website usage through distinct effects, which we organize around the Equation:

$$(1)\ \textit{Total Usage} = \textit{Usage Frequency} \times \textit{Usage Intensity}$$

Throughout, total usage is measured using visits or page impressions. Usage frequency, measured by a quantity (unique visitors or visits), captures how many users or visits a website has. Usage intensity, measured by visits per unique visitor or page impressions per visit, captures how intensively they engage while on the website. Table W6 in Web Appendix B maps each element of the identity to its measured metrics for both implementations. This distinction matters because GDPR-related mechanisms need not affect both facets in the same direction, as the following subsections show.

*Expected Effect of the GDPR on Usage Frequency and Total Usage*

We expect the GDPR to reduce usage frequency and, assuming usage intensity does not fully offset this loss, total website usage.

One reason is restricted user acquisition. Data-driven acquisition channels such as display and social advertising, email marketing, or retargeting, depend on data collected across users and websites. Prior research shows that privacy regulation reduces the effectiveness of data-driven advertising and imposes disproportionate costs on firms with fewer resources (Goldfarb and Tucker 2011; Campbell et al. 2015). The GDPR may thus reduce the number of users reaching affected websites.

A second reason is privacy salience. Consent requests make websites' data-collection practices more visible and introduce an explicit decision at the point of access. Some users may leave rather than accept or configure data processing, particularly when they distrust the website or derive limited value from using it.

Greater transparency and control may moderate these effects by reassuring users, so that websites with high usage value or strong trust may be less affected or even gain usage.

*Competing Effects of the GDPR on Usage Intensity*

The effect of the GDPR on usage intensity is theoretically ambiguous because two forces operate in opposing directions.

The first is a retention–effect. Websites use personal data to personalize content, recommendations, communications, and advertising. If the GDPR reduces the availability or usability of such data, the quality of personalization may decline, which materially affects consumption and engagement (Lee and Hosanagar 2019). Degraded retention therefore predicts fewer visits per unique visitor, fewer page impressions per visit, and shorter visit duration.

The second is an audience-composition effect. The users most likely to stop visiting may be those who derive relatively little value from a website or have low trust in it. If these marginal users leave, the website's post-GDPR user base will contain a greater share of highly engaged users, increasing the average usage intensity. This logic is consistent with evidence that privacy regulation alters the composition of firms' users and customers (Aridor et al. 2023). Remaining users may additionally consolidate activity on websites where consent has already been granted, consistent with sunk-cost reasoning (Arkes and Blumer 1985).

Because our data are aggregated across users, we cannot separately identify changes in individual behavior from changes in user composition; however, we can reveal whether website-level effects on the two facets tend to reinforce or offset one another. Accordingly, we do not predict a uniform effect on usage intensity, but examine whether intensity gains offset frequency losses, and vice versa.

*Comparison of the Effect of Consent Friction versus Privacy Salience*

Two competing demand-side explanations may account for changes in usage: consent friction and privacy salience.

Consent friction refers to the procedural cost that consent interfaces impose at website entry: the additional clicks, reading, and decisions required before content can be accessed, independent of any privacy considerations the interface may raise. If consent friction were a principal driver, its effect should be concentrated around the introduction of consent interfaces and be strongest at website entry, diminishing as users habituate (diagnostics in Table 1).

Unlike consent friction, privacy salience need not fade after the initial implementation period. Major enforcement actions and their public coverage can renew attention to websites' data practices. Responses may therefore deepen during the DPA enforcement wave rather than dissipate as users become familiar with consent interfaces.

Because the effects of consent friction, privacy salience, and restricted user acquisition overlap, we treat the related analyses as diagnostic (Table 1).

*Consideration of Boundary Conditions*

The effects of the GDPR should vary systematically with websites' business models, resources, content, audiences, and regulatory environments. We organize this heterogeneity around four broader boundary conditions.

First, data reliance: websites relying more heavily on advertising, third-party tracking, and related data-driven activities may face greater disruption to acquisition, attribution, personalization, and monetization. We therefore expect more negative effects on usage frequency among such websites. We interpret the available measures as indicators of a data-intensive business model rather than of any single activity.

Second, website resources and prominence may shape both firms' ability to absorb the regulatory effect and users' responses to it. Larger websites have stronger brands, more direct traffic, richer first-party data, and greater compliance resources, predicting smaller losses. At the same time, highly prominent websites may attract greater public scrutiny. The size relationship may therefore be nonlinear.

Third, usage value and data sensitivity shape the privacy calculus. Users tolerate privacy costs more readily when a website provides important or difficult-to-substitute services, whereas discretionary use is easier to forgo. Greater sensitivity of the personal data associated with the website's core service may increase privacy concerns, but consent and control may be particularly reassuring precisely there. The expected relationship is therefore ambiguous.

Fourth, regulatory exposure affects both the magnitude and persistence of the GDPR effect. Websites operating under stricter pre-GDPR privacy regimes may already have adapted and therefore experience a smaller incremental change. We measure strictness using the European Commission's (2008) assessment. Conversely, greater enforcement capability and public enforcement coverage may increase the regulation's salience, causing responses to persist or deepen rather than fade. The sign of the resulting relationship is an empirical question. Finally, exposure varies directly with a website's pre-GDPR share of EU users, so usage losses should increase monotonically in EU-user exposure.

*Table 1. Theoretical Mechanisms, Boundary Conditions, and Empirical Diagnostics*

| Mechanism or boundary condition | Principal theoretical implication | Operationalization | Main empirical diagnostics | Tested in |
|---|---|---|---|---|
| **Restricted user acquisition** | Reduced effectiveness of data-driven acquisition lowers usage frequency and total usage | Homepage ad intensity; tracking intensity | Unique visitors; visits; heterogeneity by data reliance | Table 2; Figure 2; Web Appendix H; Results, "Potential Mechanisms" |
| **Consent friction** | Usage losses are concentrated at website entry and around the introduction of consent interfaces; effects may diminish with habituation | Bounce rate; bouncing visits; own-cookie consent banner adoption timing; mobile share | Entry-related outcomes; usage path around consent-interface adoption; temporal effect path | Results, "Potential Mechanisms"; Web Appendix J; Web Appendix H |
| **Privacy salience and control** | Greater salience may discourage privacy-concerned users, whereas credible control may reassure others; enforcement can renew salience over time | Distribution of website-specific effects; data-type sensitivity; hedonic versus utilitarian orientation; temporal effect path | Positive, negative, and null website-specific effects; heterogeneity by usage value and sensitivity; non-cumulative estimates | Table 2; Figure 2; Web Appendices H, I, and J |
| **Retention versus audience composition** | Weaker personalization lowers usage intensity, whereas the exit of marginal users raises average intensity among the post-GDPR audience | Visits per unique visitor; page impressions per visit; visit duration; websites gaining versus losing unique visitors | Relationship between effects on usage frequency and usage intensity | Table 2; Results, "Potential Mechanisms" |
| **Data reliance** | Websites with more data-intensive business models experience larger usage losses | Homepage ad intensity; tracking intensity | Heterogeneity in usage frequency and total usage | Figure 2; Web Appendix H |
| **Website resources and prominence** | Compliance capacity may attenuate losses, while prominence may increase scrutiny | Website size | Size-decile profile and high-versus-low comparisons | Results; Figure 2; Web Appendices A and H |
| **Usage value and data sensitivity** | High usage value may reduce abandonment; sensitive data may increase privacy costs or strengthen reassurance from control | Hedonic versus utilitarian orientation; data-type sensitivity | Heterogeneity in website-specific usage effects | Figure 2; Web Appendix H |
| **Regulatory exposure** | Prior adaptation attenuates the incremental effect, whereas stronger enforcement renews salience and may amplify responses | Regulatory strictness; enforcement capability; pre-GDPR EU-user share | Heterogeneity across regulatory environments; persistence of the temporal effect path; within-website exposure gradient by EU-user share | Figure 2; Web Appendices H, I, and J; Web Appendices E and G |

Notes. Metrics as defined in Table W6 in Web Appendix B. The listed operationalizations correspond to the terminology used in the Results and Web Appendix. The empirical diagnostics provide evidence consistent with the proposed mechanisms but do not necessarily identify them separately.

### *Summary of Theoretical Mechanisms, Boundary Conditions, and Empirical Diagnostics*

Table 1 summarizes the framework and the operationalization of each mechanism and boundary condition. The central expectation is an average decline in usage frequency and total usage; the remaining analyses assess competing explanations and identify where the effects are concentrated.

## *DESCRIPTION OF EMPIRICAL STUDY*

### *Description of the Set-Up of the Empirical Study*

The GDPR provides a valuable setting for quantifying the effect of a privacy law on online usage behavior: treated and control observations can be distinguished directly from the law's applicability. The law is binding for all websites based in EU countries. It applies to any activity performed on personal data from users located in the EU, even if data processing takes place outside the EU. Penalties for noncompliance are severe, motivating compliance.

Based on the GDPR's applicability structure, our analysis focuses on website–user interactions ("observations"), with "treatment" defined as the GDPR's applicability to a focal observation. Thus, we categorize observations after the enforcement of the GDPR into treatment and control groups, as outlined in Figure 1: the GDPR applies to Cells 1-3 (treatment group), whereas non-EU users visiting non-EU websites constitute the control group (Cell 4). In line with Article 3 of the GDPR, we use the website's server location (retrieved from https://check-host.net) to determine the respective website's data processing location and GDPR's applicability.

An observation is "treated" under the GDPR regardless of whether or how the website complied with the regulation in practice; we thus measure the "intention-to-treat" effect of the GDPR.

Some websites may voluntarily comply with the GDPR (Bradford 2020), a phenomenon we refer to as the spillover effect. The Robustness section and Web Appendix E address the spillover effect and show that our data support classifying Cell 3 as treated. Overall, our categorization is in line with other studies that use non-EU users as a control group (Jia et al. 2021, Miller et al. 2026).

We use the GDPR enforcement date (May 25th, 2018) to construct a before-and-after analysis, comparing the treatment group to the control group to quantify GDPR's effect. This approach allows us to construct a GSC estimator (Xu 2017) that rests upon several critical assumptions, which we outline below.

*Figure 1. Scope of GDPR and Assignment of Website-Instances to Treatment and Control Group*

| | | *Base of User* | |
|---|---|---|---|
| | | **EU Users** | **Non-EU Users** |
| *Base of Website* | EU Website | GDPR applies = Treatment Group *(Cell 1)* | GDPR applies = Treatment Group *(Cell 2)* |
| | Non-EU Website | GDPR applies = Treatment Group *(Cell 3)* | GDPR does not apply = Control Group *(Cell 4)* |

■ Treatment Group □ Control Group

*Overview of Data*

*Description of sample.* This study uses data from SimilarWeb for the top 1,000 websites—as listed in Alexa Top Sites in April 2018—across 13 countries: 2 non-EU and 11 EU. SimilarWeb draws on a diversified and rich global user panel to measure online usage behavior. The websites in our sample are from diverse industries (Figure W1 in Web Appendix A), audiences, and sizes (measured by SimilarWeb ranks).

Usage metrics are observed for users accessing each website from the EU country where it has the largest following and for users accessing from the United States (US-only if a website appears in no EU top-1000 list). These data span between July 1st, 2017, and December 31st, 2019—nearly 11 months before and more than 19 months after the GDPR's enforcement (our longest estimation window ends at 18 months post-enforcement).

*Discussion of measurement error due to non-consenting users*. One potential concern is that GDPR may make it more difficult to measure the activity of users who do not consent to be tracked. Any residual GDPR-induced measurement error would concentrate in the treatment group, so our estimates are best read as an upper bound on the true behavioral response. A comparison with ground-truth data collected without relying on user consent (AGOF) confirms a negative effect that is statistically indistinguishable from our SimilarWeb estimates (Web Appendix D). We therefore treat measurement error as a bounded and explicitly acknowledged limitation rather than ruling it out categorically.

*Description of sample construction.* We start with data on the top 1,000 websites for each of the 13 countries (Table W2 in Web Appendix A.2). After removing duplicate websites, our sample includes 7,332 unique websites.

We consider each website's non-EU and EU users separately and refer to each combination of a website with one of the two user groups as a "website-instance." For a non-EU website such as "nzz.ch" (Switzerland), only the website instance corresponding to its EU users is treated, while the instance corresponding to its non-EU users is in the control group (Figure 1).

We then apply three screening steps—a minimum-traffic threshold (1,000 weekly visits), no zero-visit month, and no unexplained traffic changes—as detailed in Table W2 in Web Appendix A.2. Web Appendix C shows that our results are not sensitive to the traffic threshold.

This procedure results in a final sample of 9,758 website-instances belonging to 6,387 unique websites (8,057 treated and 1,701 control website-instances; Table W3 in Web Appendix A.2).

SimilarWeb does not report unique visitor information for websites with fewer than 5,000 unique visitors per month. The unique visitor analysis, therefore, uses the subsample of website-instances with reported unique visitors in every month of the observation period: 7,965 website-instances (Table W2 in Web Appendix A.2), corresponding to 5,447 treatment websites.

*Description of variables.* Table W6 in Web Appendix B defines our variables of interest, which correspond to total usage, usage frequency, and usage intensity. Following Equation (1), weekly visits decompose into monthly unique visitors (available only at monthly resolution) and weekly visits per unique visitor; weekly page impressions decompose into weekly visits and page impressions per visit.

Due to large differences in the values of each metric across websites and countries, we convert all total usage and usage frequency metrics (by adding 1 to avoid zero values) to their natural logarithms to capture relative (i.e., percentage) effects. Table W4 in Web Appendix A reports summary statistics and pairwise correlations on the final sample of 9,758 website-instances. Figure W2 in Web Appendix A shows the raw pre- and post-GDPR trends of weekly visits for the treatment and control groups.

*Description of Approach to Analyze Data*

We examine GDPR's effect on the various total usage metrics using a GSC estimator (Xu 2017). We estimate a separate GSC model for each treated website-instance and each metric; GDPR's effect can thus vary freely across websites, forming the basis for our decomposition and heterogeneity analysis. We estimate GDPR's effect on total usage and usage frequency metrics directly, and derive its effect on usage intensity from these estimates (see below).

*Description of methodology to analyze total usage and usage frequency metrics.* Following Xu (2017), we estimate the following regression model using the GSC to determine GDPR's effect on online usage behavior:

$$(2)\; ln\left(Y_{m,t,wi}+1\right) = \delta_{m,wi,t} * Treated_{wi,t} + \alpha_{m,wi} + \gamma_{m,t} + \lambda'_{m,wi} * f_{m,t} + \epsilon_{m,wi,t}$$

where m indexes the total usage or usage frequency metric for $Y$ website-instance wi in week t (when the dependent variable is unique visitors, the scale of t is one month). $Treated_{wi,t}$ is a binary indicator variable: 1 if, in week t, website-instance wi needs to consider GDPR; otherwise, 0. The term $\delta_{m,wi,t}$ represents the GDPR's effect on metric m for website-instance wi in week t, calculated as the gap between the observed outcome and the imputed counterfactual. Because we run separate models for each cumulative post-treatment period (e.g., 3 months, 18 months), we define $\delta_{m,wi}$ as the average treatment effect across the respective weeks within that specific window. The term $\alpha_{m,wi}$ is the website-instance fixed effect, and the term $\gamma_{m,t}$ is the time fixed effect. The term $f_t = [f_{1,t}, f_{2,t}, \ldots, f_{r,t}]'$is a vector of r unobserved common factors for metric m in week t, while $\lambda'_{wi} = [\lambda_{1,wi}, \lambda_{2,wi}, \ldots, \lambda_{r,wi}]'$ denotes the corresponding factor loadings for website-instance wi. Although the same factors influence the treated and control groups, each website-instance can have different factor loadings ($f_{m,t} \in = \mathbb{R}^r$, $\lambda_{m,wi} \in \mathbb{R}^r$). We select the number of factors r by cross-validation, allowing between 1 and min (5, N-1) factors, where N is the number of website-instances included in the corresponding estimation sample.

In all our model specifications, we impose additive two-way fixed effects: website-instance fixed effects absorb time-constant cross-sectional differences, and time fixed effects absorb common intertemporal changes (Xu 2017). $\epsilon_{m,t,wi}$ represents an unobserved idiosyncratic shock for metric m in week t for website-instance wi. We use Equation (2) and the control and treatment group assignment described in Figure 1 to calculate the treatment effect $\delta_{m,wi,t}$ for every website-instance wi, metric m, and post-treatment period t.

To trace the treatment effect over time, we re-estimate the model for post-treatment windows of 3, 6, 9, 12, and 18 months after the GDPR's enforcement, capturing GDPR's short- and long-term effects.

The GSC estimator constructs, for each treated website-instance, a synthetic control whose pre-treatment trajectory closely matches that of the treated instance, thereby relaxing the parallel-trends assumption (Abadie et al. 2010; Xu 2017). Following Abadie (2021), we build each donor pool from five control website-instances that belong to the same industry as the treated instance and exhibit the highest pre-treatment correlations with it (Web Appendix E.5 examines donor pool composition).

We then determine GDPR's effect on a website's treated traffic, referred to as $\Delta_{m,w}$. For websites with one treated website-instance, the website-level effect equals the corresponding proportional instance effect, exp($\delta_{m,wi}$) - 1; for websites with two treated website-instances, it is the weighted average of the two proportional instance effects exp($\delta_{m,wi}$)-1 with the relative average sizes of the two treated website-instances in the entire pre-treatment period as weights. This weighting aggregates the instance-level proportional effects according to each treated instance's pre-treatment share of the website's treated traffic.

*Description of methodology to analyze usage intensity metrics.* Each usage intensity metric is the ratio of a total usage metric to a usage frequency metric (Table W6 in Web Appendix B). Because the treatment effects estimated with Equation (2) are on the log scale, we first convert them to proportional effects. We then construct a standardized index for the GDPR's effect on each usage intensity metric by taking the ratio of these matched window-level proportional effects. (p indexes the post-treatment period, w the website):

$$\Delta\, Visits\ per\ Unique\ Visitor_{p,w} = \frac{1 + \Delta\, Visits_{p,w}}{1 + \Delta\, Unique\ Visitors_{p,w}} - 1,\ \Delta\, Page\ Impressions\ per\ Visit_{p,w} = \frac{1 + \Delta\, Page\ Impressions_{p,w}}{1 + \Delta\, Visits_{p,w}} - 1.$$

Web Appendix B derives the formulas.

## *RESULTS OF EMPIRICAL STUDY*

### *GDPR's Effect on Total Usage*

In the 3 months following GDPR's enforcement, treatment websites had 4.88% fewer mean weekly visits than control websites (see Table 2, Panel A). Over time, this decrease became even more substantial: by 18 months post-GDPR, mean weekly visits to treatment websites were 10.02% lower. The deepening is not an artifact of cumulating windows: re-estimated on non-cumulative periods, the effect peaks in months 7-9, coinciding with the CNIL's €50 million Google fine, and the median effect shows no recovery (Web Appendix I).

Because we estimate a separate treatment effect for each website, we characterize not only the average effect but also the entire distribution; the latter is unavailable in designs that recover only a single (average) effect. Three months post-GDPR, half the websites (50.06%) show no statistically detectable effect on weekly visits; by 18 months, only 21.17% remain undetected, while 54.87% show significantly negative and 23.96% significantly positive effects ($p < .05$; Table 2, Panel A; Figure W3 in Web Appendix A). Two implications follow. First, analyses ending a few months after enforcement miss most of GDPR's effect. Second, the GDPR widened the distribution rather than shifting it uniformly downward, and the persistence of a substantially positive affected share is informative: the privacy-salience channel predicts reallocation (see "Potential Mechanisms").

The results for page impressions are comparable: a 3.12% decrease (95% CI 2.37%–3.87%) at 3 months, a 9.28% decrease at 18 months, and an almost identical distributional shift (Table 2, Panel A; Web

Appendix A). These findings resonate with Goldberg et al. (2024), who report an 11.7% decline in recorded page views at four months and estimate that the corresponding real effect lies between 0% and -8.1%. Lefrere et al. (2026), studying only news websites, report precisely estimated nulls for total traffic; our larger, cross-industry sample detects the decline.

Together, these results support our expectation of an overall negative effect on total usage, with pronounced heterogeneity across websites.

*Decomposition of GDPR's Effect into Usage Frequency and Usage Intensity*

First, to understand the drivers of the overall decline in visits, we evaluate the relative trajectories of usage frequency (unique visitors) and usage intensity (visits per unique visitor) for the subsample of websites with available unique-visitor data. This analysis focuses on GDPR's effect after 3 and 18 months (see Table 2, Panel B).

1. *Unique Visitors*: Unique visitors decreased by 0.77%, three months post-GDPR, and by 6.61% after 18 months. Over time, total usage declines due to user loss.
2. *Visits per Unique Visitor*: Visits per unique visitor fell modestly, and the gap narrowed over time (-1.62% at 3 months to -0.59% at 18 months), indicating that remaining users were engaged nearly as much as before. The loss is one of users, not of engagement per user. We assess which mechanisms account for these patterns in the Potential Mechanisms section below.

Next, we divided the websites into two groups based on whether they experienced an increase or a decrease in unique visitors (see Table 2, Panel B). By 18 months, 38.27% of websites gained unique visitors and 61.73% lost them. The direction of the average effect on visits per unique visitor is opposite across the two groups,-and the opposing responses strengthen over time: Websites that gained unique visitors experienced a 6.46% average decrease in visits per unique visitor at 3 months, widening to 9.09% by 18 months. Conversely, websites that lost unique visitors experienced increases of 2.56% and 4.77%, respectively.

Next, we decomposed the effect of the GDPR on page impressions into its effects on visits (usage frequency) and page impressions per visit (usage intensity). Average page impressions per visit increased by 1.97% (3 months) and 2.15% (18 months). The same offsetting directions hold (Table 2, Panel B): among websites with increased visits, page impressions per visit decreased (-4.58% at 18 months); among websites with decreased visits, page impressions per visit increased (+5.53%).

Usage intensity thus partly counteracts usage frequency: among websites losing unique visitors, remaining users visited more often and engaged more per visit, while user-gaining websites experienced the opposite. This pattern also weighs against measurement error. GDPR-induced loss of cross-visit user identification would inflate unique visitor counts and mechanically depress per-user intensity (Lefrere et al. 2026). We observe the opposite, falling unique visitors alongside stable or rising intensity, which is inconsistent with this specific user-identification measurement error.

Prior studies have not decomposed total usage into frequency and intensity. Hence, they could not have revealed these opposing trends. Yet their findings hint at them: Lefrere et al. (2026) find page views per user decline by roughly 3.9%, only in the long run, mirroring the slow build-up we document; and Goldberg et al. (2024) find recorded page views per visit increase by 4.50% at 4 months, the same compensating intensity pattern we find among websites losing visitors.

*Heterogeneity of GDPR's Effect Across Website Characteristics*

The previous section showed that GDPR's impact varied widely across websites. We now characterize this variation using website size (Table W5 in Web Appendix A) and six additional pre-GDPR website characteristics (Figure 2). The six characteristics capture users' relationship to content (data-type sensitivity; hedonic versus utilitarian orientation), the website's dependence on data (ad reliance; tracking intensity), and its institutional context (regulatory strictness; DPA enforcement capability). We interpret all estimates as descriptive associations: pre-GDPR characteristics are not randomly assigned, so the splits identify where GDPR effects concentrate, not causal moderation. Full definitions and decompositions appear in Web Appendix H.

*Table 2. Summary of the Effect of GDPR on Total Usage (Panel A) and Its Decomposition into Usage Frequency and Usage Intensity (Panel B)*

**Panel A. Total usage across post-treatment periods**

| Metric | Statistic | Post-Treatment Period | | | | |
|---|---|---|---|---|---|---|
| | | 3 months | 6 months | 9 months | 12 months | 18 months |
| Visits | Mean | -4.88% | -7.22% | -9.07% | -9.57% | -10.02% |
| | Median | -3.49% | -5.54% | -7.54% | -8.24% | -8.91% |
| | Share sig. positive | 17.74% | 19.97% | 20.16% | 20.80% | 23.96% |
| | Share sig. negative | 32.20% | 42.81% | 50.48% | 53.25% | 54.87% |
| | Share not significant | 50.06% | 37.22% | 29.36% | 25.95% | 21.17% |
| Page Impressions | Mean | -3.12% | -4.83% | -6.33% | -6.48% | -9.28% |
| | Median | -2.75% | -3.92% | -5.44% | -6.04% | -9.29% |
| | Share sig. positive | 18.79% | 22.73% | 23.98% | 25.61% | 24.64% |
| | Share sig. negative | 27.74% | 35.83% | 42.23% | 44.90% | 51.44% |
| | Share not significant | 53.47% | 41.44% | 33.79% | 29.49% | 23.92% |

**Panel B. Decomposition into usage frequency and usage intensity**

| Component | Metric | Websites | 3 months | | | 18 months | | |
|---|---|---|---|---|---|---|---|---|
| | | | Share of Websites | Median Effect | Mean Effect | Share of Websites | Median Effect | Mean Effect |
| ***Visits = Unique Visitors × Visits per Unique Visitor*** | | | | | | | | |
| Usage Frequency | Unique Visitors | – | – | -1.24% | -0.77% | – | -6.65% | -6.61% |
| Usage Intensity | Visits per Unique Visitor | All treatment websites | 100.00% | -2.62% | -1.62% | 100.00% | -2.81% | -0.59% |
| | | Websites gaining unique visitors | 46.60% | -6.19% | -6.46% | 38.27% | -9.37% | -9.09% |
| | | Websites losing unique visitors | 53.40% | +1.23% | +2.56% | 61.73% | +1.39% | +4.77% |
| ***Page Impressions = Visits × Page Impressions per Visit*** | | | | | | | | |
| Usage Frequency | Visits | – | – | -3.49% | -4.88% | – | -8.91% | -10.02% |
| Usage Intensity | Page Impressions per Visit | All treatment websites | 100.00% | +0.56% | +1.97% | 100.00% | +0.28% | +2.15% |
| | | Websites gaining visits | 40.69% | -2.92% | -2.44% | 33.30% | -4.87% | -4.58% |
| | | Websites losing visits | 59.31% | +3.07% | +5.05% | 66.70% | +3.11% | +5.53% |

Notes: Panel A reports the effect of GDPR on total usage across the five post-treatment periods; Panel B decomposes the effect into usage frequency and usage intensity (Total Usage = Usage Frequency × Usage Intensity; Table W6 in Web Appendix B) for the 3- and 18-month periods. We estimate one GSC model per treated website-instance and aggregate into a website-level effect via a traffic-share-weighted average. Mean and median effects are computed across all treated websites. In Panel A, the shares are the fractions of websites with significantly positive, significantly negative, and no significant effect ($p < .05$). In Panel B, websites are classified as gaining or losing users (visits) by the sign of the website's estimated GSC treatment effect on the usage frequency metric; "Share of Websites" is the fraction of treated websites in each group. N = 5,790 treated websites for visits and page impressions (Tables W2 and W3 in Web Appendix A.2); unique-visitor-based metrics use the subsample of websites whose unique visitors are reported in every month of the observation period (5,447 treated websites; Table W2 in Web Appendix A.2). Because Panel B is multiplicative at the website level and the frequency metrics use the unique visitor subsample, the averaged percentages are not additive across the decomposition.

*Website size.* We divided websites into deciles by global, country, and industry ranks; Table W5 in Web Appendix A reports the industry-rank decile profile (global and country ranks yield similar results). These size patterns are descriptive: we cannot fully disentangle the resource- and compliance-based explanation from differential measurement quality across the size distribution, although the non-monotone profile and the near-absence of direct SimilarWeb integration among large websites leave little room for a measurement account (Web Appendix D).

The size effect is non-monotone. The smallest websites suffered by far the most: the bottom decile lost 21.49% of visits after 18 months. But the largest websites were not spared: the top decile declined by 9.04% at 18 months, more than the mid-sized deciles, which were least affected (deciles two through five: -5.8% to -7.3% at 18 months).

Re-estimating without the 1,000-weekly-visits threshold yields even more negative effects, confirming that low-traffic websites were hit hardest (Web Appendix C).

The vulnerability of small websites aligns with prior evidence of regulatory burdens concentrated among smaller players; only the decile partition reveals the upper tail, and the pattern is consistent with pressure toward increased market concentration (Campbell et al. 2015; Goldberg et al. 2024; Johnson et al. 2023; Peukert et al. 2022).

Users of larger, more valued websites have stronger reasons to continue, and larger websites more readily obtained consent (Campbell et al. 2015; Goldberg et al. 2024); that the largest websites nevertheless declined more is consistent with public scrutiny concentrating on the most prominent data collectors (see “Potential Mechanisms”).

Figure 2 compares websites with high versus low values of the six characteristics across the five post-treatment windows. Two patterns organize the results: the heterogeneity concerns magnitude rather than direction (both groups decline for every characteristic), and the gaps are persistent—none disappear by 18 months.

*Content*. The content characteristics provide the clearest signal. Hedonic websites declined substantially more than utilitarian websites from the outset, with a persistent gap (-15.2% versus -8.1% at 18 months). Websites in privacy-sensitive industries declined less across all horizons, and the gap widened (-7.3% versus -12.4% at 18 months). These patterns point to a privacy-calculus mechanism: users move away from discretionary, easily substitutable content while remaining with websites whose services justify the data relationship. The smaller decline among privacy-sensitive websites is consistent with a privacy-control channel (reassurance where data are most sensitive). Because data-type sensitivity is measured at the industry level, this pattern reflects a between-industry association (Web Appendix H).

*Business model*. The raw splits are counterintuitive: ad-light websites declined more than ad-heavy websites at every horizon (-10.6% versus -7.5% at 18 months), and low-tracking websites more than high-tracking websites (-9.7% versus -7.8% at 18 months). Rather than a simple exposure account, ad reliance and tracking intensity appear to proxy for commercial maturity: ad- and tracker-heavy websites are larger, more professionalized, and better able to implement consent and adapt. Conditional on size, EU-user share, and industry, the ad-reliance and tracking gaps attenuate, whereas the hedonic and sensitivity associations persist (Web Appendix H).

*Regulatory environment.* Websites facing stricter regulations declined somewhat more (-9.7% versus -7.7% at 18 months). Websites answering to better-resourced data-protection authorities were indistinguishable in the first months (-4.3% versus -4.1%), but a gap emerges as the decline deepens. Thus, institutional pressure gradually amplifies GDPR's effects rather than discontinuously upon enforcement.

Figure 2 concerns weekly visits; Web Appendix H repeats the analysis for the remaining metrics. The heterogeneity operates primarily through usage frequency: where a characteristic separates websites, it separates them by how many users they lose, not by how intensively remaining users engage. For example, hedonic websites lose 12.5% of their unique visitors within 18 months, compared with 4.0% for utilitarian websites. Usage-intensity effects are generally close to zero, with one exception: among hedonic websites, visits per unique visitor increase in the unconditional split; the pattern is not robust to conditioning or precision weighting, and we do not interpret it (Web Appendix H).

*Figure 2. Comparison of GDPR's Effect on Weekly Visits for Values of Six Website Characteristics*

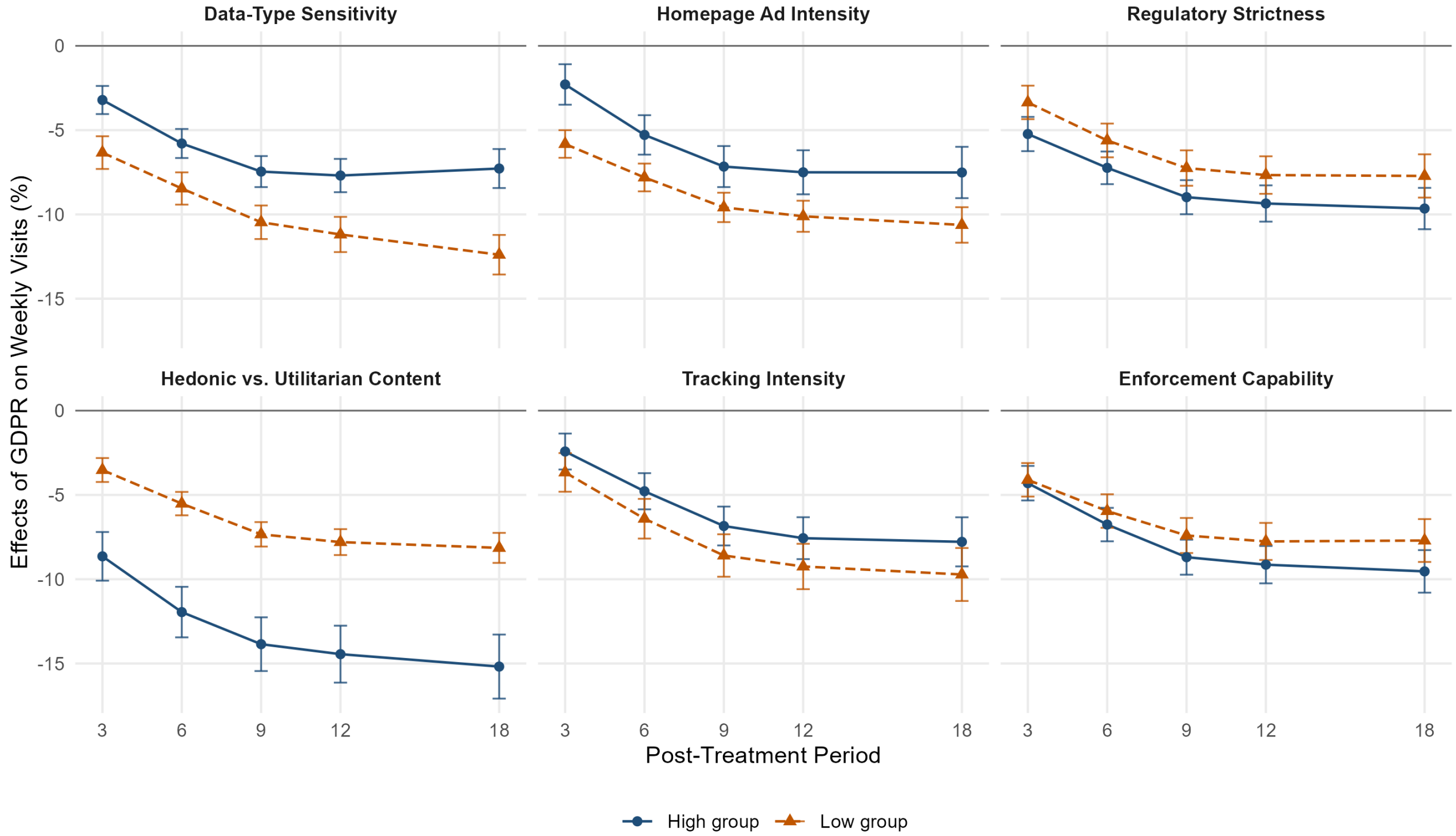

Notes: Each panel shows the average effect of GDPR on weekly visits, in percent, for websites with high versus low values of a single pre-GDPR website characteristic, across the five cumulative post-treatment periods. Effects are website-level GSC estimates; error bars = 95% confidence intervals. Websites are split at the median of each characteristic, except homepage ad intensity (top quartile) and content orientation (High = hedonic). The splits are descriptive, not causal. Group means equal the entries of Table W23 in Web Appendix H.

*Treatment intensity: EU-user share*. The most direct characteristic is a website's exposure to the regulation itself: the share of its pre-GDPR traffic from EU users. Because websites with different EU-user shares differ systematically in domicile, industry, and size, cross-website comparisons conflate exposure with composition. We therefore exploit variation within websites: for the 1,104 non-EU websites with both EU-user and non-EU-user traffic (Web Appendices G and E), we compare each website's EU traffic with its own non-EU traffic, so that any website-level effect affects both streams alike. Grading this contrast by EU-user share (276 websites per quartile) yields a strictly monotone exposure effect in the quartile medians: in group means, the lowest quartile shows no significant effect at 3 months (-0.9%), while the highest quartile loses 19.2%. Unlike the descriptive splits, this contrast has a causal interpretation: a confounding effect would have to differentially affect the same website's EU and non-EU traffic, in proportion to EU exposure (for the same geographic-source identification using comparable SimilarWeb data, see Congiu et al. 2022). Any spillover on non-EU traffic would shrink the gap, so the estimates are conservative (Robustness). In short, the more a website depended on EU users, the more usage it lost—providing strong evidence that the decline follows the regulation itself, rather than generic unobserved shocks hitting particular websites.

*Potential Mechanisms*

The evidence assembled so far—the aggregate decline and its deepening (Table 2, Panel A), the decomposition (Table 2, Panel B), heterogeneity (Figure 2), and the exposure gradient—allows us to assess the mechanisms (Table 1). We consider each in turn.

*Consent friction*. Websites responded to the regulation visibly and quickly: the share of websites displaying a cookie consent banner jumped by 23 percentage points at the enforcement of the GDPR (from 38.9% to 61.9%; Web Appendix J). Yet the friction these interfaces create explains little of the decline. The weeks around a website's own cookie consent banner adoption carry a usage cost on the order of one to two percent, an association we do not read causally, and privacy-policy adoption, which changes no interface, shows no effect at all, an informative placebo test. Bounce rates, the most direct behavioral trace of users turned away at the door, remain essentially flat (median change of -0.6 to -1.0 percentage points), while the number of bouncing visitors falls in step with total traffic: websites lost arrivals, not arrivals-who-bounce. Mobile-heavy websites, where consent dialogs are most intrusive, declined no more than desktop-oriented ones (equivalence tests; Web Appendix H). Pre- and post-GDPR adopters of consent infrastructure are statistically indistinguishable across all post-treatment periods (Web Appendix J). If consent friction drove the decline, usage should have dropped most sharply at enforcement and recovered as users habituated. We observe the opposite: no break at the enforcement cookie consent banner wave, and the decline deepens for another year. Consent friction is real, but it shaped the transition, not the trajectory.

*Restricted user acquisition*. The anatomy of the decline is consistent with restricted user acquisition: the loss runs through usage frequency; websites shed unique visitors, while the users who remain engage, if anything, more intensively; and every characteristic gap in Figure 2 operates through usage frequency rather than intensity (Web Appendix H). Together with the flat visit duration and the unchanged bounce rate (Web Appendix J), this pattern is difficult to reconcile with degraded retention or with consent friction, which would surface as users turned away at the door rather than as fewer arrivals. What the usage frequency pattern cannot do is separate restricted acquisition from privacy salience, because both operate by reducing the number of users. Distinguishing a mechanical loss of acquisition channels from users' deliberate re-sorting would require data linking traffic sources or individual consent choices to subsequent visits, which our data do not permit.

*Privacy salience.* Three patterns suggest that part of the response reflects users deliberately re-sorting their usage as the regulation made data relationships salient. First, the GDPR produced winners: by 18 months, roughly a quarter of websites show significantly positive effects, while the share without a detectable effect more than halves. Usage was reallocated rather than uniformly suppressed. Second, the sorting follows content: hedonic websites lose the most, privacy-sensitive industries lose the least, and the largest websites, the most scrutinized data collectors, decline more than mid-sized ones. Third, the timing tracks the enforcement regime: the decline deepens through the first enforcement wave, peaking in months 7-9 around the first major fine, whose association with usage operated EU-wide rather than only in France (Web Appendix J). The deepening at the fine is not a second wave of consent friction: consent-interface adoption concentrated at enforcement and rose only gradually thereafter, with no discrete jump around the fine (Web Appendix J). The salience at work is thus the sustained attention that enforcement and its coverage maintain over data practices, acting upstream on whether a user visits, rather than the momentary awareness resolved at the cookie consent banner. These patterns are consistent with a privacy-salience mechanism operating through the privacy calculus (Dinev and Hart 2006).

*Privacy control.* The positive side of the distribution bears on the privacy control effect (Table 1). The websites that saw usage increase over 18 months, the smaller declines in privacy-sensitive industries, and the rising intensity among remaining users are the patterns this effect predicts. Because privacy control and the reassuring branch of privacy salience both operate through trust, our website-level data cannot distinguish between the two channels.

*Retention.* We find little evidence supporting degraded retention as an important lasting mechanism: duration per visit is essentially unchanged (Web Appendix J), and pages per visit rise, most among websites losing users (Table 2, Panel B).

The persistence of the decline sharpens this picture. Websites' compliance activity largely normalized within months (Johnson et al. 2023), yet usage never recovered, which reflects a persistent effect rather than a temporary transition cost. We return to the acquisition-versus-salience limitation in the Discussion.

## *ROBUSTNESS OF RESULTS OF EMPIRICAL STUDY*

### *Identifying Assumptions and Robustness Tests*

The validity of the GSC estimator relies on several key identifying assumptions. Table W16 in Web Appendix G summarizes the assumptions, challenges, and the full set of tests; here, we highlight the most consequential.

*Donor Pool*. The donor pool must yield a good synthetic control group (Abadie et al. 2010; Abadie 2021). We match each treated website-instance to five industry peers by pre-treatment correlation, and confirm the effect is unchanged when we drop the industry requirement, match on EU-user share, or use ten controls instead of five (Web Appendix E.5).

*No Anticipation.* The treatment coincides with the GDPR's enforcement (May 25th, 2018). Because a few websites may have complied early or late (Perunicic 2018), we re-estimate after removing the 30 days before and after enforcement; both the short- and long-term effects on weekly visits are unchanged (Web Appendix I).

*No Interference (SUTVA).* The control group, non-EU websites' non-EU user traffic (Cell 4 in Figure 1), must be unaffected by the GDPR. The central threat is that non-EU websites voluntarily adopt uniform GDPR-compliant standards, thereby spilling over onto the control group.

First, a panel-differences estimator (Eichenbaum et al. 2024) uses each website as its own control (the same calendar week of the previous year), relying on no cross-website control group; it reproduces our main estimate to within 0.2 percentage points at 12 months (-9.76% versus -9.57% in Table 2; Web Appendix G.3).

Second, a within-website difference-in-differences comparing non-EU websites' EU-user traffic (Cell 3) with their own non-EU user traffic (Cell 4) yields slightly more negative effects (Web Appendix G.1), the opposite of what a contaminated control group would produce.

Further tests in Table W16—dose-response in pre-GDPR EU exposure and a regression of control-group traffic on it—are consistent with limited spillover (Web Appendix E.3). Any residual spillover would attenuate the measured gap, so our estimates are conservative. The converse threat—EU websites relocating to non-EU domains—is addressed with a stricter control group of confirmed non-EU-domain websites, again finding no significant difference (Web Appendix E.4).

*Unconfoundedness:* We finally assume the GDPR did not coincide with other major changes in confounding factors. We find no differential EU versus non-EU changes in internet speed, internet access, or device usage (Web Appendix F.1), and an inventory of contemporaneous privacy regulations and salience events (e.g., the CCPA) confirms that none binds the control group in a way confounded with the enforcement date (Web Appendix F.2). As in any non-experimental design, we cannot fully rule out unobserved contemporaneous factors.

*Estimator and Specification Robustness*

Table W16 in Web Appendix G reports five further estimator- and specification-robustness checks, all pointing the same way. The effect is unchanged under the original synthetic control estimator (Web Appendix G.2) and under a pooled GSC (common factor structure across all website-instances; within 3 percentage points at every time horizon; Web Appendix G.3). It is robust to functional form: OLS on log(Y), Poisson, and Negative Binomial (addressing the log(Y+1) transformation and structural zeros) all overlap the baseline, as does re-estimating at monthly rather than weekly resolution (Web Appendices G.4 and G.5). Finally, in-time placebo tests with fake enforcement dates from October 2017 to April 2018 produce effects indistinguishable from zero (Web Appendix G.6).

## *GENERAL DISCUSSION*

We analyzed the impact of the GDPR on online usage behavior, using data from 6,387 websites across 24 industries over 18 months post-GDPR. Covering over 2.6 trillion website visits, our data enabled a detailed exploration of the effect's heterogeneity and its development over time.

Overall, the GDPR reduced usage, and the effect deepened over time—weekly visits fell by 4.88% at 3 months and 10.02% at 18 months post-GDPR, with page impressions showing similar patterns—yet nearly a quarter of websites ultimately showed significant positive effects. The average decline itself confirms the direction documented in prior work (Goldberg et al. 2024; Lefrere et al. 2026); our contribution lies in decomposing that average and documenting its persistence. Decomposing total usage revealed opposing effects: usage frequency declined increasingly over time while the intensity decline moderated, and websites that gained (lost) users experienced decreasing (increasing) intensity—websites either struggle to attract users or to keep them engaged. Effects also varied systematically with the mechanisms' proxies: declines concentrated among hedonic websites and the smallest websites (consistent with increased market concentration), were attenuated where data are most sensitive, and deepened with enforcement, not with the GDPR compliance deadline.

Because usage metrics such as visits and page impressions are key revenue drivers for e-commerce and ad-financed websites (Zhou et al. 2025), the persistent usage decline points to the trade-off privacy laws bring to the fore: designed to strengthen privacy protection, they can also impose business costs that may ultimately be passed on to users.

*Theoretical Implications*

Together, our findings provide an important assessment of how privacy laws such as the GDPR alter user interactions with websites. Though we find an overall decline in usage frequency, the heterogeneity of the effects across websites suggests that rather than uniformly reducing their browsing, users may be reallocating some of their online activity, potentially spending more time on websites they trust or perceive as offering greater value-(for related evidence that GDPR transparency can positively affect consumer engagement and consent, see Godinho de Matos and Adjerid 2022). This potential reallocation is consistent with two usage frequency-effect channels—restricted user acquisition and enforcement-driven salience—while the evidence provides little support for consent friction and degraded retention as lasting drivers. Post-GDPR, users may also prefer websites where they have already "paid" a cost in data or granted consent (Arkes and Blumer 1985).

Our findings also underscore the importance of a long-term perspective: the effects intensified well into the second year. In contrast to Lefrere et al. (2026), who find a rebound in third-party cookie usage, we observe no rebound in total usage or usage frequency—the challenges may be longer-lasting than previously thought. Only usage intensity rebounds, suggesting that the remaining users engage more deeply with websites. That the decline deepens rather than fades contradicts a simple adaptation account and is consistent with enforcement-driven privacy salience.

*Managerial Implications*

Our findings carry implications for all stakeholders in the online ecosystem. Websites face persistent rather than transitory losses, concentrated among smaller players, and diverge in which effect they lose. So mitigation strategies could explore targeting the failing effect: potential acquisition remedies (e.g., rebuilding consented acquisition channels such as contextual advertising, owned media, and opt-in email) where users are lost; engagement remedies where usage intensity falls; and revenue diversification where ad-financed models come under pressure. Advertisers face smaller reachable audiences but can exploit heterogeneity by combining large, resilient websites with niche websites that exhibit higher average usage intensity and thus, potentially, more attentive remaining users. Users gained control and awareness of the data "price" of free services, purchased with some consent friction. Regulators, finally, should weigh the persistence and unevenness of the effects: one-size-fits-all rules risk disproportionately burdening smaller businesses as GDPR-style regulations spread.

*Limitations and Directions for Future Research*

Our results rest on the assumption that our control group was unaffected by GDPR. While we find substantial support for this assumption, particularly when analyzing data 18 months post-GDPR, a moderate spillover effect may have occurred during the earlier stages of GDPR implementation (within the first 3 months post-GDPR), particularly for websites with a high share of EU users. As a result, our estimates could be a lower bound of GDPR's true effect. Two potential biases run in opposite directions: GDPR-induced measurement error would overstate the decline, whereas control-group spillover would understate it, so they exert countervailing pressures on our net estimate (Web Appendix E.6).

Moreover, a potential source of measurement error arises from non-consenting users post-GDPR. The regulation allows users to decline consent for data collection, which could create discrepancies between actual and recorded online user behavior. Researchers have addressed this in various ways: Goldberg et al. (2024) use a selection model; Lefrere et al. (2026) rely on Alexa data gathered without cookies. We similarly use SimilarWeb, which aggregates its browser extension and first-party web analytics data and is designed to account for non-consenting users (Web Appendix D). SimilarWeb yields results similar to a selection model, and consent-independent AGOF ground-truth data provide corroborating evidence against a large measurement error (Web Appendix D).

Finally, our evidence for the mechanism is constrained by the website-level nature of our data. The decline in usage frequency is consistent with both restricted user acquisition and enforcement-driven privacy salience, and the two cannot be separated at the website level. Distinguishing them conclusively would require data linking traffic sources or individual users' consent choices to their subsequent browsing—for example, panel data on new versus returning visitors. Such data would also enable future work to directly test the selection logic behind the intensity offset. Our heterogeneity estimates, though mechanism-guided, remain descriptive associations rather than causal moderation effects.

# The Impact of the General Data Protection Regulation (GDPR) on Online Usage Behavior
# Web Appendix

## Table of Contents

# WEB APPENDIX A: Additional Tables and Figures

This appendix collects the tables and figures referenced in the main text.

## A.1 Comparison with Related Empirical Studies

*Table W1. Comparison of Empirical Studies on the Impact of GDPR on Online User Behavior*

| **Study** | **Goldberg et al. (2024)**<br>**American Economic Journal** | **Lefrere et al. (2026)**<br>**Management Science** | **This Study** |
|---|---|---|---|
| **Primary question** | Effect on recorded traffic and e-commerce revenue | Technical responses and downstream consequences for content providers | Effect on usage behavior and its allocation across websites and behavioral effects |
| **Primary outcomes** | Recorded page views and revenue | Tracking and consent responses; content quantity, quality, traffic, and engagement | Visits and page impressions decomposed into usage frequency and usage intensity |
| **Main finding** | Negative on average<br>(recorded page views: -11.7% at 4 months; real-effect bounds: 0.0% to -8.1%) | No detectable effect on most downstream outcomes (only pages per user: -3.9%) | Negative on average<br>(weekly visits: -4.88% at 3 months;<br>-10.02% at 18 months;<br>~24% of websites significantly positive) |
| **Primary decomposition** | Real outcomes vs. consent-related recording | No formal behavioral decomposition | Total usage =<br>usage frequency x usage intensity |
| **What the average effect conceals** | Recorded losses versus real economic losses | Technical adaptation despite largely unchanged downstream outcomes | Reallocation across websites and offsetting frequency-intensity effects |
| **Treatment-effect object** | Average and subgroup effects | Average and response-type effects | Website-specific effects; complete distribution and positive/negative/null shares |
| **Treatment-effect heterogeneity** | Size, referral channel, regulatory strictness | Advertising intensity, website rank, post-GDPR response type | By content (hedonic vs. utilitarian; data-type sensitivity), data reliance (ad intensity; tracking intensity), institutional context (regulatory strictness; enforcement capability); website size (deciles), and EU-user-share exposure (within-website), across total usage, usage frequency, and usage intensity |
| **Temporal analysis** | ~4 months | ~18 months<br>(short-vs. long-term comparison) | ~18 months in five cumulative windows (3/6/9/12/18) plus non-cumulative dynamics and enforcement timing |

| **Study** | **Goldberg et al. (2024)**<br>**American Economic Journal** | **Lefrere et al. (2026)**<br>**Management Science** | **This Study** |
|---|---|---|---|
| **Within-website separation of EU and non-EU traffic** | No | No | Yes |
| **Main method** | Panel Differences Estimator<br>Selection model | Two-way fixed-effects Difference-in-Differences Estimator<br>Callaway-Sant'Anna | Generalized Synthetic Control Estimator<br>(website-specific effects) |
| **Data source for user behavior metrics** | Adobe Analytics | Alexa | SimilarWeb,<br>AGOF |
| **Unit of analysis** | Website(dashboard)-week-level | Website-wave-level | Website-instance-week/month-level |
| **Observation period** | January – September 2017<br>January – September 2018<br>(18 months) | April 2017 – November 2019<br>(32 months) | July 2017 – December 2019<br>(30 months) |
| **Number of observations** | 69,344<br>website-weeks | 17,577 - 22,289<br>website-months | 1,278,298<br>(website-instance-weeks) |
| **Number of industries** | 5<br>(Cross-industry online firms) | 1<br>(News and media only) | 24<br>(Standardized industry categories) |
| **Number of countries** | 1 | 7 | 13 |
| **Number of websites** | 1,084 | 909 | 6,387 |

## A.2 Sample Composition and Summary Statistics

Tables W2 and W3 show the derivation of the final sample of website-instances and their distribution by base of users and websites.

*Table W2. Derivation of the Final Sample of Website-Instances*

| | **Website-Instances with EU users** | **Website-Instances with non-EU users** | **Website-Instances Total** |
|---|---|---|---|
| **Sample** of (non-unique) websites (top 1,000 websites of 11 EU countries, CH and US) | 13,000 | 13,000 | 26,000 |
| **Step 1: Sample after removal** of duplicated and non-existent websites (e.g., fraudulent pop-ups) | 6,460 | 7,332 | 13,792 |
| **Step 2: Sample after additional removal** of website-instances with average weekly visits <1,000 | 5,774 | 5,250 | 11,024 |
| **Step 3: Final sample after additional removal** of website-instances with zero visits in at least one month or strong outliers | **5,517** | **4,241** | **9,758** |
| **Final sample (only for unique visitor analysis)** after additional removal of website-instances with monthly unique visitors <5,000 in at least one month | **5,171** | **2,794** | **7,965** |

*Notes: SimilarWeb does not report unique visitor information for websites with < 5,000 unique visitors per month. Thus, the final sample for the unique visitor analysis contains only website-instances with ≥ 5,000 unique monthly visitors throughout the observation period.*

*Table W3. Distribution of Website-Instances by Base of Users and Websites*

| | | *Base of User* | | |
|---|---|---|---|---|
| | | **EU Users** | **Non-EU Users** | **Total** |
| *Base of Website* | EU Website | 3,641 *(Cell 1)* | 2,540 *(Cell 2)* | 6,181 *(3,914 websites)* |
| | Non-EU Website | 1,876 *(Cell 3)* | 1,701 *(Cell 4)* | 3,577 *(2,473 websites)* |
| | **Total** | 5,517 *(5,517 websites)* | 4,241 *(4,241 websites)* | **9,758** *(6,387 unique websites)* |

*Notes: Cell entries are website-instances; each marginal total reports the number of website-instances, with the corresponding number of unique websites in parentheses. The shaded Cells 1-3 (instances of an EU website, or EU users of a non-EU website) constitute the treatment group, to which the GDPR applies (8,057 website-instances; 5,790 unique websites). Cell 4 (non-EU users of a non-EU website) is the control group (1,701 website-instances; 1,701 unique websites). Observation period: July 2017-December 2019 (SimilarWeb).*

Figure W1 shows how the websites in our sample are distributed across industries.

*Figure W1. Distribution of Websites across Industries*

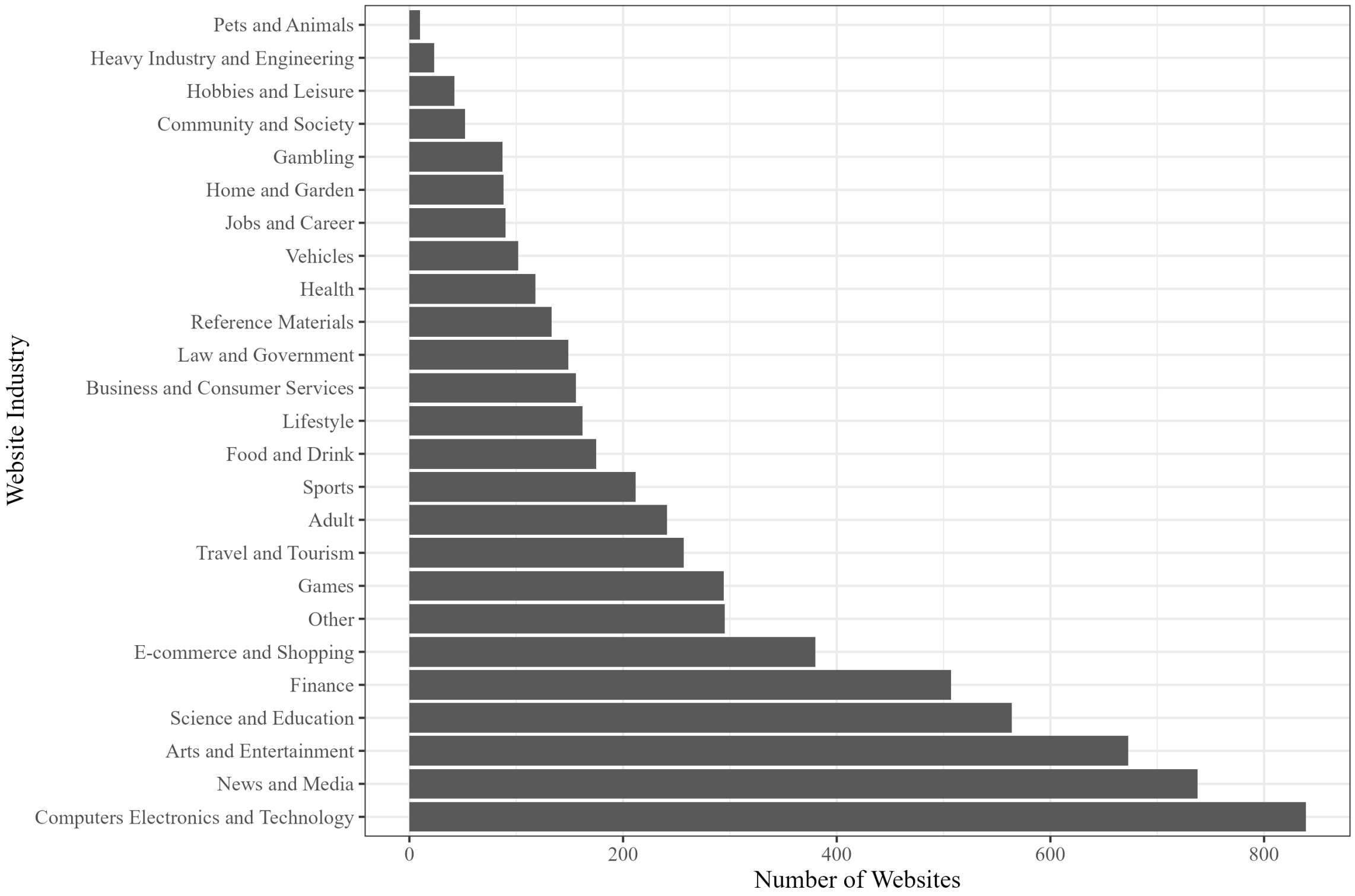

Notes: The figure reports the distribution of the N = 6,387 unique websites in the estimation sample across industries. Each website is assigned to a single industry; the top-level category of its SimilarWeb classification (e.g., "Arts and Entertainment" as named by SimilarWeb) is shown, along with a residual "Other" category for websites without a recorded classification (25 bars in total). Bars are ordered by the number of websites. Source: SimilarWeb.

We use the final sample, as defined in the "Overview of Data" section (Tables W2 and W3) to derive descriptive statistics. This final sample includes 9,758 website-instances on 6,387 unique websites with 8,057 treatment instances (Cells 1–3) and 1,701 control instances (Cell 4); observation period Jul 2017 – Dec 2019 (131 weeks; 30 months).

Panel A reports the outcome variables across their underlying observations: the 9,758 instances observed over 131 weeks yield 1,278,298 instance-week records (weekly metrics), and the 7,965 unique-visitor instances observed over 30 months yield 238,950 instance-month records.

Panel B is on the website-instance level (N = 9,758 instances; 7,965 for unique-visitor variables). Unique-visitor metrics use 7,965 instances with at least 5,000 unique visitors per month throughout the period.

*Table W4. Summary Statistics and Pairwise Correlations of Outcome Variables*

**Panel A. Outcome Variables**

| Variable | N | Mean | Median | SD | Min | P25 | P75 | Max | % Zero |
|---|---|---|---|---|---|---|---|---|---|
| Visits | 1,278,298 | 2,071,335 | 173,510 | 41,567,639 | 0 | 16,152 | 772,046 | 4,293,312,771 | 0.27% |
| Page Impressions | 1,278,298 | 15,781,940 | 740,184 | 445,857,265 | 0 | 60,759 | 3,555,225 | 48,307,793,429 | 0.27% |
| Unique Visitors | 238,950 | 2,092,592 | 432,520 | 8,524,314 | 5,000 | 96,909 | 1,619,987 | 400,282,117 | 0.00% |
| Bouncing Visits | 1,278,298 | 748,400 | 67,411 | 11,510,890 | 0 | 7,313 | 322,307 | 1,200,141,919 | 0.78% |
| Bounce Rate | 1,274,887 | 0.464 | 0.450 | 0.199 | 0.000 | 0.316 | 0.607 | 1.000 | 0.52% |
| Visits per Unique Visitor | 238,950 | 4.11 | 2.47 | 18.78 | 1.00 | 1.79 | 4.14 | 2270.46 | 0.00% |
| Page Impressions per Visit | 1,274,887 | 5.28 | 3.89 | 5.78 | 1.00 | 2.38 | 6.52 | 657.00 | 0.00% |

Notes: N = Number of observations = website-instances × time periods. Weekly metrics (visits, page impressions, bouncing visits): the 9,758 website-instances of Tables W2 and W3 × the panel's 131 weeks = 1,278,298 instance-week records. Unique-visitor metrics (unique visitors, visits-per-unique-visitor): the 7,965 instances with ≥ 5,000 unique visitors per month throughout the observation period × 30 months = 238,950 instance-month records. Bounce rate (= bouncing visits / visits) and page-impressions-per-visit (= page impressions / visits) are ratios with visits in the denominator, so they are undefined in the 3,411 instance-weeks with zero visits; their N is therefore 1,274,887 instead of 1,278,298. Bounce rate is a 0–1 share. "% Zero" is the share of observations with values equal to zero.

**Panel B. Correlation Matrix of Outcome Variables (log(Y + 1))**

| Variable | Visits | Page Impressions | Unique Visitors | Bouncing Visits | Bounce Rate | Visits per Unique Visitor | Page Impressions per Visit |
|---|---|---|---|---|---|---|---|
| Visits | 1.00 | | | | | | |
| Page Impressions | 0.97 | 1.00 | | | | | |
| Unique Visitors | 0.95 | 0.87 | 1.00 | | | | |
| Bouncing Visits | 0.98 | 0.92 | 0.96 | 1.00 | | | |
| Bounce Rate | -0.15 | -0.34 | 0.03 | 0.02 | 1.00 | | |
| Visits per Unique Visitor | 0.25 | 0.33 | -0.08 | 0.14 | -0.47 | 1.00 | |
| Page Impressions per Visit | 0.10 | 0.35 | -0.04 | -0.04 | -0.79 | 0.33 | 1.00 |

Notes: Pearson correlations of log(Y + 1)-transformed outcome metrics, the same transformation used in the regressions. Computed across website-instances, pairwise-complete: N = 9,758 for pairs of fully available metrics, and N = 7,965 for pairs involving unique visitors or visits-per-unique-visitor.

## A.3 Distribution and Significance of GDPR's Effect over Time

Figure W2 plots the average logged weekly visits series for the treatment and control groups before and after the GDPR's enforcement.

*Figure W2. Average (Logged) Number of Weekly Visits between Treatment and Control Group and Pre- and Post-GDPR Period*

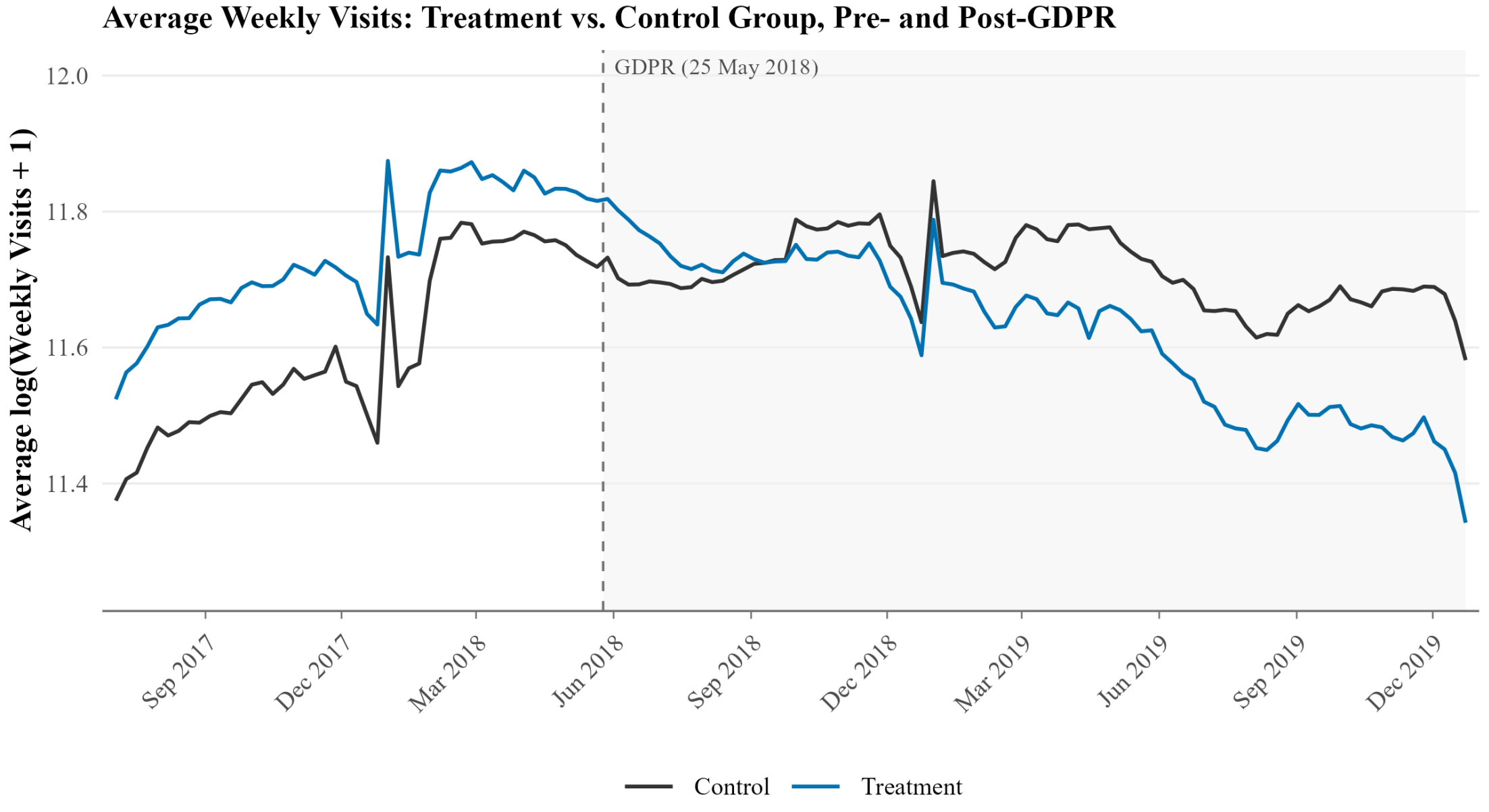

Notes: The blue line is the treatment group (EU website-instances, Cells 1-3) and the black line is the control group (Cell 4); each plot shows the average log(weekly visits + 1) across instances by week. The dashed line marks the GDPR's enforcement date (May 25, 2018), and the shaded region marks the post-GDPR period.

Figure W3 summarizes the website-level GDPR effect for weekly visits, unique visitors, and page impressions. Panel A shows the distribution of the effect over the five post-treatment periods; Panel B shows the share of websites with a statistically significant positive or negative effect.

*Figure W3. Distribution and Significance of the GDPR Effect over Time*

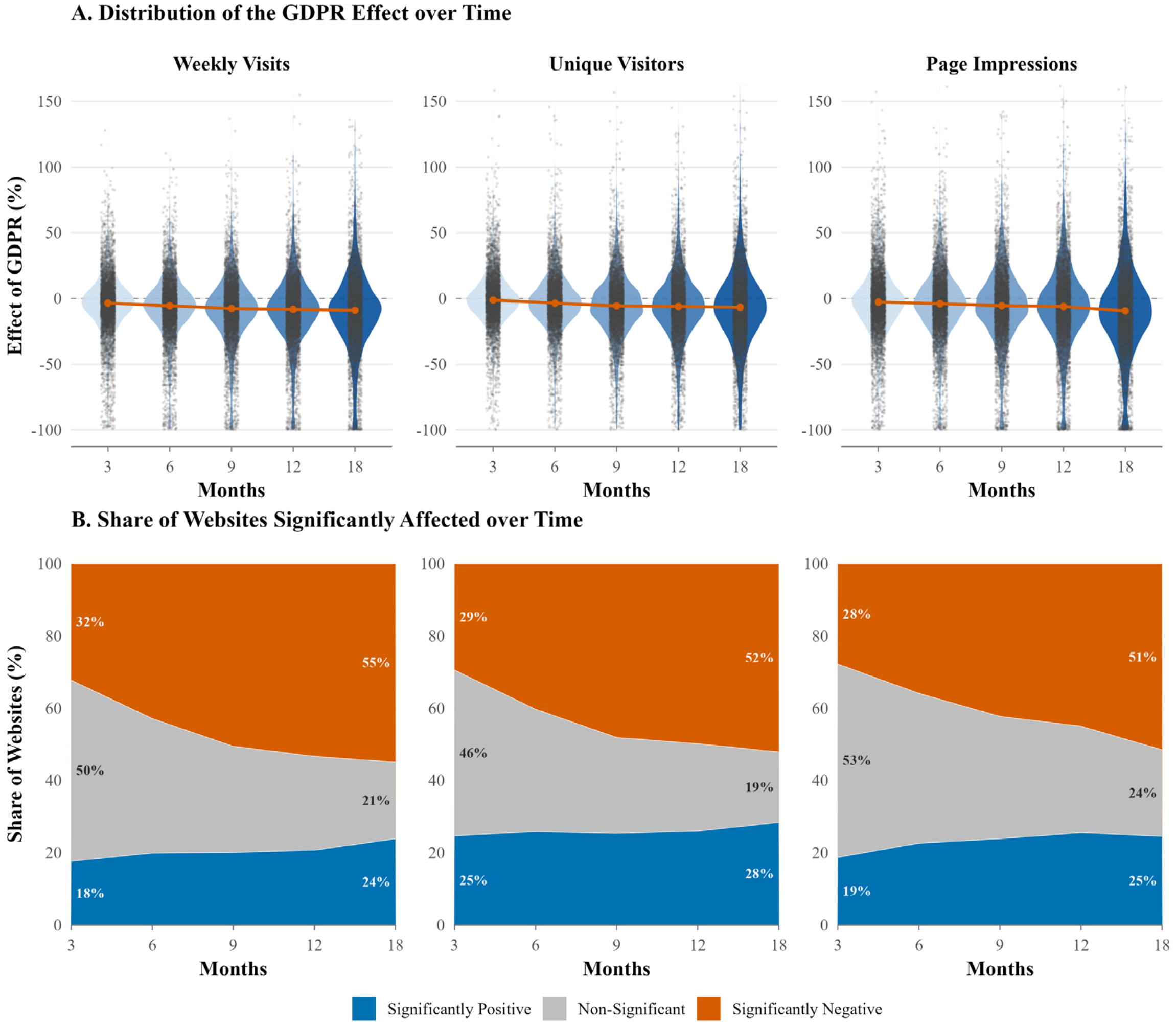

Notes: Each website's GDPR effect is its website-specific Generalized Synthetic Control estimate, expressed as a percentage change, for the five cumulative post-treatment periods (3, 6, 9, 12, and 18 months). Panel A shows the distribution of these website-level effects: each violin represents the effect distribution in a given period (the vertical axis is truncated for readability), the points are individual websites, and the orange line connects the period medians. Panel B classifies websites into three groups relative to all websites: the share of significantly negatively affected websites in orange is the number of websites statistically significantly negatively affected by GDPR ($p < .05$) relative to all websites; the share of significantly positively affected websites in blue depicts the number of websites statistically significantly positively affected by GDPR ($p < .05$) relative to all websites; the remaining websites in gray are not statistically significantly affected ($p \geq .05$).

## A.4 Distribution of GDPR's Effect across Website Size

Table W5 reports the GDPR's average effect on weekly visits across deciles of industry rank, a proxy for website size.

*Table W5. Distribution of GDPR's Effect on Weekly Visits across Deciles of Industry Rank*

| | **Post-Treatment Period** | | | | |
|---|---|---|---|---|---|
| **Industry Rank Decile** | **3 months** | **6 months** | **9 months** | **12 months** | **18 months** |
| 1 | -3.74% | -5.60% | -7.56% | -8.50% | -9.04% |
| 2 | -2.02% | -3.78% | -5.43% | -6.06% | -6.49% |
| 3 | -2.26% | -3.70% | -5.15% | -5.63% | -5.82% |
| 4 | -2.96% | -4.76% | -5.89% | -6.33% | -6.55% |
| 5 | -1.82% | -4.08% | -5.86% | -6.57% | -7.25% |
| 6 | -4.30% | -6.46% | -8.64% | -9.37% | -10.39% |
| 7 | -6.23% | -8.89% | -10.89% | -11.29% | -11.51% |
| 8 | -6.21% | -8.47% | -10.29% | -10.63% | -11.31% |
| 9 | -5.41% | -8.22% | -10.62% | -10.74% | -10.31% |
| 10 | -13.89% | -18.29% | -20.38% | -20.55% | -21.49% |
| Top 50% | -2.56% | -4.38% | -5.98% | -6.62% | -7.03% |
| Bottom 50% | -7.21% | -10.07% | -12.16% | -12.52% | -13.00% |

Notes: This table shows the distribution of the average effect of GDPR (in %) on weekly visits per website across deciles of industry rank and post-treatment periods. Industry rank is SimilarWeb's traffic-based rank of a website within its industry category (24 categories, e.g., News and Media), measured before GDPR enforcement. We rank websites within their industry so that the size deciles compare each website with its industry peers and are not driven by differences in typical traffic levels across industries; deciles based on global and country rank yield similar results. The 10% largest websites pre-GDPR (i.e., the ones with the lowest rank numbers) are part of the 1$^{st}$ decile, while the 10% smallest websites pre-GDPR are part of the 10$^{th}$ decile. N = 5,790 treated websites (~579 per decile).

## WEB APPENDIX B: Derivation of Usage-Intensity Metrics

Every usage intensity metric is the ratio of a total usage metric and a usage frequency metric (Table W6).

*Table W6. Relationship between Total Usage, Usage Frequency, and Usage Intensity Metrics*

| Total Usage Metric | Usage Frequency Metric | Usage Intensity Metric |
|---|---|---|
| **Weekly Visits** | **Monthly Unique Visitors** | **Visits per Unique Visitor** |
| *Number of visits to a website in a week* (Measure of website's traffic volume) | *Number of unique users visiting a website in a month* (Measure of website's reach) | *Average number of visits per unique visitor* |
| **Weekly Page Impressions** | **Weekly Visits** | **Page Impressions per Visit** |
| *Number of pages visited on a website in a week by the entire user base* (Measure of website's ability to spark engagement) | *Number of visits to a website in a week* (Measure of website's traffic volume) | *Average number of pages viewed per visit* |

For example, we calculate in Equation (W1) the visits per unique visitor on a website w by dividing the visits on website w by the unique visitors on website w; page impressions per visit follow the same logic in Equation (W2):

$$\text{(W1)}\ \ \text{Visits per Unique Visitor}_\text{w} = \frac{\text{Visits}_\text{w}}{\text{Unique Visitors}_\text{w}}$$

$$\text{(W2)}\ \text{Page Impressions per Visit}_\text{w} = \frac{\text{Page Impressions}_\text{w}}{\text{Visits}_\text{w}}$$

We use the derivation for the estimated GDPR effects instead of using the underlying observations to estimate it. Thus, we estimate each of the two effects (i.e., total usage and usage frequency) separately, at its own temporal resolution. Then, we apply Equations (W1) and (W2) to those estimates. We estimate the effects on visits and on page impressions on the weekly panel of 9,758 website-instances (5,517 with EU users and 4,241 with non-EU users) belonging to 6,387 unique websites, and the effect on unique visitors on the monthly panel of 7,965 website-instances (5,171 with EU users and 2,794 with non-EU users), because SimilarWeb reports unique visitors only for websites with at least 5,000 unique visitors per month (Table W2); the observation period is July 2017 to December 2019 throughout. Therefore we formed Equation (W12) for the 7,965 website-instances for which both input effects are available, and the corresponding expression for page impressions per visit for the 9,758 website-instances for which both input effects are available. Because we measure the effect on the log scale, it is invariant to whether the underlying outcome is aggregated weekly or monthly, so combining effects estimated at different resolutions requires no re-aggregation of the raw data.

This appendix derives the usage-intensity treatment-effect formulas from these definitions. We use Equations (W1) and (W2) to examine GDPR's effect on the usage intensity metrics in the following manner:

1. We use Equation (2) in the main text to calculate GDPR's effect (Δ) for the total usage and usage frequency metrics. Our methodology to derive this effect, Δ, captures the relative changes of each metric for each website *w* and each post-treatment period *p*.

2. We then incorporate GDPR's effect, Δ, for all the metrics for each post-treatment period *p* in Equations (W1) and (W2). As we capture the relative changes for the metrics, we can incorporate GDPR's effect, Δ, for each post-treatment period *p* for each website *w* by multiplying the metrics' baseline (counterfactual) values with $(1 + \Delta_{p,w})$. Because the same baseline enters the numerator and denominator, it cancels in the final expression, so the derived effect does not depend on which baseline we use.

3. Finally, we rearrange the equations to isolate GDPR's effect, Δ, on the usage intensity metric of interest.

We now use the described process to derive GDPR's effect on "Visits per Unique Visitor". We derive the "Page Impressions per Visit" accordingly. We begin by incorporating GDPR's effect, Δ, for post-treatment period *p* for website *w* for all metrics within Equation (W1):

$$\text{(W3) } \text{Visits}_{p,w} = \text{Visits}_{w} * \left(1 + \Delta\,\text{Visits}_{p,w}\right)$$

$$\text{(W4) } \text{Unique Visitors}_{p,w} = \text{Unique Visitors}_{w} * \left(1 + \Delta\,\text{Unique Visitors}_{p,w}\right)$$

$$\text{(W5) } \text{Visits per Unique Visitor}_{p,w} = \text{Visits per Unique Visitor}_{w} * \left(1 + \Delta\,\text{Visits per Unique Visitor}_{p,w}\right)$$

Incorporating GDPR's effect Δ (Equations (W3)-(W5)) into Equation (W1) yields:

$$\text{(W6) } \text{Visits per Unique Visitor}_{p,w} = \frac{\text{Visits}_{p,w}}{\text{Unique Visitors}_{p,w}}$$

$$\text{(W7) } \text{Visits per Unique Visitor}_{w} * (1 + \Delta\,\text{Visits per Unique Visitor}_{p,w}) = \frac{\text{Visits}_{w} * (1 + \Delta\,\text{Visits}_{p,w})}{\text{Unique Visitors}_{w} * (1 + \Delta\,\text{Unique Visitors}_{p,w})}$$

Rearranging Equation (W7) leads to:

$$\text{(W8) } 1 + \Delta\,\text{Visits per Unique Visitor}_{p,w} = \frac{\text{Visits}_{w} * (1 + \Delta\,\text{Visits}_{p,w})}{\text{Unique Visitors}_{w} * \left(1 + \Delta\,\text{Unique Visitors}_{p,w}\right) * \text{Visits per Unique Visitor}_{w}}$$

$$\text{(W9) } \Delta\,\text{Visits per Unique Visitor}_{p,w} = \frac{\text{Visits}_{w} * (1 + \Delta\,\text{Visits}_{p,w})}{\text{Unique Visitors}_{w} * \left(1 + \Delta\,\text{Unique Visitors}_{p,w}\right) * \text{Visits per Unique Visitor}_{w}} - 1$$

$$\text{(W10) } \Delta\,\text{Visits per Unique Visitor}_{p,w} = \frac{\text{Visits}_{w}}{\text{Unique Visitors}_{w}} * \frac{1}{\text{Visits per Unique Visitor}_{w}} * \frac{1 + \Delta\,\text{Visits}_{p,w}}{1 + \Delta\,\text{Unique Visitors}_{p,w}} - 1$$

We use the definition in Equation (W1) to replace the ratio of visits to unique visitors in Equation (W10) by $\text{Visits per Unique Visitor}_{\text{w}}$, which yields:

$$\text{(W11) } \Delta\,\text{Visits per Unique Visitor}_{\text{p,w}} = \text{Visits per Unique Visitor}_{\text{w}} * \frac{1}{\text{Visits per Unique Visitor}_{\text{w}}} * \frac{1 + \Delta\,\text{Visits}_{\text{p,w}}}{1 + \Delta\,\text{Unique Visitors}_{\text{p,w}}} - 1$$

Finally, rearranging Equation (W11) yields:

$$\text{(W12) } \Delta\,\text{Visits per Unique Visitor}_{\text{p,w}} = \frac{1 + \Delta\,\text{Visits}_{\text{p,w}}}{1 + \Delta\,\text{Unique Visitors}_{\text{p,w}}} - 1$$

In Equation (W12), we use the estimated GDPR effects on the corresponding total usage and usage frequency metrics, i.e., visits and unique visitors, to determine GDPR's effect on the visits per unique visitor. Equation (W12) corresponds to the visits-per-unique-visitor effect formula used in the main text: applying the same derivation to Equation (W2) yields GDPR's effect on page impressions per visit as used in the main text.

We use these equations to determine GDPR's effect on the visits per unique visitor and page impressions per visit for each website and each post-treatment period *p* (i.e., a cumulative effect covering 3, 6, 9, 12, and 18 months post-GDPR's enforcement). We apply the same derivation to any ratio of two metrics whose GDPR effects we estimated separately; we use it in Web Appendix J to construct the effects on the bounce rate from the effects on bouncing visits and weekly visits.

## WEB APPENDIX C: Robustness Check Concerning Website Selection

We remove from our main sample all website-instances with fewer than 1,000 average weekly visits (Step 2 in Table W2). Because this threshold could affect our estimates, we recompute GDPR's effect on weekly visits for a sample without this removal, i.e., one for which we perform only Steps 1 and 3. Removing the threshold makes GDPR's effect more negative in every post-enforcement period; the difference is statistically significant ($p < .05$) in all periods except at three months: at 18 months, the reduction in weekly visits deepens from 10.02% in our original sample to 11.80% without the threshold. Because low-trafficked websites experience stronger negative effects than more-trafficked ones, our main results are a conservative estimate of GDPR's effect on website selection.

*Table W7. Summary of GDPR's Effect on Weekly Visits with and without Using the Threshold for Website Selection of 1,000 Average Weekly Visits*

| | | **Post-Treatment Period** | | | | |
|---|---|---|---|---|---|---|
| | | **3 months** | **6 months** | **9 months** | **12 months** | **18 months** |
| **Original Sample Obtained with Selection Threshold (N=9,758 website-instances)** | Mean: | -4.88% | -7.22% | -9.07% | -9.57% | -10.02% |
| | Median: | -3.49% | -5.54% | -7.54% | -8.24% | -8.91% |
| **New Sample Obtained without Selection Threshold (N=11,760 website-instances)** | Mean: | -5.77% | -8.51% | -10.66% | -11.31% | -11.80% |
| | Median: | -4.02% | -6.03% | -8.29% | -8.92% | -9.90% |
| **p-value of t-test:** | | 0.070 | 0.009** | 0.002** | 0.001*** | 0.004** |

Notes: The original sample is obtained after the removal of duplicate and non-existent websites (Step 1 in Table W2), website-instances with average weekly visits < 1,000 (Step 2), and website-instances with zero visits in at least one month or strong outliers (Step 3). The new sample without a selection threshold is obtained by performing only Steps 1 and 3 but omitting Step 2 in Table W2. N denotes website-instances. Significance level: *** 0.1%-level ** 1%-level * 5%-level.

## WEB APPENDIX D: Robustness Checks Concerning SimilarWeb Data

The GDPR could introduce measurement error if SimilarWeb's recorded behavior diverges from actual behavior post-GDPR through either of two channels. On the user side, a rising share of users declining to consent to tracking would widen the gap between actual and recorded behavior. On the data-generating side, previous research documented that GDPR changed the third-party trackers and first-party analytics that servers deploy and share (Peukert et al. 2022; Miller et al. 2026; Johnson et al. 2023; Aridor et al. 2023), which could alter what SimilarWeb records independently of any change in user behavior. We acknowledge this possibility.

It is worth being explicit about what such an error would imply. Because the data-sharing changes are themselves GDPR-induced, they would concentrate among treated website-instances while largely sparing the control group that anchors our counterfactual; undercounting treated websites' post-GDPR traffic would then make their measured decline exceed their true decline. Any such error would therefore bias us toward overstating the GDPR effect, so our estimates are best read as an upper bound on the true behavioral response.

The relevant question is whether our measurement accounts for the entire observed decline or only part of it. We address it in two steps: a conceptual discussion of SimilarWeb's four data channels, and an empirical validation against an independent second provider, AGOF. Both indicate that measurement change is unlikely to account for our findings, though we cannot rule it out entirely.

### D.1 Conceptual Discussion of Measurement Error in SimilarWeb Data

SimilarWeb (www.similarweb.com), our primary source, provides standardized usage metrics for the 6,387 websites across 24 industries, summarized in Table W2 and Figure W1 in Web Appendix A. SimilarWeb states in writing that "GDPR had no impact on SimilarWeb data" and describes removing personally identifiable information at the source, collecting no IP data, and using no metric, including unique visits, that requires personal data. As an interested party, SimilarWeb's self-description cannot, on its own, establish that its data are unaffected; we report it to characterize the data-generating process and rely on the AGOF validation below as the actual test.

SimilarWeb's data combines four sources with differing GDPR exposure. Direct measurement, publishers sharing first-party analytics such as Google Analytics, is the channel most exposed to GDPR, as some publishers may curtail or stop sharing first-party analytics with SimilarWeb, plausibly more so among larger publishers facing greater regulatory scrutiny or smaller firms lacking compliance resources. We cannot verify the extent to which such tools operate cookielessly or how SimilarWeb imputes missing observations, so we bound this channel empirically through AGOF comparison and the direct-integration check reported in the conclusion.

Its browser extension panel (about 1 million users since August 2017) predates GDPR and offers users website statistics in exchange for anonymized traffic; this data source is governed by extension-install consent rather than per-website tracker consent and is largely unaffected by the GDPR. Crucially, this panel collects data directly from users' devices and is structurally analogous to the panel-based, user-side measurement infrastructure used in other studies on the GDPR (e.g., Lefrere et al. 2026, who rely on Alexa's Toolbar and over 25,000 browser extensions); it is therefore not subject to the GDPR-induced changes in server-side tracking documented above. SimilarWeb's data are thus a mix of server-side and user-side measurement, with the user-side channel providing coverage of post-GDPR behavior independent of the server-side concern. We cannot observe how heavily SimilarWeb weights this channel, so we do not claim it alone suffices. A website that withdraws from direct integration nonetheless remains observable through other data sources, and differential exits would have to occur simultaneously across multiple independent channels to change our qualitative conclusions.

Partnership data (e.g., Internet Service Providers (ISPs), measurement firms, and demand-side platforms (DSPs)) may be subject to GDPR and are the channel our AGOF comparison directly tests. Public data extraction is exempt under Recital 162 for statistical purposes.

Consistent with this argument, some EU data protection authorities (DPAs) have signaled openness to trackers that use only the website's own (first-party) data. The Irish Data Protection Commission (2020) clarified it "is unlikely that first-party analytics cookies would be considered a priority for enforcement action by the DPC," and the French and German DPAs noted that some trackers may be exempted from consent under specific conditions, such as trackers intended to generate traffic statistics (Commission Nationale de l'Informatique et des Libertés 2020; Datenschutzkonferenz 2021; Baden-Württemberg DPA 2022). Because interpretation and enforcement of the opt-in requirement differ across websites and member states, online tracking to obtain traffic statistics, as SimilarWeb and its data providers do, is much more feasible without large measurement error than other forms of tracking.

Finally, SimilarWeb's July 2019 methodology update was applied retroactively and consistently across our study period and does not coincide with GDPR's enforcement. However, as we lack access to raw data, we cannot fully verify these claims; instead, we validate them empirically against AGOF.

### D.2 Empirical Comparison of SimilarWeb and Independent AGOF Benchmark Data

To evaluate the validity of SimilarWeb data, we obtained data from the German Association of Online Research (AGOF, www.agof.de) for 72 websites in 2018. AGOF generates anonymous aggregate page-impression statistics and, for unique users, processes personal data under legitimate interest (Article 6(1)(f) of the GDPR); a user therefore defaults to being tracked and can escape only by deleting cookies or opting out, which prior evidence suggests few users do. Johnson et al. (2020) find opt-out rates for targeted advertising averaging 0.23% of all ad impressions, with similar figures in the US (0.23%) and Europe (0.26%). Such deletion would affect only the measurement of unique users, not page impressions, visits, or duration. AGOF's measurement is therefore structurally not susceptible to GDPR-induced consent effects: it measures page impressions anonymously, without processing personal data, which places this measurement outside the GDPR's jurisdiction (Article 2 GDPR), and it measures unique users under legitimate interest (Article 6(1)(f)), confirmed in writing by AGOF and its measurement provider INFOnline, where opt-out is negligible.

Externally certified by the German Media Analysis Working Group (AGMA) and the German Audit Bureau of Circulation (IVW), AGOF is the industry-standard currency for German web traffic. This characteristic makes it a strong independent benchmark for detecting whether SimilarWeb's measurement shifted at GDPR beyond genuine behavior change, even though it is not device-level user-side data. Its weekly page impressions and monthly unique users are directly comparable to SimilarWeb's weekly page impressions and monthly unique visitors. Because AGOF tracks global traffic, whereas these SimilarWeb website-instances track only German behavior, we restricted the comparison to the 65 websites with a German traffic share above 80% (obtained from SimilarWeb in April 2022); AGOF data are available only for Germany, so we cannot perform the same test elsewhere.

We conducted three sets of tests to compare the two data sources pre- and post-GDPR: a raw-data comparison of correlations and level differences, a before-after regression, and a GSC estimation. The raw-data comparison examined correlations and differences between page impressions and unique visitors in AGOF and SimilarWeb data before and after the GDPR; we observed no statistically significant differences, consistent with SimilarWeb tracking AGOF similarly before and after GDPR.

The before-after analysis used regression without a control group (since AGOF only includes German websites) to examine the effect of the "Postperiod" dummy: the post-period effects were similar in sign and magnitude for both data sources for page impressions (aggregate analysis: -4.90% AGOF vs. -6.10% SimilarWeb) and unique visitors (aggregate analysis: -2.20% AGOF vs. -1.50% SimilarWeb), differing by 1.2 and 0.7 percentage points, respectively. The GDPR effect estimation using the GSC estimator applied the Generalized Synthetic Control estimator to page impressions for both data sources from January to December 2018, obtaining 3-month and 6-month effects; the estimated GDPR effect was not statistically distinguishable between the two data sources for page impressions 3 months post-GDPR (AGOF: -3.47% vs. SimilarWeb: -1.30%) or 6 months post-GDPR (AGOF: -1.99% vs. SimilarWeb: -1.03%). The point estimates differed by 2.17 and 0.96 percentage points; in both comparisons, the SimilarWeb effect is smaller in magnitude than the AGOF benchmark, so we found no sign of the overstatement that the upper-bound reading allows for. We could not repeat this analysis for unique visitors because of insufficient pre-treatment observations.

For the 65 websites in the comparison sample, the aggregate correlation between AGOF and SimilarWeb page impressions is 0.92 pre-GDPR and 0.92 post-GDPR; for unique visitors, it rises slightly from 0.92 to 0.95 (Table W8). The average per-website correlations are likewise stable (page impressions: 0.62 pre- vs. 0.61 post-GDPR, t = 0.16, p = 0.88; unique visitors: 0.71 vs. 0.72, t = -0.23, p = 0.82), as are the median per-website correlations (page impressions: 0.70 vs. 0.70; unique visitors: 0.85 vs. 0.83). The relevant question for our purposes is not the level of these correlations (which reflects idiosyncratic website-level noise) but whether the relationship shifts at GDPR. The per-website correlations show no detectable change at enforcement. The one aggregate series that moves, unique visitors, from 0.92 to 0.95, moves toward closer agreement between the two sources, the opposite of what GDPR-induced measurement divergence would produce.

*Table W8. Correlation Between AGOF and SimilarWeb Data Pre- and Post-GDPR*

| Basis of Correlation | Number of Websites | Analyzed Variable | Number of Observations | | Pearson Correlation Coefficient | |
|---|---|---|---|---|---|---|
| | | | Pre-GDPR | Post-GDPR | Pre-GDPR (95%-CI) | Post-GDPR (95%-CI) |
| Aggregate correlation across websites | 65 | Page Impressions | 1,357 | 1,940 | 0.92 (0.91, 0.93) | 0.92 (0.91, 0.92) |
| Average correlation per website | 65 | Page Impressions | 1,357 | 1,940 | 0.62[a] | 0.61[a] |
| Median correlation per website | 65 | Page Impressions | 1,357 | 1,940 | 0.70[b] | 0.70[b] |
| Aggregate correlation across websites | 65 | Unique Visitors | 323 | 453 | 0.92 (0.91, 0.94) | 0.95 (0.94, 0.96) |
| Average correlation per website | 65 | Unique Visitors | 323 | 453 | 0.71[a] | 0.72[a] |
| Median correlation per website | 65 | Unique Visitors | 323 | 453 | 0.85[b] | 0.83[b] |

Notes: Number of observations for page impressions before (1,357) and after (1,940) GDPR enforcement = 3,297 [= (65 websites × 51 weeks) – 18]. For one website (fupa.net), we only observed AGOF and SimilarWeb data for 33 weeks (i.e., 18 weeks were missing). Number of observations for Unique Visitors before (323) and after (453) GDPR enforcement = 776 [= (65 websites × 12 months) – 4]. For one website (fupa.net), we only observed AGOF and SimilarWeb data for 8 months (i.e., 4 months were missing). [a] We report the average of the correlation per website without a 95%-Confidence Interval (95%-CI). A two-sample t-test indicated no significant statistical difference between the average correlation pre- and post-GDPR enforcement for page impressions (t = 0.16, p = 0.88) and unique visitors (t = -0.23, p = 0.82). [b] We report the median of the correlation per website without a 95%-Confidence Interval.

The level differences between the two data sources show no significant shift at GDPR's enforcement, though these level tests are noisy; we therefore rely more on the log-log regression below. The average absolute difference per website shows no significant change for page impressions (21,629,687 pre- vs. 21,771,359 post-GDPR; t = 0.01, p = 0.99) or unique visitors (587,495 vs. 450,271; t = -0.39, p = 0.70), and log-differences and relative differences yield similar results. A regression of log(1 + SimilarWeb) on log(1 + AGOF) and a post-GDPR indicator confirms this: the post-GDPR coefficient is small and not statistically significant for page impressions (-0.03, p = 0.16; N = 3,297 website-weeks) and for unique visitors (0.00, p = 0.89; N = 776 website-months), indicating no structural break in the SimilarWeb–AGOF relationship. At the website level, the before-after effects are also comparable (page impressions: -4.48% AGOF vs. -5.83% SimilarWeb; unique visitors: -1.58% vs. -1.10%; two-sample t-test p = 0.71).

### D.3 Summary and Conclusion

Together, the conceptual discussion and the empirical comparison with AGOF suggest that measurement change is unlikely to be large enough to drive our results, though we cannot rule it out. SimilarWeb is widely used in academic research, including for GDPR-related questions (e.g., Calzada and Gil 2020; Lu et al. 2020; Congiu et al. 2022; Peukert et al. 2022, Appendix A2.2, pages 4-5). Lefrere et al. (2026) reach comparable conclusions using an analogous third-party estimate from Alexa rather than SimilarWeb. Such precedent motivates but does not by itself establish validity in our setting, which is why we rely on the AGOF comparison and the direct test below.

As a direct test of the mechanism (i.e., publishers ceasing to share first-party analytics with SimilarWeb), we examined the status of direct integration on SimilarWeb's website pages and in Wayback Machine archives covering the GDPR window. Across our 6,387 websites, only 19 (0.30%) currently integrate their first-party data (Google Analytics) directly with SimilarWeb. Direct integration is similarly rare beyond our sample: only three of the 600 highest-traffic websites worldwide integrate first-party data, and none of the 100 highest-ranked treated websites in our sample do. Among the 54 sample websites with archived SimilarWeb snapshots from late 2017 and early 2018, concentrated among the largest, most prominent websites in our sample (including google.com, facebook.com, youtube.com, nytimes.com, and theguardian.com), precisely those the concern singles out, all displayed SimilarWeb's own estimated data rather than publisher-verified data, and none showed Google Analytics integration. Because direct integration is rare today and absent among the large websites we observed before GDPR, the integration-exit mechanism has little scope to operate; we therefore consider it unlikely to materially drive our results, though we cannot rule it out with certainty.

We acknowledge several limitations: First, we cannot independently verify SimilarWeb's data-processing claims. Second, no available benchmark is device-level user-side data; although AGOF's page impression and visits measurement is itself largely outside GDPR's scope, it is not device-level, so we cannot fully rule out a measurement component. Third, the AGOF comparison covers only 65 German websites, which limits its power to detect small differences. Therefore, we treat GDPR-induced measurement change as a possibility we cannot categorically rule out; the converging evidence above, the multi-source and user-side structure of SimilarWeb's data, the upper-bound direction of any bias, the AGOF validation, and the near-absence of direct integration, indicate that measurement error is unlikely to account for the patterns we document.

## WEB APPENDIX E: Robustness Checks Concerning the Control Group

Our control group comprises website-instances of non-EU websites visited by non-EU users (Cell 4 in Figure 1), which lie outside the GDPR's direct scope. The concern is that the regulation nonetheless spills over to these instances: a non-EU website serving EU users may be subject to the GDPR for processing related to those users, depending on the nature of its activities, and because differentiating EU from non-EU users is costly, some websites may voluntarily extend GDPR compliance to all of their users—the "Brussels effect" (Peukert et al. 2022). Were our control websites themselves affected, the treatment-versus-control comparison would understate the magnitude of the true effect.

We therefore examine whether non-EU control websites voluntarily complied, whether user behavior spilled across the EU/non-EU boundary, whether our within-website design is itself contaminated, and whether control-group composition or self-selection could drive the result. One conclusion recurs across these checks: no specification designed to reduce potential control-group contamination yields a statistically significantly less negative effect than the corresponding benchmark, whereas the specifications that most directly remove potentially exposed controls yield more negative estimates. This pattern indicates that any residual contamination is more likely to attenuate than to generate our findings. We therefore interpret the main estimates as a conservative lower bound on the magnitude of the GDPR's negative effect with respect to control-group contamination.

Several of the checks below are estimated on random subsamples of treatment website-instances for computational reasons; each comparison is therefore made against the benchmark specification re-estimated on the same subsample, which is why benchmark magnitudes differ across checks.

### E.1 Supply-Side Voluntary Compliance

Voluntary compliance is most likely for non-EU websites with a high share of EU users, for whom treating EU and non-EU users differently is least worthwhile; U.S.-based businesses with minimal EU contacts, by contrast, have little reason to comply (Nicolosi 2022).

Figure W4 provides a first diagnostic of where contamination, if present, should be most likely to occur. It compares GSC-estimated effects for the EU-user instances of non-EU websites (treated Cell 3 instances), splitting websites at the median pre-GDPR EU-user share. The donor pool comprises non-EU websites with no observed EU-user traffic, which should have substantially weaker incentives to extend GDPR-related changes across users.

Because EU-user instances are treated regardless of whether a website extends compliance to its non-EU users, this comparison does not directly test contamination of Cell 4. Instead, it identifies whether treatment effects differ between websites with relatively high and low EU exposure, thereby helping locate the subset in which website-wide compliance is most plausible.

At 3 months, the decline is significantly milder for high-EU-share websites than for low-EU-share websites (-0.17% versus -3.85%; two-sample t-test, p = 0.01). This heterogeneity is consistent with differences in websites' compliance responses, although the comparison does not identify the underlying mechanism. Because extending a common compliance regime to all users should be most attractive for websites with high EU exposure, the result identifies the subset in which control-group contamination is most plausible.

If the Brussels effect contaminates Cell 4 in our sample, it should therefore be most visible as a relative post-GDPR decline in the non-EU-user traffic of these high-exposure websites. Table W12 below tests this effect and finds the opposite: the non-EU-user instances of more-exposed websites grew rather than declined post-GDPR. Table W13 below further shows that removing exposed controls produces more negative estimates.

By 18 months, the difference between the two groups in Figure W4 is no longer statistically significant at the 5% level (-5.05% versus -8.58%; p = 0.08). This pattern is consistent with prior evidence that some GDPR spillovers onto non-EU users diminished after the initial implementation period (Johnson et al. 2023).

*Figure W4. GDPR's average 3-month (left panel) and 18-month (right panel) effect on weekly visits for EU-user instances of non-EU websites, by pre-GDPR EU-user share*

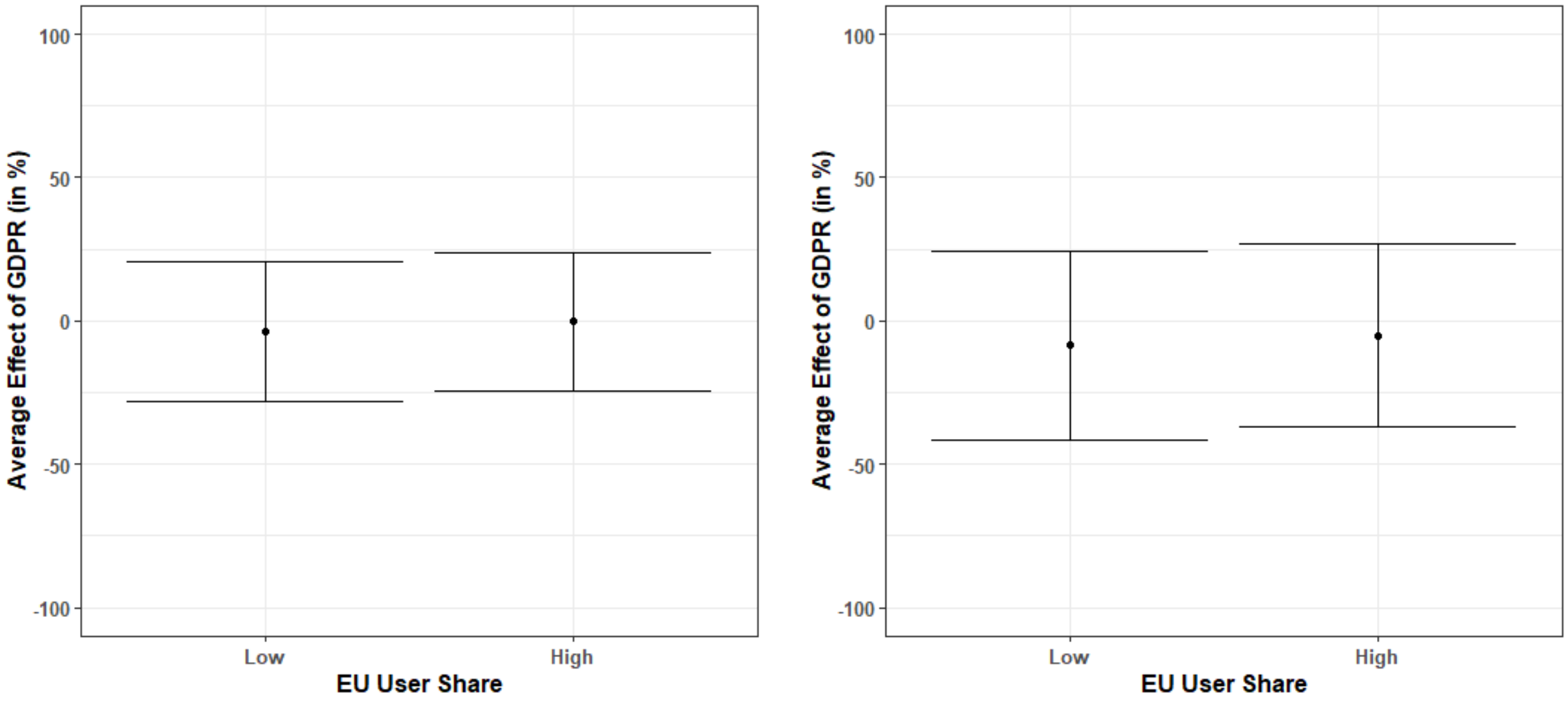

Notes: The left panel reports the 3-month mean effect for websites with a low EU-user share (-3.85%; SD = 24.34%) and a high EU-user share (-0.17%; SD = 24.16%). The difference is statistically significant in a two-sample t-test (t = -2.46, p = 0.01). The right panel reports the corresponding 18-month effects for websites with a low EU-user share (-8.58%; SD = 32.84%) and a high EU-user share (-5.05%; SD = 31.78%). The difference is not statistically significant at the 5% level (t = -1.76, p = 0.08).

We also inspected the control websites' interfaces directly. For a random sample of 85 control websites, we retrieved Internet Archive Wayback Machine snapshots at the GDPR enforcement date (May 25, 2018) and the last day of our observation period (December 31, 2019). 94.12% of these websites did not adjust their privacy-related user interfaces post-GDPR, and the remaining 5.88% made only slight adjustments that fell short of GDPR compliance. Visible interface-level voluntary compliance among the inspected non-EU websites is thus rare. Because archived interfaces do not reveal backend or other nonvisible data-processing changes, this analysis cannot rule out forms of compliance that were not visible to users.

### E.2 User-Behavior Spillover Across Cells

If the GDPR were fully applied to all users, non-EU users would behave identically whether they visited EU or non-EU websites. The descriptive averages do not exhibit this pattern. Table W9 shows that non-EU users visit EU websites (where the GDPR applies) less post-GDPR than pre-GDPR, but visit non-EU websites (where it does not) more—a behavioral difference inconsistent with a complete or dominant compliance-related spillover.

The mirror comparison shows the same pattern: Table W10 shows that on non-EU websites, EU users visit less post-GDPR while non-EU users visit more. These comparisons indicate that the decline is concentrated in the treated cells and is inconsistent with complete spillover across the EU/non-EU boundary. Because the tables report aggregate pre/post comparisons without controlling for differential trends or compositional changes, we treat them as descriptive evidence rather than estimates of the magnitude of the partial spillover.

We note that the post-GDPR increase in Cell 4 visits is also consistent with partial substitution of usage toward non-EU websites. Unlike contamination through compliance, GDPR-induced substitution would increase control traffic and could therefore inflate, rather than attenuate, cross-website treatment-versus-control comparisons. The panel-differences estimator (Web Appendix G), which does not rely on cross-website comparisons and reproduces the main result within 0.2 percentage points, provides reassuring evidence that the net bias from substitution and compliance-related contamination combined is unlikely to be large.

*Table W9. Comparison of EU- and Non-EU Visits for Non-EU Users (Cell 2 vs. Cell 4)*

| Website Location | User Location | Pre-/Post-GDPR | Average visits per week | Difference | Difference-in-Difference |
|---|---|---|---|---|---|
| EU websites | Non-EU users | Pre-GDPR | 3,159,196 | -321,937 | -555,844 |
| | | Post-GDPR | 2,837,259 | | |
| Non-EU websites | | Pre-GDPR | 2,190,523 | +233,907 | |
| | | Post-GDPR | 2,424,430 | | |

*Table W10. Comparison of EU- and Non-EU User Visits for Non-EU Websites (Cell 3 vs. Cell 4)*

| Website Location | User Location | Pre-/Post-GDPR | Average visits per week | Difference | Difference-in-Difference |
|---|---|---|---|---|---|
| Non-EU websites | EU users | Pre-GDPR | 1,578,877 | -11,320 | -245,236 |
| | | Post-GDPR | 1,567,548 | | |
| | Non-EU users | Pre-GDPR | 2,190,523 | +233,907 | |
| | | Post-GDPR | 2,424,430 | | |

## E.3 Within-Website Tests: Dose-Response and Direct Within-Control Spillover

The within-website subsample (Web Appendix G), in which the same website contributes both a treated EU-user instance and a control non-EU-user instance, holds website identity fixed. Dividing its 1,104 websites into quartiles of pre-GDPR EU-user share, we find that the mean effect in the highest-EU-share quartile (Q4, websites with between 99.4% and 99.9% EU traffic) is approximately -20% across all five windows, against -1% to -7% in the lowest quartile (Table W11).

More importantly, the median estimates exhibit a monotonic exposure-response pattern at every horizon. At three months, the median effect becomes progressively more negative, from -4.00% in Q1 to -9.21% in Q2, -13.37% in Q3, and -27.35% in Q4. At 18 months, the corresponding estimates are -8.88%, -14.56%, -15.94%, and -24.36%. The mean estimates show the same broad exposure effect, particularly the substantially larger effect in Q4, although the Q2 and Q3 means are inverted at all five post-treatment periods.

This monotonic pattern in the median estimates indicates that the estimated effect becomes more negative as a website's exposure to EU users increases, supporting our classification of EU-user website instances as treated rather than as controls. This effect runs opposite to the one in Figure W4, where the decline is milder for high-EU share websites, because the two quantities differ: Figure W4 reports the effect on the EU-user instance relative to a donor pool of non-EU websites, whereas Table W11 reports the within-website gap between a website's EU-user and non-EU user instances. Because Cell 4 traffic grows faster for more-exposed websites (Table W12), netting it out makes the within-website gap more negative as EU exposure rises.

This result also argues against website-wide contamination: uniform compliance affecting all of a website's users would compress the EU-versus-non-EU gap toward zero at every exposure level, rather than producing progressively more negative median gaps as EU exposure increases. Any compliance that reaches a website's non-EU users would attenuate the within-website gaps we report.

*Table W11. Within-Website Difference-in-Differences (DiD): Effect of GDPR by EU-Share Quartile*

| **Metric** | **Post-Treatment Period** | | | | |
|---|---|---|---|---|---|
| | **3 months** | **6 months** | **9 months** | **12 months** | **18 months** |
| **Q1 (EU share 0-19%)** | | | | | |
| Mean | -0.90% | -3.97%** | -6.28%*** | -6.85%*** | -7.49%*** |
| 95% Confidence interval | [-5.02%, +4.29%] | [-7.66%, -0.19%] | [-9.48%, -3.30%] | [-10.12%, -3.71%] | [-10.71%, -4.45%] |
| Median | -4.00%*** | -5.42%*** | -6.95%*** | -7.27%*** | -8.88%*** |
| 95% Confidence interval | [-6.24%, -1.55%] | [-7.58%, -3.58%] | [-9.69%, -4.57%] | [-9.90%, -5.09%] | [-10.86%, -7.11%] |
| **Q2 (EU share 19-88%)** | | | | | |
| Mean | -4.90%** | -8.31%*** | -10.76%*** | -11.50%*** | -14.13%*** |
| 95% Confidence interval | [-9.34%, -0.25%] | [-13.11%, -2.87%] | [-15.70%, -5.45%] | [-16.57%, -5.98%] | [-19.02%, -8.53%] |
| Median | -9.21%*** | -11.37%*** | -13.48%*** | -12.93%*** | -14.56%*** |
| 95% Confidence interval | [-12.51%, -5.45%] | [-13.87%, -7.59%] | [-15.89%, -10.60%] | [-16.44%, -10.30%] | [-16.96%, -10.97%] |
| **Q3 (EU share 89-99%)** | | | | | |
| Mean | -3.76% | -8.15%** | -8.52%** | -9.01%** | -8.17%** |
| 95% Confidence interval | [-10.80%, +4.21%] | [-14.23%, -0.62%] | [-14.52%, -1.57%] | [-15.21%, -2.06%] | [-14.12%, -2.03%] |
| Median | -13.37%*** | -13.27%*** | -15.24%*** | -15.12%*** | -15.94%*** |
| 95% Confidence interval | [-18.03%, -7.24%] | [-19.39%, -9.62%] | [-21.05%, -9.54%] | [-20.14%, -9.97%] | [-20.68%, -9.69%] |
| **Q4 (EU share 99-100%)** | | | | | |
| Mean | -19.17%*** | -21.91%*** | -20.07%*** | -20.77%*** | -20.20%*** |
| 95% Confidence interval | [-23.87%, -14.00%] | [-25.90%, -17.52%] | [-23.93%, -16.06%] | [-24.58%, -16.54%] | [-24.02%, -15.96%] |
| Median | -27.35%*** | -26.18%*** | -23.27%*** | -23.69%*** | -24.36%*** |
| 95% Confidence interval | [-30.04%, -21.50%] | [-29.73%, -22.71%] | [-27.45%, -20.62%] | [-28.48%, -20.40%] | [-27.72%, -19.89%] |
| **Overall** | | | | | |
| Mean | -7.18%*** | -10.58%*** | -11.41%*** | -12.03%*** | -12.50%*** |
| 95% Confidence interval | [-9.83%, -4.20%] | [-12.99%, -8.00%] | [-13.78%, -8.80%] | [-14.42%, -9.47%] | [-14.87%, -10.00%] |
| Median | -11.02%*** | -12.60%*** | -13.51%*** | -13.52%*** | -14.67%*** |
| 95% Confidence interval | [-13.04%, -9.03%] | [-13.98%, -10.71%] | [-15.02%, -11.79%] | [-15.21%, -11.98%] | [-16.62%, -12.66%] |
| **N (quartile)** | 276 | 276 | 276 | 276 | 276 |
| **N (websites)** | 1,104 | 1,104 | 1,104 | 1,104 | 1,104 |

Notes: The sample consists of the 1,104 non-EU websites that have both EU-user and non-EU-user traffic. "Mean" reports the average across websites and "Median" the analogous median; 95% intervals are percentile bootstrap intervals over websites. Quartiles are the sample quartiles of pre-GDPR EU-user share and are mutually exclusive and exhaustive: Q1 = [0.1%, 19.0%], Q2 = (19.0%, 88.6%], Q3 = (88.6%, 99.4%], Q4 = (99.4%, 99.9%]. The maximum EU-user share is 99.9%: because every website contributes an observed non-EU-user (control) series, none has a 100% EU-user share, so the Q4 upper bound of "100%" is rounding. Stars test each quartile estimate against the null of no effect (zero), not against other quartiles. The monotone dose-response is confirmed by an omnibus rank test: the Spearman correlation is negative and significant in every window ($\rho$ = -0.24, -0.24, -0.18, -0.18, and -0.16 at 3, 6, 9, 12, and 18 months; all $p < 1\times10^{-7}$). N = 1,104 websites (276 per quartile).

We also test the control group directly, regressing Cell 4's traffic on a post-GDPR indicator interacted with each website's pre-GDPR EU-traffic share, with website and calendar-week-of-year fixed effects. Contamination through voluntary compliance would, all else equal, make more EU-exposed controls decline more—a negative interaction. We find the opposite: the interaction is positive at every window, ranging from +0.352 at 3 months to +0.267 at 18 months (all $p < 0.001$; Table W12).

The baseline post-GDPR coefficient—the within-control effect for a website with zero pre-GDPR EU exposure—is also positive throughout, ranging from +6.1% to +32.3% across the five windows, and the positive interaction means that more-exposed controls grew even more, not less.

The control group, therefore, expanded rather than contracted over the post-treatment period, and more EU-exposed controls experienced greater rather than smaller growth. This pattern provides no evidence of the negative differential change predicted by control-group contamination. Although the regression cannot rule out a negative compliance effect that was offset by other contemporaneous trends, it indicates that such contamination does not dominate the observed development of control traffic.

*Table W12. Differential Post-GDPR Trends within Cell 4 by Pre-GDPR EU Exposure*

| | ***Post-Treatment Period*** | | | | |
|---|---|---|---|---|---|
| | **3 months** | **6 months** | **9 months** | **12 months** | **18 months** |
| **Post-GDPR** | 0.280***<br>(0.019) | 0.172***<br>(0.014) | 0.109***<br>(0.013) | 0.059***<br>(0.012) | 0.062***<br>(0.014) |
| [% effect at EU share = 0] | [+32.3%] | [+18.8%] | [+11.5%] | [+6.1%] | [+6.4%] |
| **Post-GDPR × EU share$^{pre}$** | 0.352***<br>(0.028) | 0.328***<br>(0.027) | 0.290***<br>(0.028) | 0.283***<br>(0.029) | 0.267***<br>(0.031) |
| [% effect per 10 pp of EU share] | [+3.6%] | [+3.3%] | [+2.9%] | [+2.9%] | [+2.7%] |
| **Website fixed effects** | Yes | Yes | Yes | Yes | Yes |
| **Week-of-year fixed effects** | Yes | Yes | Yes | Yes | Yes |
| **N (websites)** | 1,701 | 1,701 | 1,701 | 1,701 | 1,701 |

Notes: Dependent variable is $\ln(\text{Visits}_{w,t} + 1)$. Sample: Cell 4 only (non-EU websites with non-EU user traffic; same 1,701 control website-instances as in Table 2). EU $\text{share}_w^{pre}$ is each website's pre-GDPR share of EU-user traffic (cross-sectionally invariant within website; absorbed as a main effect by website fixed effects). Standard errors clustered at the website level in parentheses. Bracketed values report percentage interpretations: the Post-GDPR row reports the within-Cell 4 effect for a website with zero pre-GDPR EU traffic; the interaction row reports the additional growth associated with a 10-percentage-point increase in EU $\text{share}_w^{pre}$. *** $p < 0.01$, ** $p < 0.05$, * $p < 0.10$. These coefficients describe differential post-period changes within the control group and should not be interpreted as causal effects of GDPR.

## E.4 Self-Selection and Control-Group Composition

A first composition concern is strategic relocation: EU-based websites might shift data processing outside the EU to evade compliance, contaminating both groups. We build a stricter control group from websites whose domain suffixes indicate a non-EU location (e.g., ".com") and re-estimate. The effect is essentially unchanged—at 3 months, -4.56% (stricter) versus -7.64% (original), and at 18 months, -19.90% versus -18.64%—with no significant difference at either horizon ($p > 0.05$). Although the stricter control point estimate is less negative at three months, the difference from the original estimate is not statistically significant. At 18 months, the stricter-control point estimate is slightly more negative, but again not statistically distinguishable from the original estimate. We therefore find no evidence that the alternative geographic classification changes the estimated effect.

A second concern is that Swiss firms may have faced greater GDPR exposure and contemporaneous pressure from domestic data-protection reforms than US firms. Consequently, the 229 Swiss website-instances in our control group (alongside 1,472 US instances, N = 1,701) could themselves have been more affected by privacy-law activity than US controls. Re-estimating with a US-only control group yields more negative point estimates: in a two-way fixed-effects specification, the coefficient moves from -0.090 to -0.114 at 3 months and from -0.206 to -0.227 at 12 months (each reported coefficient is statistically significant at $p < 0.001$). The direction of these estimates is consistent with Swiss controls attenuating rather than inflating the estimated effect.

We can also remove EU-exposed controls entirely. Restricting the pooled GSC (Web Appendix G) to the 597 "pure" non-EU websites with zero EU-user traffic—versus 1,701 control website-instances in the main specification—yields effects more negative than the unrestricted estimate at all five horizons, by 2.5 to 12 percentage points (Table W13).

We do not read the restricted magnitude as the true effect because the 597 pure controls are compositionally distinct from the full control group. The panel-differences estimator (Web Appendix G), which eliminates the cross-website comparison altogether, lands within 0.2 percentage points of our main result. This close correspondence provides additional evidence that the main finding is not driven by cross-website differences or control-group contamination.

Taken together, the restricted-control and panel-differences results indicate that the main estimate is conservative: removing plausible sources of contamination either yields a more negative estimate or leaves it essentially unchanged.

The restricted-versus-original gap is 2.48 percentage points at 3 months and 11.95 percentage points at 18 months, though it is not monotonic across windows. This pattern is consistent with increasing attenuation among EU-exposed controls at longer horizons, thereby reinforcing the lower-bound interpretation. However, because the restricted and original control groups differ in composition, the widening cannot be attributed solely to contamination.

*Table W13. Sensitivity to Excluding High-EU-Share Websites from the Control Group: Pooled Generalized Synthetic Control (GSC) Estimates of the Effect of GDPR on Visits*

| **Metric** | | **Post-Treatment Period** | | | | |
|---|---|---|---|---|---|---|
| | | **3 months** | **6 months** | **9 months** | **12 months** | **18 months** |
| **Visits (Pooled GSC, Restricted Control)** | Mean | -5.48% | -9.54% | -14.45% | -17.16% | -24.96% |
| | 95% Confidence interval | [-6.64, -4.26] | [-11.09, -7.87] | [-16.59, -12.20] | [-19.43, -14.75] | [-27.10, -22.77] |
| | Number of treatment websites | 5,790 | 5,790 | 5,790 | 5,790 | 5,790 |
| | Number of control websites | 597 | 597 | 597 | 597 | 597 |
| **Visits (Pooled GSC, Original Control)** | Mean | -3.00% | -6.02% | -6.85% | -10.64% | -13.01% |
| | 95% Confidence interval | [-3.87, -2.10] | [-6.97, -5.13] | [-8.55, -5.10] | [-12.81, -8.32] | [-16.25, -9.57] |
| | Number of treatment websites | 5,790 | 5,790 | 5,790 | 5,790 | 5,790 |
| | Number of control websites | 1,701 | 1,701 | 1,701 | 1,701 | 1,701 |
| **Visits (Website-Level GSC)** | Mean: | -4.88% | -7.22% | -9.07% | -9.57% | -10.02% |
| | 95% Confidence interval | [-5.50, -4.23] | [-7.83, -6.57] | [-9.72, -8.36] | [-10.25, -8.81] | [-10.79, -9.18] |
| | Number of treatment websites | 5,790 | 5,790 | 5,790 | 5,790 | 5,790 |
| | Number of control websites | 1,701 | 1,701 | 1,701 | 1,701 | 1,701 |
| **Absolute Difference (Pooled Restricted vs. Original GSC)** | | 2.48 pp | 3.52 pp | 7.60 pp | 6.52 pp | 11.95 pp |

Notes: This table reports the GDPR effect on weekly visits estimated by (i) the pooled Generalized Synthetic Control (GSC) with the restricted control group of 597 pure non-EU websites (zero EU-user traffic); (ii) the pooled GSC with the original 1,701 control website-instances; and (iii) our main website-level GSC estimator (Table 2). All three specifications use the same underlying sample of 5,790 treatment websites. 95% confidence intervals are bootstrap intervals obtained by resampling websites. The Absolute Difference row reports the gap between the restricted and unrestricted pooled GSC estimates, isolating the effect of removing potentially contaminated controls.

## E.5 Donor-Pool Composition

The GSC estimator embeds two donor-pool choices: capping the matched control instances at the five with the highest pre-treatment correlation to avoid overfitting, and requiring controls to share the treatment website's industry. Our estimates are not statistically different under the alternatives examined.

Dropping the industry requirement and matching purely on pre-treatment correlation (random subset of 100 treatment website-instances) does not produce a statistically significant difference from the original matching specification: the short-term effect is -2.84% under the new specification versus -6.69% under the original, and the long-term effect is -5.65% versus -8.50% (no significant difference at the 5% level).

Matching instead on similar EU-traffic share (random subsample of 393 websites) does neither yield a statistically significant difference: -7.07% under the new specification versus -3.79% under the original short-term, and -4.21% versus -5.77% long-term (no significant difference at 5%). And doubling the donor pool from five to ten controls (a subsample of 60 treatment website-instances) yields short-term effects of -13.91% (ten controls) versus -12.62% (five) and long-term effects of -17.36% versus -14.87% (no significant difference at 5%).

None of these alternative donor-pool specifications produces an estimate that is statistically significantly less negative than its corresponding benchmark. The estimator’s industry-matching and donor-pool settings, therefore, do not appear to drive our findings.

### E.6 Summary and Conclusion

Across these checks, we find no systematic evidence that exposure of the control group to GDPR-related compliance drives our main findings. The descriptive cell comparisons are inconsistent with complete spillover; visible interface-level adaptation is uncommon among the inspected controls; and controls with greater EU exposure do not exhibit the negative differential post-GDPR pattern predicted by a simple contamination account. Estimates are also statistically indistinguishable from their corresponding benchmarks when we apply a stricter geographic classification or modify the donor-pool construction.

The specifications that most directly remove potentially exposed controls provide particularly strong evidence. Using US-only controls yields more negative point estimates, while excluding all controls with observed EU traffic produces substantially more negative estimates at every horizon. The pure controls are compositionally distinct, so we do not interpret the restricted estimates as the exact uncontaminated effect. Most importantly, the within-website panel-differences estimator closely reproduces the main result without relying on cross-website comparisons.

Taken together, the results indicate that control-group contamination is unlikely to explain the estimated decline in website usage. No robustness specification designed to reduce contamination produces a statistically significantly less negative effect, whereas several produce substantially more negative effects. The evidence, therefore, indicates that any remaining compliance-related contamination is more likely to bias the estimates towards zero than away from zero. We consequently interpret our main estimates as a conservative lower bound on the magnitude of the GDPR's negative effect on online usage with respect to control-group contamination.

This lower-bound interpretation concerns control-group contamination specifically. As discussed in the main text, residual GDPR-induced measurement error would operate in the opposite direction by inflating the estimated decline. The two potential biases, therefore, bracket the true effect in opposite directions. The AGOF validation in Web Appendix D indicates that the measurement-error component is small.

## WEB APPENDIX F: Robustness Checks Concerning Confounding Factors

Our identification does not require the absence of other changes around GDPR enforcement. A contemporaneous change threatens our estimates only if it generates a differential shift between treated and control website-instances around May 25, 2018; shocks common to both groups differ out by design. The four-cell structure additionally holds each dimension constant in turn: the Cell 2 versus Cell 4 comparison holds user location constant and compares non-EU users across EU and non-EU websites, while the Cell 3 versus Cell 4 comparison holds website location constant and compares EU and non-EU users on non-EU websites. A confounder would therefore have to generate a differential change closely aligned with the regulatory boundary rather than a general user-side or website-side trend.

### F.1 Digital Access Indicators

We examine four observable factors that plausibly affect browsing: mobile internet speed, internet-access share, laptop-usage share, and smartphone-usage share. Because these are hard to measure per website-instance, we track them by user location, proxying EU users with German users and non-EU users with U.S. users.

Table W14 shows that the observed pre- versus post-GDPR differentials are modest and run in opposite directions. Mobile-speed growth is slightly greater in Germany (+20% versus +18%), whereas internet-access growth is greater in the United States (+7.95% versus +5.49%). The latter difference has the right direction to inflate the estimated GDPR effect, as faster growth in the U.S. online population could raise control-group traffic. Reported laptop/desktop and smartphone usage shares are unchanged in both countries. Taking them together, these indicators reveal no consistent relative change favoring either user group that could readily explain the estimated decline in treated traffic.

Because Germany and the United States are proxies for broader EU and non-EU user populations, and because the indicators are available only as annual snapshots, this comparison is descriptive rather than a formal test of a discontinuity on May 25, 2018. Changes in internet access also need not translate one-for-one into website visits, so Table W11 should not be interpreted as placing a quantitative bound on confounding.

*Table W14. Descriptive Comparison of Selected Digital-Access Indicators*

| Confounding Factor | User Group | Pre-GDPR: January 2018 | Post-GDPR: January 2019 |
|---|---|---|---|
| **Average Mobile Internet Connection Speed** | EU (Germany) | 26.43 Mbps | 31.69 Mbps *(+**20%** YoY-Growth)* |
| | Non-EU (USA) | 27.22 Mbps | 32.01 Mbps *(+**18%** YoY-Growth)* |
| **Share of Population with Access to the Internet** | EU (Germany) | 91% | 96% *(+**5.49%** YoY-Growth)* |
| | Non-EU (USA) | 88% | 95% *(+**7.95%** YoY-Growth)* |
| **Share of Population Using Laptops or Desktops** | EU (Germany) | 76% | 76% *(+**0.00%** YoY-Growth)* |
| | Non-EU (USA) | 77% | 77% *(+**0.00%** YoY-Growth)* |
| **Share of Population Using Smartphones** | EU (Germany) | 75% | 75% *(+**0.00%** YoY-Growth)* |
| | Non-EU (USA) | 78% | 78% *(+**0.00%** YoY-Growth)* |

Source: Datareportal (2018, 2019).

## F.2 Contemporaneous Privacy Regulations and Salience Events

A contemporaneous privacy regulation or event threatens our identification only if it affected treated and control website-instances differentially during the July 2017 – December 2019 observation window in a manner aligned with GDPR's May 25, 2018, enforcement. Such threats include shocks to U.S. and Swiss control websites as well as non-GDPR shocks that disproportionately affect EU users or EU websites.

The plausible candidates fall into two categories — privacy regulations adopted by non-EU jurisdictions and global privacy-salience events — which we map in Table W15. We address cross-border data-transfer rules separately below.

*Table W15. Contemporaneous Privacy Regulations and Salience Events, July 2017* – December 2019

| Regulation / Event | Enactment Date | Effective Date | Jurisdiction | Scope and Mechanism | Threat to identification? | Source |
|---|---|---|---|---|---|---|
| California CCPA | Jun 28, 2018 | Jan 1, 2020 | California (US) | Opt-out for personal-information sales; no cookie consent banners or cookie walls | No. Effective date beyond panel; Opt-out regime imposes no user-facing friction | California Civil Code §1798.100 et seq.; IAPP CCPA Resource Center |
| Vermont Data Broker Law (Act 171) | May 22, 2018 | Jan 1, 2019 | Vermont (US) | Data-broker registration only; does not regulate general websites | No. Out of scope | 9 V.S.A. ch. 62; IAPP State Privacy Tracker |
| Nevada SB 220 | May 29, 2019 | Oct 1, 2019 | Nevada (US) | Narrow opt-out for "online operators"; binds only Nevada residents | No. Narrow scope | Nevada Senate Bill 220; IAPP State Privacy Tracker |
| Illinois BIPA | Oct 3, 2008 | Oct 3, 2008 | Illinois (US) | Biometric data only; pre-existing throughout panel | No. Pre-existing and narrow scope | 740 ILCS 14 |
| Swiss Schengen Data Protection Act | Mar 1, 2019 | Mar 1, 2019 | Switzerland | Criminal-justice data only | No. Out of scope | Swiss Federal Council 18.061 |
| Brazilian LGPD | Aug 14, 2018 | Sep 18, 2020 | Brazil | General data-protection law | No. Effective beyond panel; Brazil not in controls | Lei Geral de Proteção de Dados 13.709/2018 |
| EU ePrivacy Regulation (proposed) | Jan 10, 2017 (proposal) | Never enacted | EU | Proposed amendments to ePrivacy Directive on cookies and electronic communications | No. Not enacted during or after panel | European Commission COM(2017) 10 final |
| Cambridge Analytica scandal | Mar 17, 2018 | (Event) | Global | User-awareness shock; potential cross-treated/control effect | No. Affects treated and controls symmetrically. | The Guardian, March 17, 2018; New York Times, March 17, 2018 |
| Facebook data breach disclosure | Sep 28, 2018 | (Event) | Global | User-awareness shock | No. Affects treated and controls symmetrically. | Facebook Newsroom statement, September 2018 |

None of the identified regulations or events have the combination of timing, scope, and mechanism required to explain the full pattern of treatment-control divergence.

The two regulations with the broadest scope (California's CCPA and Brazil's LGPD) were enacted within our window but came into force only after our data end (January 1, 2020 and September 2020), and the ePrivacy Regulation, the obvious potential confound of the treatment group, was never enacted. Of the regulations in force during our window, none plausibly affects general website traffic: Vermont's Data Broker Law (January 2019) binds only data brokers, Nevada SB 220 (October 2019) imposes a narrow opt-out on Nevada residents, Illinois BIPA covers biometric data only, and Switzerland's Schengen Data Protection Act (March 2019) covers criminal-justice data only. Even if CCPA had been in force, its opt-out design imposes no cookie consent banner or cookie-wall friction and offers no mechanism to reduce control-group traffic.

The cross-border data-transfer environment was likewise stable in dimensions relevant to our design. In particular, the EU–U.S. Privacy Shield was in force throughout the panel.

Finally, the salience shocks — the Cambridge Analytica scandal (March 17, 2018) and the Facebook data-breach disclosure (September 28, 2018) — are global awareness events that would have affected treated and control websites symmetrically, biasing our comparison toward zero rather than toward the negative effect we observe.

Later GDPR enforcement actions, such as the CNIL's January 2019 fine against Google, are downstream components of the regulatory regime rather than external confounders. They may contribute to the evolution of the treatment effect over time, which Web Appendix I examines directly.

### F.3 Summary and Conclusion

Taken together, the descriptive digital-access indicators, legal timelines, and privacy-salience analyses provide no evidence that the identified contemporaneous factors explain the estimated GDPR effect. The observed digital-access differentials are modest and do not consistently favor either user group. The privacy laws effective during the panel either predate GDPR, apply to narrow activities or populations, or occur too late to explain the initial treatment-control divergence. Global privacy-salience events were not confined to either treatment or control jurisdictions and, if they affected treated and control website-instances similarly, would differ out by design rather than generate the observed differential break.

These checks cannot rule out every possible unobserved contemporaneous shock. They do show, however, that none of the principal observable candidates has the timing, scope, and mechanism required to reproduce the full pattern of results across user locations, website locations, and post-treatment periods.

## WEB APPENDIX G: Robustness Check Concerning the Choice of the Estimator

Table W16 summarizes the identification assumptions underlying the main-text estimates, the corresponding challenges and solutions, and the robustness results, with pointers to the appendix sections that report each test in full.

*Table W16. Identification Assumptions, Challenges, Solutions, and Results*

| Identification Assumption | Challenge | Solution | Supports Robustness of Results |
|---|---|---|---|
| Composition of the Donor Pool | Identify control websites with similar characteristics as treatment websites | Five industry-matched control instances chosen by highest pre-treatment correlation (robustness: no-industry match; EU-share match; ten controls) | Yes: no significant difference across donor-pool specifications (GSC; App E.5) |
| No Anticipation of Treatment | Early- or Late-GDPR compliance of treatment websites | Window analysis, which removes 30 days pre- & post-GDPR's enforcement | Yes: no significant difference (GSC; App I) |
| No Interference (SUTVA) | Voluntary GDPR compliance of control websites (uniform privacy interface for all users) | Analysis of control websites based on share of EU traffic | Yes: 3-month gap vanishes by 18 months (GSC; App E, H) |
| | | Analysis of non-EU user interface on control websites | Yes: 94% of inspected controls unchanged (App E) |
| | | Exclusion of control websites with any EU-user traffic | Yes: estimates more negative; any bias is toward zero (Pooled GSC; App E) |
| | | Regression of control-group traffic on Post-GDPR × pre-GDPR EU-user share | Yes: interaction positive (all $p < 0.001$); opposite of contamination (TWFE; App E) |
| | Spillover to user behavior: non-EU users react to GDPR awareness | Analysis of EU websites and non-EU websites for non-EU users | Yes: changes present but limited in magnitude (GSC; App E) |
| | | Analysis of EU users and non-EU users for non-EU websites | Yes: slightly stronger DID effects; limited spillover (GSC, App E; DID, App G.1) |
| | Residual contamination of any form (incl. algorithm feedback from EU onto non-EU traffic) | Dose-response: within-website DID by pre-GDPR EU-user-share quartile | Yes: effect scales with exposure (Q1 -0.9% n.s. to Q4 -19.2% at 3 months) (DID; App E) |
| | | Panel-differences estimator: each website vs. its own prior year (no control group) | Yes: within 0.2 pp of main estimate (App G.3) |
| | Self-selection of EU websites to control group(non-compliance of treatment websites) | Analysis of strategic geographic shifts of website server locations | Yes: no significant difference with stricter control group (GSC; App E) |
| Unconfoundedness | Coincidence of GDPR's enforcement with other major changes in confounding factors | Analysis of differences in internet speed, share of population with internet access, laptop, and smartphone across EU and non-EU users | Yes: no substantial differences between user groups (App F) |
| | | Inventory of contemporaneous privacy regulations and salience events (timing, scope, mechanism) | Yes: none binds the control group in a way confounded with enforcement (App F) |
| | Swiss control websites subject to national privacy-law activity | Exclusion of Swiss websites from the control group (US-only controls) | Yes: estimates ≈2 pp more negative; bias toward zero (TWFE; App E) |
| GSC Estimator & Specification Validity | Sensitivity to the synthetic-control estimator | Re-estimate with the original synthetic control estimator (synth) instead of GSC (gsynth) | Yes: no significant difference between estimators (synth vs. gsynth; App G.2) |

| Identification Assumption | Challenge | Solution | Supports Robustness of Results |
|---|---|---|---|
| | Sensitivity to per-instance versus pooled estimation | Pooled GSC with a common interactive fixed-effect structure across all website-instances | Yes: within 3 pp at every horizon; CIs overlap at four of five windows (Pooled GSC; App G.3) |
| | log(Y+1) transformation and structural zeros (Chen and Roth 2024) | Re-estimate with OLS on log(Y), Poisson, and Negative Binomial | Yes: CIs overlap the log(Y+1) baseline at every horizon (App G.4) |
| | Mixed weekly and monthly temporal scales | Re-estimate the pooled GSC at monthly instead of weekly resolution | Yes: estimates statistically unchanged (App G.5) |
| | Spurious effects in the absence of treatment | In-time placebo tests with fake enforcement dates (October 2017 - April 2018) | Yes: effects indistinguishable from zero at every placebo date (App G.6) |

Notes: SUTVA: Stable Unit Treatment Value Assumption. GSC: Generalized Synthetic Control Estimator. DID: Difference-in-Differences Estimator. TWFE: Two-Way Fixed-Effects Estimator. All robustness results are directionally consistent with the main reported effects.

To measure GDPR's effect on online usage behavior, we use the generalized synthetic control (GSC) estimator of Equation (2) as our main specification in the paper. As a robustness check on the choice of estimator, we complement it with a difference-in-differences (DID) estimator applied to a within-website design, and then subject the estimator to a battery of further checks reported in the subsections below.

## G.1 Within-Website Difference-in-Differences (DiD) Estimator

The DID design uses a sub-sample of 1,104 websites. It is smaller than our main sample because it requires that the treatment and control website-instances belong to the same website: we assign non-EU website-instances with EU users to the treatment group (Figure 1, Cell 3) and non-EU website-instances with non-EU users to the control group (Figure 1, Cell 4). Figure W5 plots the average log(visits + 1) for the two groups and confirms similar trends pre-GDPR but diverging trends post-GDPR; parallel pre-treatment trends on weekly visits are the identifying assumption of this analysis. Using the regression model in Equation (W13), we estimate the treatment effect on weekly visits for each website-instance wi and trace its evolution over time by re-running the analysis across progressively longer post-enforcement windows: 3 months (through August 25, 2018, covering weeks 1–60), then 6 (weeks 1–73), 9 (weeks 1–86), 12 (weeks 1–99), and 18 months (weeks 1–125). These windows span GDPR's short-term (3-month) and long-term (18-month) effects.

$$(W13)\ \ln\left(Y_{q,t,wi} + 1\right) = \beta_{0,q,wi} + \beta_{1,q,wi} * EU_{wi} + \beta_{2,q,wi} * Postperiod_t + \beta_{3,q,wi} * Treated_{t,wi} + \epsilon_{q,t,wi}$$

$Y_{q,t,wi}$: Value of usage behavior metric *q* in week *t* on website-instance *wi*

$EU_{wi}$: EU-Dummy, i.e., binary variable for which a value of 1 indicates that the users of website-instance *wi* are EU-based, else 0

$Postperiod_t$: Post-period-Dummy, i.e., binary variable for which a value of 1 indicates that the observation in week *t* lies in the post-treatment period, else 0

$Treated_{t,wi}$: = $EU_{wi}$ * $Postperiod_t$; Treatment-Dummy, i.e., binary variable for which a value of 1 indicates that in week *t*, website-instance *wi* needs to consider GDPR, otherwise 0

$\epsilon_{q,t,wi}$: Error term for usage behavior metric *q* in week *t* for website-instance *wi*

Table W17 reports the DID estimates alongside the GSC estimates from Table 2; the number of treatment and control websites is 1,104. The DID estimator returns slightly but consistently larger (more negative) effects than the GSC estimator, both for GDPR's short-term effect (3 months post-enforcement: mean DID -7.18% versus GSC -4.88%; median DID -11.02% versus GSC -3.49%) and its long-term effect (18 months: mean DID -12.50% versus GSC -10.02%; median DID -14.67% versus GSC -8.91%). These magnitudes resemble Goldberg et al. (2024), who found an 11.7% drop in recorded page views and a 13.3% drop in recorded revenue after GDPR's enforcement in a DID analysis that used the same websites' prior-year traffic as the control group.

*Table W17. Comparison of the Effect of GDPR on Weekly Visits Obtained from a Difference-in-Differences (DID) Estimator and a Generalized Synthetic Control (GSC) Estimator*

| **Metric** | | **3 months** | **6 months** | **9 months** | **12 months** | **18 months** |
|---|---|---|---|---|---|---|
| **Visits (DID Estimator)** | Mean | -7.18% | -10.58% | -11.41% | -12.03% | -12.50% |
| | Median | -11.02% | -12.60% | -13.51% | -13.52% | -14.67% |
| | Number of treatment websites | 1,104 | 1,104 | 1,104 | 1,104 | 1,104 |
| | Number of control websites | 1,104 | 1,104 | 1,104 | 1,104 | 1,104 |
| **Visits (GSC Estimator)** | Mean | -4.88% | -7.22% | -9.07% | -9.57% | -10.02% |
| | Median | -3.49% | -5.54% | -7.54% | -8.24% | -8.91% |
| | Number of treatment websites | 5,790 | 5,790 | 5,790 | 5,790 | 5,790 |
| | Number of control websites | 1,701 | 1,701 | 1,701 | 1,701 | 1,701 |
| | p-value of t-test: | 0.12 | 0.01** | 0.08 | 0.06 | 0.06 |

Notes: The table summarizes GDPR's effect on weekly visits obtained from a difference-in-differences (DID) estimator using non-EU website-instances with EU users as the treatment group and non-EU website-instances with non-EU users as the control group. The table shows the mean and median values of the change in visits due to GDPR across all websites. The table compares the DID results with our main results obtained from a generalized synthetic control (GSC) estimator, as shown in Table 2. The number of treatment and control websites is 1,104. Significance level: *** 0.1%-level ** 1%-level * 5%-level.

This within-website difference-in-differences corresponds to the geographic-source identification of Congiu et al. (2022), who report comparable GDPR effects on traffic using the same SimilarWeb data. The consistently larger DID estimates also bound the spillover concern raised in the main text. Non-EU websites that serve both EU users (treatment) and non-EU users (control) may choose to adopt GDPR for the entire website — for example, to avoid the cost of treating EU and non-EU users differently — which would contaminate the control group; in the presence of such a spillover, treatment and control traffic would grow more similar post-enforcement, and the estimated effect would shrink. Instead, we find consistently larger, not smaller, effects, which supports the suitability of the control group. Because a direct within-website control is unavailable for every treatment instance, we cannot provide DID estimates for the full treatment sample and retain the GSC estimator — which constructs a synthetic control for each treatment instance — as our main specification; we therefore cannot rule out that some control instances complied with GDPR. Thus, we conclude that any spillover onto the control group could be present but is limited in size.

*Figure W5. Average log(visits + 1) for Non-EU Websites with EU User Traffic (Treatment Group) and Non-EU Websites with Non-EU User Traffic (Control Group)*

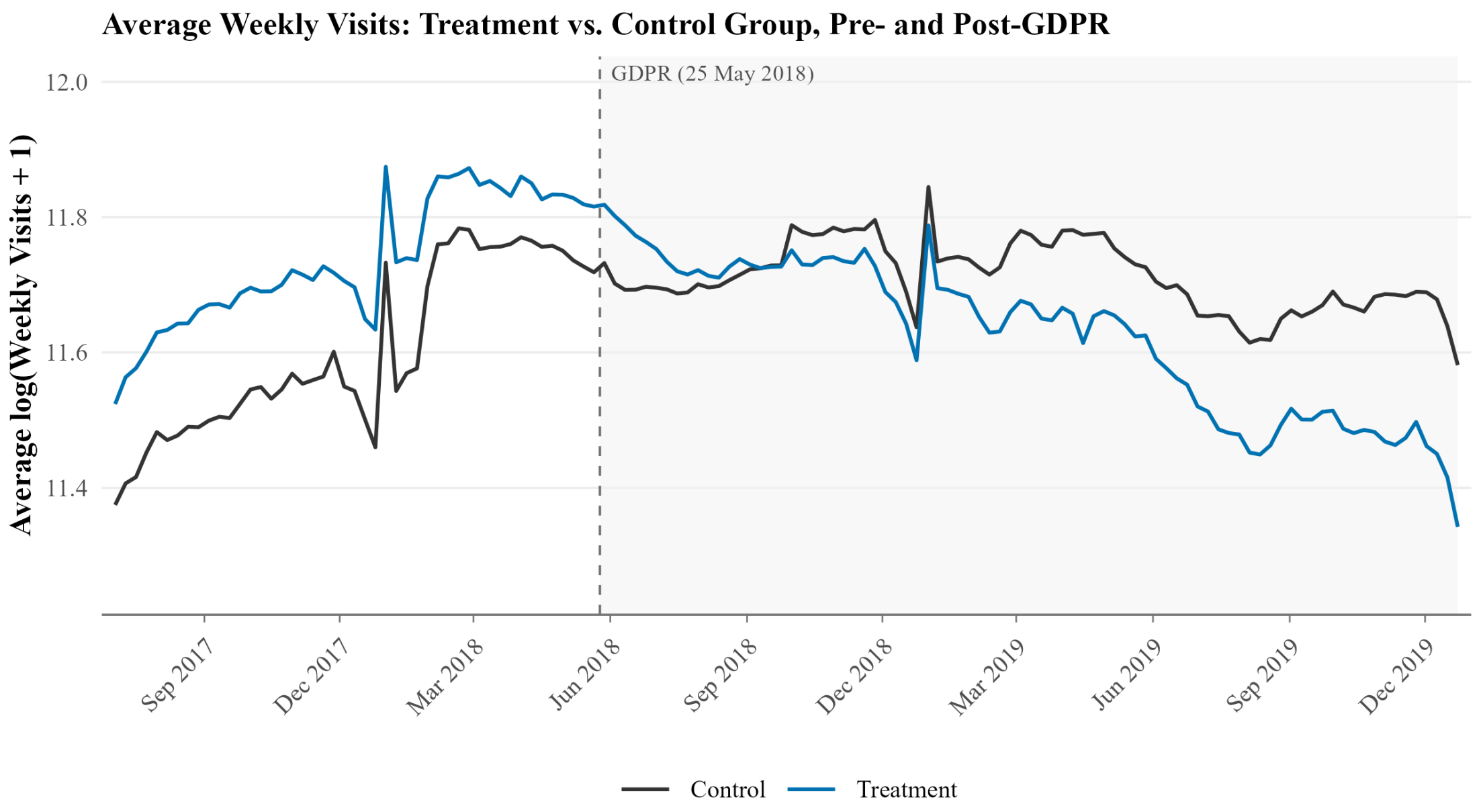

Notes: The blue line is the treatment group (EU website-instances, Cells 1-3) and the black line is the control group (Cell 4); each plots the average log(weekly visits + 1) across instances by week. The dashed line marks the GDPR's enforcement date (May 25, 2018), and the shaded region marks the post-GDPR period.

## G.2 Alternative Synthetic Control Estimator

We implement the GSC estimator (Xu 2017) through the R package gsynth, using nonparametric bootstrap inference with 1,000 replications. To check that the choice of package does not drive our estimates, we re-estimate GDPR's effect on weekly visits for a random subsample of 3,455 website-instances using the original synth package and compare it to gsynth. The two agree: the short-term effects differ insignificantly (average effect for synth: -4.96%; for gsynth: -6.88%; no significant difference at the 5% level), as do the long-term effects (average effect for synth: -18.39%; for gsynth: -19.29%; no significant difference at the 5% level).

## G.3 Alternative Estimators: Pooled GSC and Panel Differences

We subject the estimator to two further checks that alter the source of counterfactual variation. First, we estimate a pooled GSC model that jointly fits a common interactive fixed-effect structure at the website-instance level across all treated and control instances, assuming a shared latent factor structure with the number of factors r* selected by cross-validation — a methodologically distinct alternative to our website-instance-specific approach. The two estimators land within 3 percentage points of one another at every post-treatment horizon, with a mean absolute difference of 1.87 percentage points across the five windows; their empirical 95% confidence intervals overlap at four of the five windows (6, 9, 12, and 18 months) and miss by only approximately 0.4 percentage points at 3 months. Both are negative and grow more negative over time. At 12 and 18 months, the pooled estimator is somewhat more negative, suggesting that our website-instance-specific estimates are, if anything, conservative at longer horizons. Both estimators use the same underlying sample of 5,790 treatment and 1,701 control websites, and the consistency across two methodologically distinct approaches supports the robustness of our main findings.

Second, to remove any reliance on a cross-website control group, we estimate a panel-differences specification (Eichenbaum et al. 2024) in which each website serves as its own control, using the corresponding week of the previous year as the counterfactual. The specification regresses log visits on a website fixed effect, a calendar-week-of-year fixed effect that absorbs seasonality by netting out the average level for each calendar week, and a binary post-GDPR indicator equal to 1 for all weeks after May 25, 2018; its coefficient is the average treatment effect of GDPR, with standard errors clustered at the website level. Unlike Equation (2) in the main text, which operates at the website-instance level, this specification operates at the website level, summing a website's EU and non-EU instances before taking logs, so any spillover between its EU and non-EU traffic is internalized by construction. Applied to the same 5,790 treatment websites over calendar-year half-blocks (CW27–52 of 2018 versus CW27–52 of 2017, roughly the first six months of enforcement; that block pooled with CW1–21 of 2019 versus CW1–21 of 2018, roughly twelve months), the estimates converge tightly with our main results: −9.76% at twelve months against Table 2's −9.57% (a gap of 0.2 percentage points), and −5.70% at roughly six months, which falls between the 3-month (−4.88%) and 6-month (−7.22%) windows of Table 2.

*Table W18. Comparison of the Effect of GDPR on the Weekly Visits Obtained from a Pooled and an Individual Generalized Synthetic Control (GSC) Estimator (Panel A) and Obtained from a Panel-Differences Estimator and an Individual Generalized Synthetic Control (GSC) Estimator (Panel B)*

**Panel A. Pooled vs. Individual (Website-Level) GSC Estimator**

| **Metric** | | **3 months** | **6 months** | **9 months** | **12 months** | **18 months** |
|---|---|---|---|---|---|---|
| **Visits (Pooled GSC)** | Mean | -3.00% | -6.02% | -6.85% | -10.64% | -13.01% |
| | 95% Confidence interval | [-3.87, -2.10] | [-6.97, -5.13] | [-8.55, -5.10] | [-12.81, -8.32] | [-16.25, -9.57] |
| | Number of treatment websites | 5,790 | 5,790 | 5,790 | 5,790 | 5,790 |
| | Number of control websites | 1,701 | 1,701 | 1,701 | 1,701 | 1,701 |
| **Visits (Website-Level GSC)** | Mean: | -4.88% | -7.22% | -9.07% | -9.57% | -10.02% |
| | 95% Confidence interval | [-5.50, -4.23] | [-7.83, -6.57] | [-9.72, -8.36] | [-10.25, -8.81] | [-10.79, -9.18] |
| | Number of treatment websites | 5,790 | 5,790 | 5,790 | 5,790 | 5,790 |
| | Number of control websites | 1,701 | 1,701 | 1,701 | 1,701 | 1,701 |
| **Absolute Difference** | | 1.88 pp | 1.20 pp | 2.22 pp | 1.07 pp | 2.99 pp |

Notes: Panel A compares the GDPR effect on weekly visits estimated by our website-instance-specific generalized synthetic control (GSC) estimator (Table 2) and a pooled GSC estimator that jointly estimates a common interactive fixed-effect structure across all treated and control website-instances. Both estimators use the same underlying sample of 5,790 treatment websites and 1,701 control websites; the website-instance-specific estimator aggregates effects to the website level, whereas the pooled estimator is estimated at the website-instance level. 95% confidence intervals are bootstrap intervals obtained by resampling websites.

**Panel B. Panel-Differences vs. Individual (Website-Level) GSC Estimator**

| **Metric** | | **~6 months** | **~12 months** |
|---|---|---|---|
| **Visits (Panel Differences Estimator)** | Mean | -5.70% | -9.76% |
| | 95% Confidence interval | [-7.20, -4.17] | [-11.22, -8.28] |
| | Number of treatment websites | 5,790 | 5,790 |
| **Visits (Website-Level GSC Estimator)** | Mean | -7.22% | -9.57% |
| | 95% Confidence interval | [-7.83, -6.57] | [-10.25, -8.81] |
| | Number of treatment websites: | 5,790 | 5,790 |
| | Number of control websites: | 1,701 | 1,701 |
| **Absolute Difference** | | 1.52 pp | 0.19 pp |

Notes. Panel B compares the effect of GDPR on weekly visits estimated by the panel-differences estimator (Eichenbaum et al. 2024), in which each website serves as its own control against its own traffic in the same calendar week one year earlier, and our main website-level GSC estimator (Table 2). The panel-differences estimator uses no cross-website control group, so spillover across treated and control websites cannot bias the estimates. The previous-year counterfactual assumes no secular year-over-year growth in website traffic; positive secular growth would make the estimator understate the decline, and the tight convergence with the GSC estimate bounds this concern. For the panel-differences estimator, 95% confidence intervals are based on standard errors clustered at the website level; for the website-level GSC estimator, 95% confidence intervals are bootstrap intervals obtained by resampling websites.

## G.4 Alternative Functional Forms

Chen and Roth (2024) show that the log(Y + 1) transformation can distort treatment-effect estimates when zero outcomes constitute a meaningful share of the distribution, as in bilateral trade flows or individual earnings. Our setting is materially different: zeros are vanishingly rare. Across the 289,248 website-instance weeks in this appendix's panel (1,104 websites × 2 instances × 131 weeks), only 593 (0.21%) have zero weekly visits, and the unique-visitor subsample contains no zero months at all by construction, because the ≥5,000-unique-visitors-per-month reporting requirement keeps every retained instance-month strictly positive. With the distribution mass near 170,000 weekly visits — five orders of magnitude above the Chen and Roth (2024) zone of concern for very low Y — the +1 adjustment has no practical effect on our estimates.

To confirm the non-distortment of treatment-effect estimates, we re-estimate the per-website DID on the 1,104 within-website websites under four functional forms: OLS on log(Y + 1) (our baseline, which replicates the DID Estimator row of Table W17 exactly), OLS on log(Y) using positive observations only (dropping the +1 adjustment entirely), Poisson, and Negative Binomial, aggregating the per-website coefficients across websites as the cross-sectional mean (Panel A) or median (Panel B) of $(\exp(\beta_i) - 1) \times 100$ with bootstrap 95% confidence intervals. All four yield negative effects that increase monotonically in magnitude, and at 6, 9, 12, and 18 months, all four are statistically significant at the 5% level for both mean and median aggregations. Thus, the finding that GDPR yields a sustained and growing decline in EU-user traffic is fully robust to the functional form.

The OLS log(Y) specification, which drops the observations Chen and Roth (2024) flag, lands within roughly 1.4 percentage points of the baseline at every horizon, directly confirming that the +1 adjustment does not drive our estimates. The cross-specification spread is modest — at 12 months, mean effects range from −8.53% (Poisson and Negative Binomial) to −12.03% (OLS log(Y + 1)) and medians from −12.05% to −13.52% — with Poisson and Negative Binomial producing identical point estimates (they share the same per-website systematic mean structure; the dispersion parameter affects only the confidence intervals) and slightly smaller magnitudes that reflect the level-weighting these specifications place on larger websites. The single non-significant cell is the Poisson/Negative Binomial mean at 3 months (−2.54%, 95% CI [−5.3, +0.2] for Poisson and [−5.2, +0.4] for Negative Binomial); the corresponding medians are significant (−8.58%, CI [−10.3, −6.5]), and all four confidence intervals overlap at every post-treatment period in both panels.

*Table W19. Robustness of the GDPR Effect on Weekly Visits to Alternative Functional Forms*

| **Panel A: Mean across 1,104 websites** | | | | | | |
|---|---|---|---|---|---|---|
| **Functional Form** | | **Post-Treatment Period** | | | | |
| | | **3 months** | **6 months** | **9 months** | **12 months** | **18 months** |
| **OLS log(Y+1)** | Mean<br>95% CI | -7.18%<br>[-9.8, -4.2] | -10.58%<br>[-13.0, -7.8] | -11.41%<br>[-13.7, -8.8] | -12.03%<br>[-14.4, -9.4] | -12.50%<br>[-14.8, -10.1] |
| **OLS log(Y)** | Mean<br>95% CI | -5.97%<br>[-8.5, -3.5] | -9.34%<br>[-11.5, -7.0] | -9.89%<br>[-12.2, -7.4] | -10.58%<br>[-12.9, -8.3] | -11.12%<br>[-13.4, -8.8] |
| **Poisson** | Mean<br>95% CI | -2.54%<br>[-5.3, +0.2] | -6.83%<br>[-9.1, -4.5] | -7.63%<br>[-9.9, -5.2] | -8.53%<br>[-10.8, -6.0] | -9.41%<br>[-11.6, -7.3] |
| **Negative Binomial** | Mean<br>95% CI | -2.54%<br>[-5.2, +0.4] | -6.83%<br>[-9.1, -4.3] | -7.63%<br>[-9.9, -5.3] | -8.53%<br>[-10.9, -5.9] | -9.41%<br>[-11.5, -7.1] |
| **All CI overlap** | | Yes | Yes | Yes | Yes | Yes |
| **Panel B: Median across 1,104 websites** | | | | | | |
| **OLS log(Y+1)** | Median<br>95% CI | -11.02%<br>[-13.0, -9.0] | -12.60%<br>[-13.9, -10.7] | -13.51%<br>[-15.0, -11.8] | -13.52%<br>[-15.1, -11.9] | -14.67%<br>[-16.7, -12.6] |
| **OLS log(Y)** | Median<br>95% CI | -9.64%<br>[-11.7, -8.2] | -11.88%<br>[-13.3, -10.3] | -12.58%<br>[-14.1, -10.8] | -12.87%<br>[-14.6, -11.3] | -13.79%<br>[-15.8, -12.1] |
| **Poisson** | Median<br>95% CI | -8.58%<br>[-10.3, -6.5] | -10.60%<br>[-12.5, -8.9] | -11.41%<br>[-12.9, -9.9] | -12.05%<br>[-13.8, -10.4] | -12.94%<br>[-15.0, -11.3] |
| **Negative Binomial** | Median<br>95% CI | -8.58%<br>[-10.3, -6.5] | -10.60%<br>[-12.5, -8.9] | -11.41%<br>[-12.8, -10.0] | -12.05%<br>[-13.8, -10.4] | -12.94%<br>[-14.8, -11.3] |
| **All CI overlap** | | Yes | Yes | Yes | Yes | Yes |

Notes: This table reports the GDPR effect on EU-user weekly visits estimated via per-website DiD on the Web Appendix G sample (1,104 non-EU websites with both EU- and non-EU-traffic instances), aggregated identically to the DID Estimator row of Table W17. For each website, a within-website DiD coefficient is estimated using the corresponding functional form: OLS on log(Y+1), OLS on log(Y) (positive observations only; 0.27% of observations dropped globally), Poisson, or Negative Binomial. Per-website coefficients are then aggregated across the 1,104 websites as the cross-sectional mean (Panel A) or median (Panel B) of $(\exp(\beta_i) - 1) \times 100$, with 95% bootstrap confidence intervals. The OLS log(Y+1) Mean row replicates the DID Estimator row of Table W17 exactly. Poisson and Negative Binomial yield identical point estimates because the per-website systematic mean structure is identical; the dispersion parameter affects only the confidence intervals. The "All CI overlap" row at the bottom of each panel indicates whether the four specifications' 95% confidence intervals share a common region at each horizon; every cell reads "Yes," showing substantial overlap in the uncertainty intervals across specifications.

## G.5 Temporal Aggregation: Weekly vs. Monthly Resolution

SimilarWeb reports unique visitors at the monthly level only, while it reports visits weekly. To verify that this difference in temporal resolution does not affect our estimates, we re-estimate the pooled GSC specification using monthly-aggregated visits instead of weekly visits and compare the resulting effects with their weekly-resolution counterparts. Absolute differences between the weekly and monthly point estimates range from 0.24 to 3.22 percentage points across the five post-treatment periods — at most one percentage point at 3, 6, and 9 months and roughly three percentage points at 12 and 18 months — and the 95% bootstrap confidence intervals overlap at every post-treatment period. Together with the similarity of the point estimates, this indicates that temporal aggregation does not alter the qualitative conclusion of a sustained negative GDPR effect.

*Table W20. Robustness of the GDPR Effect on Weekly Visits to Temporal Aggregation: Pooled GSC at Weekly vs. Monthly Resolutions*

| **Metric** | | **Post-Treatment Period** | | | | |
|---|---|---|---|---|---|---|
| | | **3 months** | **6 months** | **9 months** | **12 months** | **18 months** |
| **Weekly Visits (Pooled GSC)** | Mean<br>95% Confidence interval | -3.00%<br>[-3.87, -2.10] | -6.02%<br>[-6.97, -5.13] | -6.85%<br>[-8.55, -5.10] | -10.64%<br>[-12.81, -8.32] | -13.01%<br>[-16.25, -9.57] |
| **Monthly Visits (Pooled GSC)** | Mean<br>95% Confidence interval | -2.09%<br>[-3.04, -1.14] | -5.05%<br>[-6.12, -3.96] | -6.61%<br>[-7.95, -5.34] | -7.57%<br>[-9.58, -5.55] | -9.79%<br>[-13.03, -6.43] |
| **Absolute Difference** | | 0.91 pp | 0.97 pp | 0.24 pp | 3.07 pp | 3.22 pp |
| **95% CI overlap** | | Yes | Yes | Yes | Yes | Yes |
| **Number of treatment websites** | | 5,790 | 5,790 | 5,790 | 5,790 | 5,790 |
| **Number of control websites** | | 1,701 | 1,701 | 1,701 | 1,701 | 1,701 |

Notes: This table reports the GDPR effect on visits estimated by the pooled GSC at two temporal resolutions on the same Table 2 sample. The weekly-resolution specification uses ln(Weekly Visits + 1) as the outcome; the monthly-resolution specification aggregates weekly visits to monthly within each (website, year-month) cell and uses ln(Monthly Visits + 1) as the outcome. The 95% CI overlap row indicates whether the weekly and monthly intervals share any interval.

## G.6 In-Time Placebo Tests

Finally, we verify that our estimator does not detect spurious effects in the absence of treatment. Using the same matched-control difference-in-differences estimator with week fixed effects as our within-website analysis — for each website, $\log(1 + \text{visits}) = \gamma \cdot \text{EU} + \beta \cdot \text{Treated}$ + week fixed effects, estimated on its EU-traffic series against five correlation-matched non-EU control websites — and aggregating across the 1,104 websites, we assign a placebo enforcement date to each pre-enforcement month and re-estimate using only pre-enforcement data. At every placebo enforcement date from October 2017 through April 2018, the estimated effect on weekly visits is statistically indistinguishable from zero: the point estimates range from +12.8% to −3.1%, all 95% confidence intervals contain zero, and all p-values exceed 0.05. We conclude that our estimates are not an artifact of the estimator, although we had to omit placebo dates in August and September 2017, for which the number of pre-treatment weeks is too small to render the per-website estimates reliable.

*Table W21. In-Time Placebo Test on the Within-Website Subsample (Web Appendix G): Effect on Weekly Visits at Fake GDPR Pre-enforcement Treatment Dates.*

| Placebo treatment date | Pre-treatment weeks | Mean % effect [95% CI] | p-value |
|---|---|---|---|
| Oct 2017 | 15 | +12.83 [-0.54, 31.36] | 0.071 |
| Nov 2017 | 19 | +5.07 [-3.07, 15.04] | 0.249 |
| Dec 2017 | 23 | +1.72 [−3.97, 8.96] | 0.615 |
| Jan 2018 | 27 | −0.31 [−5.13, 6.39] | 0.869 |
| Feb 2018 | 31 | −1.40 [−6.66, 5.88] | 0.611 |
| Mar 2018 | 35 | −2.60 [−6.35, 2.12] | 0.252 |
| Apr 2018 | 39 | −3.07 [−5.81, 0.20] | 0.060 |

Notes: Estimator = per-website matched-control difference-in-differences with week fixed effects, log(1+visits) = $\gamma \cdot$EU + $\beta \cdot$Treated + week FE, estimated separately for each of the 1,104 within-website overlap websites on its EU-traffic series and five correlation-matched non-EU control websites; each placebo assigns a fake enforcement date and uses pre-enforcement data only. 95% confidence intervals and two-sided p-values from a website-level nonparametric bootstrap. Placebo dates in August and September 2017 (7 and 11 pre-treatment weeks, respectively) are omitted because too few pre-treatment weeks make the per-website estimates unreliable.

## WEB APPENDIX H: Heterogeneity of GDPR's Effect Across Website Characteristics

This appendix characterizes heterogeneity in the website-level effects of the GDPR. It defines the characteristics and their interpretation (H.1), makes the unconditional heterogeneity the centerpiece (H.2), asks whether the strongest patterns survive adjustment for website size, EU-user share, and industry composition (H.3), summarizes supplementary inference and robustness (H.4), and states the findings (H.5). Because the characteristics are pre-treatment but not randomly assigned, every result is a descriptive association that locates where the GDPR effect concentrates, not a causal moderator.

### H.1 Website Characteristics and Interpretation

The generalized synthetic control procedure outlined in the paper produces an estimated treatment effect for each treated website-instance, and after aggregation, for each treated website, outcome metric, and post-treatment period. This appendix regresses those website-level effects, not the average treatment effects reported in the manuscript, on pre-GDPR website characteristics. Table W22 lists the characteristics, what each captures, its source and construction, the definition of the "High" group with the resulting split, and coverage as a share of the underlying sample.

*Table W22. Website Characteristics: Definition, Construction, and Coverage*

| Characteristic | What it captures | Source (date) | "High" group (share of covered) | Coverage |
|---|---|---|---|---|
| Hedonic (vs. utilitarian) | Content orientation | LLM-ensemble classification (pre-GDPR content) | Hedonic; 26.6% | 100.0% |
| Data-type sensitivity | Sensitivity of the personal data typically involved in the industry's core service | Industry-level data-sensitivity index | Above-median industry sensitivity (median 45); 46.4% | 100.0% |
| Homepage ad intensity | Advertising dependence of the website | Internet Archive homepage snapshots (April 2018) | ≥5 rendered ad slots (≈ top quartile); 28.4% | 91.4% |
| Tracking intensity | Third-party tracker load | WhoTracks.me census (April 2018), Ghostery TrackerDB | Above-median (4 trackers); 45.7% | 47.5% |
| Regulatory strictness | Stringency of the responsible data-protection regime | EC Flash Eurobarometer 226 (2008), Survey of 4,835 data controllers | Above-median score (2.53); 45.0% | 81.2% |
| Enforcement capability | Resources of the responsible data-protection authority | DPA budgets 2017, EDPB report (2020) | Above-median budget/capita (€0.30); 48.3% | 81.2% |
| Mobile share | Mobile orientation of the audience | SimilarWeb (pre-GDPR weeks) | Above-median (44%); 50.8% | 100% |
| Website size | Traffic scale | SimilarWeb traffic rank (pre-GDPR) | Larger (industry rank below median of 10.5); 50.0% | 98.9% |

Notes. Coverage is the share of the 5,790 treated websites in our main sample; unique visitor analyses use the corresponding 5,447-website sample. All splits are at the median except homepage ad intensity (≥5 rendered ad slots) and the binary hedonic classification. Data-type sensitivity is an industry-level index: every website inherits its industry's data-sensitivity score, so the measure does not vary within industries. Regulatory strictness and enforcement capability are defined only for majority-EU user websites; tracking intensity only for websites matched in the WhoTracks.me census. Hedonic websites include, e.g., youtube.com, netflix.com, spotify.com; utilitarian websites include google.com, wikipedia.org, linkedin.com.

*Construction.* Content orientation (hedonic vs. utilitarian) is assigned by an ensemble of three separate, context-isolated runs of the same large language model (Claude Opus 4.8), each applying a common pre-specified written codebook to every website from only the website's domain and the model's background knowledge of it as of 2018, with the label set by majority vote. The runs saw no traffic data, estimated effects, or regression results, and did not see one another's labels. Because the classification depends only on the identity of the website, it cannot encode the outcomes under study; inter-run agreement is high (Fleiss' $\kappa = 0.97$, 98% unanimous, 26.6% hedonic).

Data-type sensitivity is an industry-level index of how sensitive the personal data are that an industry's core service typically involves: we score each of the 151 industry categories in our taxonomy on a 1-to-100 scale, anchored by the kinds of data the GDPR singles out for heightened protection. That is, industries built on special-category data under Article 9 receive the highest scores (health and its sub-fields at 95, adult content at 90), followed by industries handling financial and identity data (finance, 85) and transactional data (e-commerce, 75), while industries that collect little personal data in the course of their service sit at the floor (heavy industry, engineering, agriculture, and pets, at 10); every website inherits its industry's score, so the measure is constant within industries and its conditional specification therefore cannot include industry fixed effects.

We measure homepage ad intensity by retrieving archived homepage snapshots from the Internet Archive's Wayback Machine for the last pre-GDPR capture (April 2018): we count the distinct rendered advertising slots on each captured homepage and label a website high-intensity if it shows at least five, approximately the top quartile of the covered distribution.

Tracking intensity is the number of distinct third-party trackers a website loaded in the April 2018 WhoTracks.me census, with trackers identified through the Ghostery TrackerDB, split at the median of four.

Regulatory strictness is a country-level stringency score for the responsible data-protection regime from the 2008 EC Flash Eurobarometer 226 survey of 4,835 data controllers (European Commission 2008), and enforcement capability is the responsible authority's 2017 budget per capita (European Data Protection Board 2020). We assign each to a website through its majority EU-user country. Thus, we define both only if we can identify a single responsible regime and authority.

Mobile share is the pre-GDPR average share of a website's visits originating from mobile devices (SimilarWeb), split at the median (44%).

Website size is the website's pre-GDPR SimilarWeb traffic rank within its industry; "larger" denotes the higher-ranked half (an industry rank below the median of 10.5).

*Coverage.* Coverage varies across characteristics, from complete (hedonic orientation, data-type sensitivity, mobile share) to approximately 48% for tracking intensity, which we limit to websites matched in the WhoTracks.me census. This incompleteness does not threaten the analysis. Each characteristic is examined one at a time on its own covered subsample, so a website missing one measure still contributes to every other comparison, and nothing is dropped list-wise. The quantity being compared, the website-instance level GDPR effect, is estimated in the first stage for every treated website regardless of which characteristics are observed, so coverage determines only which websites enter a given High–Low split, not the effects being split. The missingness is not selective on the outcome: among covered websites, the mean GDPR effect tracks the full-sample mean closely, within about 0.3 percentage points for homepage ad intensity and website size, and within roughly one to two percentage points for the institutional measures and the tracking subsample, at both the three- and eighteen-month horizons, so restricting to covered websites shifts composition only marginally.

For regulatory strictness and enforcement capability, the restriction is moreover definitional rather than incidental, since a single responsible regime and authority exist only for majority-EU-user websites. We therefore report coverage transparently and read each comparison as descriptive of its covered subsample, without imputing values or extrapolating to websites where a characteristic cannot be measured.

## H.2 Unconditional Heterogeneity

Table W22 catalogs eight characteristics, of which two are treated apart from the standard High-Low comparison. Website size is examined separately in the main text: because its association with the GDPR effect is nonmonotonic, a median High-Low split would be misleading, and we characterize it instead through its full decile profile. We report mobile share in the tables below, but it shows little consistent association with overall usage, and enters the analysis mainly as supplementary evidence for the consent-friction mechanism. Figure W6 accordingly plots the High- and Low-group effect trajectories for the six focal characteristics, communicating the dynamic patterns more efficiently than the coefficient tables that follow, and Tables W23 and W24 report the full High-versus-Low descriptive results, mobile share included, for all five outcomes.

*Figure W6. Unconditional GDPR Effect on Weekly Visits, High vs. Low Group, by Characteristic and Post-Treatment Period*

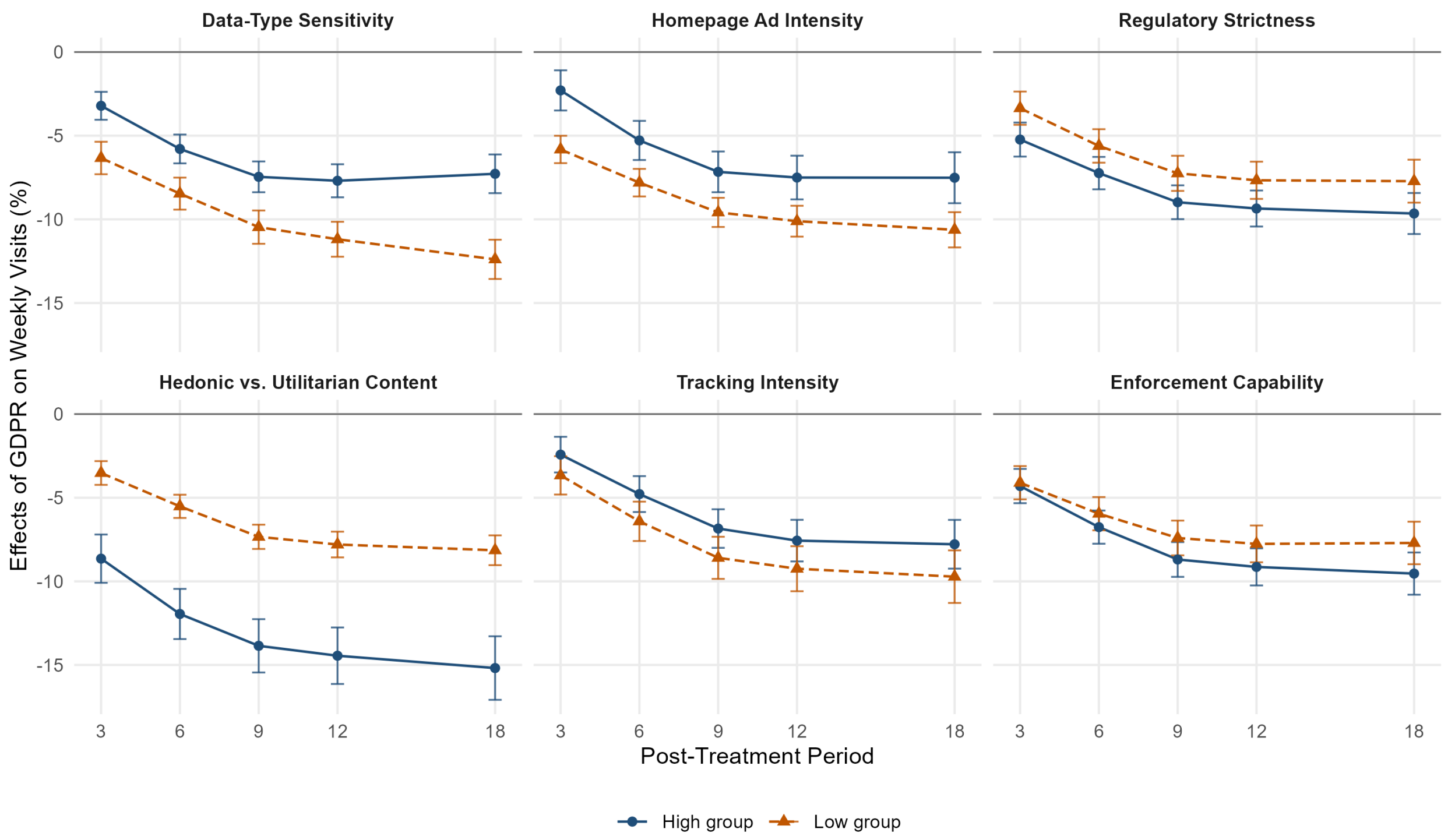

Notes: Each panel shows the average effect of GDPR on weekly visits, in percent, for websites with high versus low values of a single pre-GDPR website characteristic, across the five cumulative post-treatment periods (3, 6, 9, 12, and 18 months after enforcement). Effects are website-level GSC estimates with error bars showing 95% confidence intervals for the group means. Websites are split at the median for each characteristic, except for homepage ad intensity (High = at least five display-ad slots; top quartile) and content orientation (High = hedonic; Low = utilitarian). The splits are descriptive, not causal. Group means are equal to the corresponding entries in Table W23 (weekly visits).

*Table W23. Unconditional GDPR Effect on Total Usage and Usage Frequency (%), High vs. Low*

**Panel A. Weekly visits**

| Characteristic | Group | Share | 3m | 6m | 9m | 12m | 18m |
|---|---|---|---|---|---|---|---|
| Hedonic | High | 26.6% | -8.64 | -11.95 | -13.85 | -14.45 | -15.18 |
| | Low | 73.4% | -3.52 | -5.52 | -7.34 | -7.80 | -8.14 |
| Data-type sensitivity | High | 46.4% | -3.21 | -5.79 | -7.46 | -7.69 | -7.28 |
| | Low | 53.6% | -6.33 | -8.46 | -10.47 | -11.19 | -12.39 |
| Homepage ad intensity | High | 28.4% | -2.29 | -5.28 | -7.16 | -7.50 | -7.51 |
| | Low | 71.6% | -5.82 | -7.81 | -9.58 | -10.11 | -10.63 |
| Tracking intensity | High | 45.7% | -2.42 | -4.78 | -6.84 | -7.57 | -7.78 |
| | Low | 54.3% | -3.66 | -6.41 | -8.59 | -9.25 | -9.72 |
| Regulatory strictness | High | 45.0% | -5.23 | -7.24 | -8.98 | -9.35 | -9.65 |
| | Low | 55.0% | -3.36 | -5.61 | -7.25 | -7.67 | -7.72 |
| Enforcement capability | High | 48.3% | -4.30 | -6.76 | -8.69 | -9.14 | -9.54 |
| | Low | 51.7% | -4.10 | -5.96 | -7.41 | -7.76 | -7.71 |
| Mobile share | High | 50.0% | -4.07 | -6.38 | -8.41 | -8.80 | -9.32 |
| | Low | 50.0% | -5.70 | -8.06 | -9.73 | -10.33 | -10.71 |

**Panel B. Unique visitors**

| Characteristic | Group | Share | 3m | 6m | 9m | 12m | 18m |
|---|---|---|---|---|---|---|---|
| Hedonic | High | 26.6% | -4.87 | -8.22 | -10.54 | -11.43 | -12.51 |
| | Low | 73.4% | +0.19 | -1.74 | -3.72 | -4.17 | -3.97 |
| Data-type sensitivity | High | 47.0% | +0.55 | -1.70 | -3.48 | -3.67 | -3.12 |
| | Low | 53.0% | -1.95 | -4.66 | -7.28 | -8.41 | -9.71 |
| Homepage ad intensity | High | 29.0% | +1.48 | -0.80 | -2.56 | -3.01 | -3.04 |
| | Low | 71.0% | -1.77 | -4.18 | -6.46 | -7.18 | -7.73 |
| Tracking intensity | High | 46.2% | +0.92 | -1.66 | -3.90 | -4.66 | -4.86 |
| | Low | 53.8% | +0.25 | -2.40 | -4.78 | -5.61 | -5.93 |
| Regulatory strictness | High | 44.7% | -1.14 | -3.52 | -5.81 | -6.42 | -6.36 |
| | Low | 55.3% | +0.63 | -1.80 | -3.97 | -4.55 | -5.06 |
| Enforcement capability | High | 48.4% | -0.11 | -2.92 | -5.47 | -6.33 | -6.64 |
| | Low | 51.6% | -0.21 | -2.24 | -4.15 | -4.49 | -4.70 |
| Mobile share | High | 49.4% | -0.29 | -2.65 | -5.17 | -5.76 | -6.31 |
| | Low | 50.6% | -1.25 | -3.87 | -5.81 | -6.59 | -6.91 |

**Panel C. Weekly page impressions**

| Characteristic | Group | Share | 3m | 6m | 9m | 12m | 18m |
|---|---|---|---|---|---|---|---|
| Hedonic | High | 26.3% | -7.86 | -10.52 | -12.58 | -12.97 | -16.88 |
| | Low | 73.7% | -1.86 | -2.84 | -3.94 | -3.95 | -6.32 |
| Data-type sensitivity | High | 45.6% | -0.69 | -2.53 | -3.59 | -3.42 | -6.05 |
| | Low | 54.4% | -5.16 | -6.76 | -8.63 | -9.05 | -12.07 |
| Homepage ad intensity | High | 28.4% | -0.72 | -2.85 | -4.37 | -4.48 | -7.63 |
| | Low | 71.6% | -3.91 | -5.45 | -6.91 | -7.08 | -9.64 |
| Tracking intensity | High | 45.9% | -1.44 | -3.29 | -4.81 | -5.10 | -7.91 |
| | Low | 54.1% | -2.81 | -4.66 | -6.32 | -6.66 | -9.90 |
| Regulatory strictness | High | 45.4% | -2.36 | -3.63 | -4.97 | -4.86 | -6.97 |
| | Low | 54.6% | -1.98 | -3.27 | -4.40 | -4.50 | -8.92 |
| Enforcement capability | High | 49.1% | -1.79 | -3.33 | -4.72 | -4.67 | -7.68 |
| | Low | 50.9% | -2.50 | -3.53 | -4.60 | -4.65 | -8.37 |
| Mobile share | High | 49.8% | -2.64 | -4.34 | -6.02 | -5.95 | -9.10 |
| | Low | 50.2% | -3.50 | -5.22 | -6.53 | -6.91 | -9.43 |

Notes: Entries are the mean of the website-level percentage effects, within each group; groups are defined as in Table W22. Entries are unconditional (not adjusted for size, EU-user share, or industry). Share is each group's percentage of the characteristics covered on websites, with the High and Low groups summing to 100.

*Table W24. Unconditional GDPR Effect on Usage Intensity (%), High vs. Low*

**Panel A. Visits per unique visitor**

| Characteristic | Group | Share | 3m | 6m | 9m | 12m | 18m |
|---|---|---|---|---|---|---|---|
| Hedonic | High | 26.4% | -1.19 | -1.22 | -0.05 | +0.52 | +1.77 |
| | Low | 73.6% | -1.77 | -1.63 | -1.29 | -1.29 | -1.42 |
| Data-type sensitivity | High | 47.4% | -1.67 | -1.53 | -0.70 | -0.34 | +0.41 |
| | Low | 52.6% | -1.58 | -1.52 | -1.21 | -1.24 | -1.49 |
| Homepage ad intensity | High | 29.1% | -1.49 | -2.00 | -1.54 | -1.34 | -0.86 |
| | Low | 70.9% | -1.55 | -1.32 | -0.75 | -0.64 | -0.40 |
| Tracking intensity | High | 46.2% | -1.44 | -1.00 | -0.33 | -0.17 | +0.13 |
| | Low | 53.8% | -1.52 | -1.68 | -1.15 | -0.95 | -0.56 |
| Regulatory strictness | High | 44.7% | -1.56 | -1.27 | -0.50 | -0.15 | +0.32 |
| | Low | 55.3% | -2.12 | -2.29 | -1.89 | -1.92 | -1.72 |
| Enforcement capability | High | 48.0% | -1.87 | -1.84 | -1.18 | -0.76 | -0.32 |
| | Low | 52.0% | -1.87 | -1.82 | -1.35 | -1.46 | -1.26 |
| Mobile share | High | 49.7% | -1.85 | -1.33 | -0.53 | -0.32 | -0.08 |
| | Low | 50.3% | -1.39 | -1.72 | -1.40 | -1.31 | -1.08 |

**Panel B. Page impressions per visit**

| Characteristic | Group | Share | 3m | 6m | 9m | 12m | 18m |
|---|---|---|---|---|---|---|---|
| Hedonic | High | 26.2% | +0.63 | +1.47 | +1.79 | +2.52 | -0.46 |
| | Low | 73.8% | +2.43 | +3.29 | +4.20 | +5.02 | +3.04 |
| Data-type sensitivity | High | 46.3% | +3.37 | +4.25 | +5.21 | +6.02 | +3.56 |
| | Low | 53.7% | +0.77 | +1.60 | +2.17 | +2.97 | +0.94 |
| Homepage ad intensity | High | 28.7% | +1.86 | +3.02 | +3.82 | +4.45 | +1.75 |
| | Low | 71.3% | +2.11 | +2.77 | +3.45 | +4.29 | +2.29 |
| Tracking intensity | High | 46.0% | +1.39 | +1.94 | +2.80 | +3.34 | +0.97 |
| | Low | 54.0% | +1.10 | +2.36 | +3.10 | +3.88 | +1.22 |
| Regulatory strictness | High | 45.3% | +3.21 | +4.53 | +5.55 | +6.51 | +4.62 |
| | Low | 54.7% | +1.47 | +2.32 | +3.14 | +4.04 | -0.16 |
| Enforcement capability | High | 48.6% | +2.54 | +4.03 | +5.05 | +6.06 | +3.22 |
| | Low | 51.4% | +1.99 | +2.65 | +3.46 | +4.31 | +0.85 |
| Mobile share | High | 50.0% | +2.03 | +2.87 | +3.71 | +4.71 | +2.66 |
| | Low | 50.0% | +1.92 | +2.78 | +3.46 | +4.06 | +1.65 |

*Notes. As in Table W23; intensity metrics are ratios (visits per unique visitor; page impressions per visit).*

Three substantive patterns organize these results:

*Hedonic orientation***.** Content orientation is the strongest and most persistent source of heterogeneity. Hedonic websites lose substantially more usage than utilitarian websites on every total usage and usage frequency metric, and the gap widens over time: on weekly visits, the hedonic group moves from about -8.6% at three months to -15.2% at eighteen months, against -3.5% to -8.1% for the utilitarian group.

*Usage frequency versus intensity.* Most of the heterogeneity in total usage is associated with the number of users rather than with usage per user: the High-Low differences are largest for unique visitors among the decomposition components and attenuate for the intensity ratios, where usage intensity among remaining users sometimes offsets part of the decline.

*Secondary patterns***.** High-homepage-ad-intensity websites show smaller short-run declines, with the High-Low gap narrowing through nine months before widening again; regulatory strictness and enforcement capability show only small High-Low gaps. Data-type sensitivity shows a stable ordering: websites in more data-sensitive industries experience smaller usage declines than those in less-sensitive industries, on visits, unique visitors, and page impressions, on weekly visits to about three percentage points smaller at three months and five at eighteen, and the gap is roughly constant through twelve months before widening at eighteen. Because the measure is defined at the industry level, however, we cannot separate this effect from the one of industry composition; the conditional model for this characteristic omits industry fixed effects, so we read it as a between-industry association rather than a within-industry effect.

### H.3 Conditional Association Summary

A reasonable remaining question is whether the raw High-Low differences merely reflect website size, EU-user share, or industry composition. Table W25, therefore, reports the conditional associations from regressing the website-level effects on each characteristic together with website-size-decile fixed effects, EU-user share, and industry fixed effects (with the focal characteristic varying within industry). Data-type sensitivity is the exception: because the industry-level index does not vary within industries, its conditional model omits the industry fixed effects, so its coefficient reflects a between-industry comparison net only of size and EU-user share. The table shows the primary unweighted-OLS estimand at 3 and 18 months with 95% confidence intervals; a dagger (†) marks associations that remain significant after a Benjamini–Hochberg adjustment for multiple comparisons across the seven characteristics within each outcome and horizon. Hedonic orientation and data-type sensitivity enter as binary “high indicators” (coefficient in percentage points); the other characteristics enter as standardized continuous variables (per one-standard-deviation increase).

*Table W25. Conditional Associations, Unweighted OLS, at 3 and 18 Months (coefficient [95% CI])*

**Panel A. Total usage and usage frequency**

| Characteristic | Weekly visits 3m | Weekly visits 18m | Unique visitors 3m | Unique visitors 18m | Page impressions 3m | Page impressions 18m |
|---|---|---|---|---|---|---|
| Hedonic | -3.98† [-7.1, -0.9] | -8.97† [-12.8, -5.2] | -1.58 [-4.0, +0.8] | -4.65† [-7.8, -1.5] | -2.79 [-6.3, +0.7] | -8.41† [-12.5, -4.3] |
| Data-type sensitivity | +4.13† [+2.8, +5.5] | +5.88† [+4.2, +7.6] | +2.49† [+1.2, +3.8] | +5.23† [+3.5, +7.0] | +5.89† [+4.4, +7.4] | +6.49† [+4.6, +8.4] |
| Homepage ad intensity | +1.96† [+1.0, +3.0] | +1.12 [-0.1, +2.3] | +1.66† [+0.8, +2.5] | +1.29 [+0.2, +2.4] | +0.75 [-0.4, +1.9] | +0.62 [-0.8, +2.0] |
| Tracking intensity | +0.65 [-0.7, +1.9] | +1.10 [-0.6, +2.8] | +0.26 [-0.8, +1.3] | +0.36 [-1.1, +1.8] | +0.58 [-0.9, +2.1] | +1.01 [-0.9, +2.9] |
| Regulatory strictness | -0.28 [-2.8, +2.2] | +0.00 [-2.4, +2.4] | -0.81 [-2.5, +0.9] | -1.06 [-3.3, +1.1] | +0.18 [-1.7, +2.1] | +1.37 [-0.7, +3.4] |
| Enforcement capability | +0.47 [-1.6, +2.6] | +0.10 [-2.1, +2.3] | -0.08 [-1.5, +1.4] | -0.63 [-2.8, +1.5] | +0.43 [-1.2, +2.1] | +0.51 [-1.4, +2.4] |
| Mobile share | -0.91 [-2.0, +0.2] | -1.26 [-2.7, +0.1] | +0.06 [-0.9, +1.0] | -0.50 [-1.7, +0.7] | -1.04 [-2.3, +0.2] | -0.37 [-1.9, +1.2] |

**Panel B. Usage intensity**

| Characteristic | Visits/user 3m | Visits/user 18m | Pages/visit 3m | Pages/visit 18m |
|---|---|---|---|---|
| Hedonic | -1.79 [-5.6, +2.0] | -5.21 [-10.5, +0.1] | -1.80 [-5.1, +1.5] | -4.59 [-8.9, -0.2] |
| Data-type sensitivity | +1.65 [+0.1, +3.2] | +3.77† [+1.8, +5.8] | +2.60† [+1.2, +4.0] | +2.39† [+0.5, +4.3] |
| Homepage ad intensity | +0.03 [-1.0, +1.1] | -0.65 [-2.0, +0.7] | -1.58† [-2.6, -0.6] | -0.06 [-1.3, +1.1] |
| Tracking intensity | -0.32 [-1.5, +0.8] | +0.17 [-1.6, +1.9] | -0.05 [-1.5, +1.4] | -0.84 [-2.7, +1.0] |
| Regulatory strictness | +0.91 [-0.1, +1.9] | +1.28 [+0.0, +2.5] | +1.02 [-0.4, +2.5] | +2.36 [-1.1, +5.8] |
| Enforcement capability | +0.68 [-0.3, +1.6] | +0.65 [-0.3, +1.6] | +0.25 [-1.0, +1.5] | +0.89 [-1.9, +3.6] |
| Mobile share | -0.33 [-1.6, +1.0] | +0.65 [-1.1, +2.4] | +0.38 [-0.8, +1.5] | +2.09† [+0.7, +3.5] |

*Notes. Each cell reports the unweighted-OLS coefficient (percentage points) with its 95% confidence interval, from a regression of website-level effects on the characteristic, website-size-decile fixed effects, EU-user share, and industry fixed effects — except data-type sensitivity, whose regression omits industry fixed effects because the industry-level index does not vary within industries. Hedonic orientation and data-type sensitivity are binary "high indicators"; the remaining characteristics are standardized (per one-SD). † marks Benjamini–Hochberg-significant coefficients ($q < .05$) across the seven characteristics within each outcome and horizon.*

Precision weighting yields the same qualitative conclusions for the principal total-usage findings, particularly the negative hedonic association. Several weaker associations attenuate, and the hedonic page-impressions-per-visit coefficient changes sign. We therefore use unweighted OLS as the primary average-website estimand and do not interpret unstable intensity associations in a directional sense.

## H.4 Supplementary Inference and Robustness

Supplementary analyses, including equivalence (TOST; Lakens et al. 2018; Tryon and Lewis 2008) and standardized-effect-size results, do not alter the ordering of the evidence, and the Benjamini–Hochberg adjustment that carries the multiplicity information is reflected in the daggers of Table W25. The principal conclusions are robust to the adjustment specification, as summarized in Table W26.

*Table W26. Robustness of the Principal Conditional Associations*

| Check | Result |
|---|---|
| EU-share quintile fixed effects (vs. linear control) | Principal coefficients essentially unchanged |
| Hedonic and data-type sensitivity jointly included | Both persist: hedonic remains negative, sensitivity remains positive |
| Unanimous hedonic classification only | Nearly identical estimates |
| Precision weighting (WLS) | Hedonic total usage result persists; usage intensity results less stable |
| Alternative size adjustment | Conclusions unchanged |

*Notes. In the joint model (hedonic orientation and the data-type-sensitivity High indicator entered together, with website-size-decile fixed effects and EU-user share; industry fixed effects omitted because the sensitivity index is industry-level), both associations persist on weekly visits: the hedonic coefficient is -3.6 at 3 months and -5.0 at 18 months, and the sensitivity coefficient is +3.9 and +5.6 (all p < .001); the pattern is the same for unique visitors and page impressions. Full TOST, Cohen's d, weight-diagnostic, flexible-EU, and unanimous-label tables are available from the authors on request.*

## H.5 Summary and Conclusion

First, content orientation is the strongest source of heterogeneity. Hedonic websites experience substantially larger declines in visits, unique visitors, and page impressions, and the gap grows over time. This result remains after adjusting for website size, EU-user share, and industry composition and is stable across the principal robustness checks.

Second, homepage ad intensity is associated with smaller short-run declines, but the relationship weakens over time and under precision weighting. Tracking intensity, regulatory strictness, enforcement capability, and mobile share show little robust association with total usage.

Third, data-type sensitivity is associated with smaller usage declines: websites in more data-sensitive industries lose less usage on visits, unique visitors, and page impressions, and the association is statistically robust. However, because the index varies only across industries, its conditional model cannot include industry fixed effects, so we cannot separate the pattern from broader industry composition. Thus, we should interpret it as a between-industry association.

## WEB APPENDIX I: Non-Cumulative Period Effects and the Timing of the Decline

Because cumulative post-treatment windows average early and late responses, they may mask the evolution of the GDPR's effect. The 18-month estimate, for example, averages the effect over the entire period from enforcement until month 18. This appendix therefore decomposes the cumulative estimates into non-cumulative period effects covering months 1-3, 4-6, 7-9, 10-12, and 13-18. The decomposition uses the same website-level GSC estimates and synthetic counterfactuals that we showed in Table 2. It shows how the magnitude of the estimated effect evolves across successive post-treatment periods and how cumulative windows smooth this variation.

*Table W27. Cumulative vs. Non-Cumulative Window Effects of GDPR on Weekly Visits*

| Metric | | 3 months | 6 months | 9 months | 12 months | 18 months |
|---|---|---|---|---|---|---|
| **Visits — Cumulative GSC (1–k)** | Mean | -4.88% | -7.22% | -9.07% | -9.57% | -10.02% |
| | 95% CI (mean) | [-5.50, -4.23] | [-7.83, -6.57] | [-9.72, -8.36] | [-10.25, -8.81] | [-10.79, -9.18] |
| | Median | -3.49% | -5.54% | -7.54% | -8.24% | -8.91% |
| | 95% CI (median) | [-4.05, -2.95] | [-6.11, -5.09] | [-8.09, -6.93] | [-8.75, -7.70] | [-9.64, -8.33] |
| | Treatment websites | 5,790 | 5,790 | 5,790 | 5,790 | 5,790 |
| | Control websites | 1,701 | 1,701 | 1,701 | 1,701 | 1,701 |
| **Visits — Non-cumulative GSC (period)** | Period | 1–3 m | 4–6 m | 7–9 m | 10–12 m | 13–18 m |
| | Mean | -4.88% | -8.18% | -10.68% | -8.59% | -7.97% |
| | 95% CI (mean) | [-5.50, -4.23] | [-8.89, -7.43] | [-11.50, -9.80] | [-9.61, -7.56] | [-9.11, -6.74] |
| | Median | -3.49% | -7.43% | -10.88% | -10.29% | -10.68% |
| | 95% CI (median) | [-4.05, -2.95] | [-7.93, -6.89] | [-11.57, -10.22] | [-11.24, -9.47] | [-11.92, -9.77] |
| | Treatment websites | 5,790 | 5,790 | 5,790 | 5,790 | 5,790 |
| | Control websites | 1,701 | 1,701 | 1,701 | 1,701 | 1,701 |
| **Difference (Non-cumulative − Cumulative)** | Δ Mean | — | −0.96*** | −1.61*** | +0.98*** | +2.05*** |
| | Δ Median | — | −1.89*** | −3.34*** | −2.05*** | −1.77*** |

Notes: Effects are reported as percentage changes; Δ values are percentage-point differences between the non-cumulative and cumulative estimates. The 95% bootstrap confidence intervals are obtained by resampling websites, and Δ = non-cumulative − cumulative is tested by paired bootstrap on the same websites (*** $p < 0.001$). Both specifications draw on the same sample of 5,790 treated websites and 1,701 control website-instances and the same synthetic counterfactuals. The cumulative effect over months 1–k combines the corresponding period-specific effects through month k. Because that combination is formed at the website level, before conversion to percentages and cross-website aggregation, the displayed cumulative summaries need not equal weighted averages of the displayed non-cumulative summaries. Column headings denote the endpoint of each post-treatment period: the cumulative estimates cover months 1–k, and the non-cumulative estimates only the period ending at month k, as shown in the period row.

Three findings emerge. First, the period-specific effects become progressively more negative over the first three quarters. The mean period effect grows from -4.88% (months 1–3) to -8.18% (months 4–6) and reaches its most negative value of -10.68% during months 7–9. This period includes the first major GDPR enforcement action, the CNIL's January 2019 fine against Google, but we should not interpret the temporal coincidence as establishing that enforcement caused the aggregate decline (Web Appendix J).

Second, the median period effect remains close to -10.5% from months 7–9 onward (-10.88%, -10.29%, and -10.68%), indicating little recovery for the typical website. In contrast, the mean becomes less negative after month 9 (to -7.97% in months 13–18), implying that the apparent recovery is concentrated in the upper portion of the cross-sectional distribution rather than reflecting broad recovery across websites.

Third, the cumulative and non-cumulative estimates differ significantly at every horizon beyond the first (paired bootstrap, all $p<.001$), consistent with cumulative windows smoothing meaningful time variation in the period-specific effects. During the build-up phase in months 4–9, the cumulative means understate the magnitude of the contemporaneous period effects. After month 9, however, the cumulative mean is slightly more negative than the corresponding non-cumulative mean because it continues to average over the earlier, more negative periods. The non-cumulative median remains more negative than the cumulative median throughout all subsequent periods. We should therefore interpret the cumulative estimates in Table 2 as smoothed summaries rather than uniformly conservative estimates of the later-period effects.

These period-specific dynamics also provide context for the consent-management evidence in Web Appendix J. Cookie consent banner adoption increased sharply around GDPR enforcement, whereas the estimated usage effect deepened from months 1-3 through months 7-9 and remained substantial thereafter. The differing temporal patterns suggest that temporary interface changes surrounding GDPR implementation are unlikely to fully account for the sustained usage decline. This comparison is suggestive rather than causal: cookie consent banner adoption does not directly measure user friction, and the period decomposition does not identify the precise onset of the usage effect within the first three months.

Finally, the estimates are not driven by observations immediately surrounding GDPR enforcement. Re-estimating the analysis after removing observations from the 30 days before and after May 25, 2018, leaves the estimated effects on weekly visits essentially unchanged across all horizons (Table W28); if anything, the estimates become slightly more negative. This result is consistent with the no-anticipation evidence reported in Table W16.

*Table W28. Summary of GDPR's Effect on Weekly Visits with and without a 30-Day-Window around May 25, 2018*

| **Weekly Visits** | **Original Results** (entire observation period) | | **Robustness Results** (omitting 30 days pre- and post-GDPR) | |
|---|---|---|---|---|
| | **Median** | **Mean** | **Median** | **Mean** |
| **3 months** | -3.49% | -4.88% | -4.55% | -5.49% |
| | **p-value of t-test:** 0.26 *No significant difference between the original and robustness results* | | | |
| **6 months** | -5.54% | -7.22% | -6.38% | -7.91% |
| | **p-value of t-test:** 0.17 *No significant difference between the original and robustness results* | | | |
| **9 months** | -7.54% | -9.07% | -8.24% | -9.74% |
| | **p-value of t-test:** 0.20 *No significant difference between the original and robustness results* | | | |
| **12 months** | -8.24% | -9.57% | -8.78% | -10.09% |
| | **p-value of t-test:** 0.35 *No significant difference between the original and robustness results* | | | |
| **18 months** | -8.91% | -10.02% | -9.56% | -10.30% |
| | **p-value of t-test:** 0.65 *No significant difference between the original and robustness results* | | | |

Notes: The original results use the entire observation period. The robustness results omit observations from the 30 days immediately before and after May 25, 2018. None of the differences between the original and robustness estimates is statistically significant at the 5% level. $* p < .05$, $** p < .01$, $*** p < .001$.

Taken together, the non-cumulative estimates show that the magnitude of the GDPR effect continued to evolve after the initial three-month period. The mean and median effects became progressively more negative through months 7–9, and the median remained near its maximum magnitude thereafter. Cumulative estimates accurately summarize the overall post-treatment decline but smooth this dynamic pattern: they understate the contemporaneous decline during the build-up phase and partly obscure the persistence of the effect for the typical website. The decomposition establishes the evolution of the effect across multi-month periods but does not identify its precise onset within the first post-treatment quarter.

## WEB APPENDIX J: Consent Management Adoption, Enforcement Timing, and Post-Arrival Behavior

This appendix provides diagnostic evidence on three potential components of the GDPR response: websites' adoption of visible consent mechanism infrastructure (J.1), the timing of national enforcement actions (J.2), and changes in bouncing behavior and visit duration (J.3); Section J.4 draws the three strands together. These analyses help assess whether the timing and behavioral composition of the usage decline are consistent with particular mechanisms. They do not separately identify the causal contributions of consent friction, enforcement, privacy salience, changes in user acquisition, or user reallocation. The analyses are therefore most informative about whether simple mechanisms account for the observed timing and behavioral patterns, rather than about the contribution of any single mechanism to the aggregate treatment effect.

### J.1 Consent Mechanism Adoption

We track websites' visible interface responses related to consent using two independent sources. These measures capture observable privacy policies, cookie consent banners, and consent-management platforms; they do not measure all legal, technical, or backend dimensions of GDPR compliance.

First, we match the crawl data of Degeling et al. (2019) to 3,674 websites in our sample. Table W29 reports monthly adoption rates during 2018: the share displaying a privacy policy rises gradually from 67.3% in January 2018 to 90.3% in December 2018, without a comparably large break at the time of GDPR enforcement. By contrast, the share displaying a cookie consent banner jumps by 23 percentage points around GDPR enforcement, from 38.9% in the pre-GDPR May crawl to 61.9% in the post-GDPR May crawl, and reaches 66.1% by December 2018.

*Table W29. Privacy Policy and Cookie Consent Banner Adoption in 2018 (N = 3,674 Matched Websites)*

| **Month** | **Privacy policy (%)** | **Δ (pp)** | **Cookie consent banner (%)** | **Δ (pp)** |
|---|---|---|---|---|
| Jan 2018 | 67.3 | | 38.1 | |
| Feb 2018 | 71.0 | +3.7 | 38.2 | +0.2 |
| Mar 2018 | 74.7 | +3.7 | 38.6 | +0.4 |
| Apr 2018 | 76.6 | +1.9 | 38.6 | +0.0 |
| May 2018 (pre-GDPR) | 77.2 | +0.6 | 38.9 | +0.3 |
| May 2018 (post-GDPR) | 80.8 | +3.6 | 61.9 | +23.0 |
| Jun 2018 | 83.0 | +2.2 | 61.9 | +0.0 |
| Jul 2018 | 83.2 | +0.1 | 61.9 | +0.0 |
| Aug 2018 | 84.4 | +1.3 | 65.9 | +4.0 |
| Sep 2018 | 87.7 | +3.3 | 65.9 | +0.0 |
| Oct 2018 | 88.2 | +0.5 | 66.1 | +0.2 |
| Nov 2018 | 90.2 | +2.0 | 66.1 | +0.0 |
| Dec 2018 | 90.3 | +0.1 | 66.1 | +0.0 |

Notes. Adoption is measured from the longitudinal crawls of Degeling et al. (2019), who monitored the 500 most-visited websites in each EU member state with monthly snapshots during 2018, including one crawl shortly before and one shortly after May 25, 2018; we match N = 3,674 websites of our treatment sample to their crawl data by domain. Entries report the share of matched websites on which the crawl identifies the indicated feature; Δ is the percentage-point change relative to the preceding observation, computed from unrounded shares and therefore not always equal to the difference of the rounded entries shown. The two May observations correspond to the crawls conducted before and after May 25, 2018. Because the crawls access each website from an EU vantage point, entries reflect the interface shown to EU visitors. The measures capture the visible presence of a privacy policy or cookie consent banner, not its legal adequacy, design, intrusiveness, or users' responses to it; identical consecutive entries indicate that no change was detected between crawls. Two matched websites lack SimilarWeb traffic data and are excluded from the usage analyses in Table W32, which therefore use N = 3,672.

Second, we observe consent-management-platform (CMP) adoption monthly from May 2017 to December 2019 for 4,385 matched websites (WhoTracks.me). Table W30 shows cumulative first-detected CMP adoption rising from 0.7% in May 2017 to 5.5% by May 2018 and 22.2% by December 2019, with the sharpest monthly increases in May and June 2018.

We next describe how the timing of visible consent-mechanism adoption relates to website usage. Table W31 groups websites by when they first displayed a cookie consent banner (Degeling et al. 2019), before the GDPR (N = 1,428), after the GDPR (N = 1,000), or never (N = 881), and reports the mean estimated effect on weekly visits within each group. Websites that ever displayed a cookie consent banner show significantly smaller declines than those that never did: the gap is about 8 percentage points over the first six months and widens to roughly 12 percentage points by 18 months (all $p < 0.001$), so it does not fade. The two adopter groups, however, are statistically indistinguishable in every window (differences below half a percentage point; all $p > 0.72$), so the operative contrast is cookie consent banner adoption versus non-adoption, not the timing of adoption. Because cookie consent banner adoption correlates with website size and likely proxies for differences in compliance capacity, business model, and user composition, we read this pattern as descriptive selection rather than a causal effect of the cookie consent banner or its timing.

*Table W30. Cumulative First-Detected CMP Adoption (N = 4,385 Matched Websites)*

| Month | CMP (%) | Δ (pp) |
|---|---|---|
| 2017-05 | 0.7 | — |
| 2017-06 | 0.8 | +0.1 |
| 2017-07 | 0.9 | +0.0 |
| 2017-08 | 1.0 | +0.1 |
| 2017-09 | 1.0 | +0.0 |
| 2017-10 | 1.0 | +0.0 |
| 2017-11 | 1.0 | +0.0 |
| 2017-12 | 1.1 | +0.1 |
| 2018-01 | 1.6 | +0.4 |
| 2018-02 | 2.0 | +0.4 |
| 2018-03 | 2.3 | +0.4 |
| 2018-04 | 2.9 | +0.5 |
| 2018-05 | 5.5 | +2.6 |
| 2018-06 | 7.6 | +2.1 |
| 2018-07 | 8.2 | +0.6 |
| 2018-08 | 9.1 | +0.9 |
| 2018-09 | 10.1 | +1.0 |
| 2018-10 | 10.9 | +0.8 |
| 2018-11 | 12.5 | +1.6 |
| 2018-12 | 13.0 | +0.5 |
| 2019-01 | 13.5 | +0.6 |
| 2019-02 | 14.3 | +0.8 |
| 2019-03 | 15.7 | +1.3 |
| 2019-04 | 17.0 | +1.3 |
| 2019-05 | 17.9 | +0.9 |
| 2019-06 | 18.2 | +0.3 |
| 2019-07 | 19.2 | +1.0 |
| 2019-08 | 19.8 | +0.6 |
| 2019-09 | 20.2 | +0.3 |
| 2019-10 | 20.7 | +0.5 |
| 2019-11 | 21.2 | +0.5 |
| 2019-12 | 22.2 | +1.0 |

Notes. Consent-management-platform (CMP) adoption is measured from the monthly WhoTracks.me census, which records the third-party services detected on each website; we match N = 4,385 websites of our treatment sample to the census by domain. A website is coded as having a CMP in a given month when at least one recognized consent-management service (e.g., OneTrust, Cookiebot, Didomi, TrustArc, Evidon, iubenda, or the IAB Europe Transparency and Consent Framework, classified via Ghostery's TrackerDB) is detected among its third-party scripts. Entries report the cumulative share of matched websites with a CMP detected in the indicated month or any earlier month; Δ is the percentage-point increase relative to the preceding month, computed from unrounded shares and therefore not always equal to the difference of the rounded entries shown. The window spans the first available census month (May 2017) through December 2019. Detection depends on WhoTracks.me's coverage and classification and does not capture proprietary or otherwise unrecognized consent interfaces; matched websites are somewhat larger and less EU-concentrated than unmatched ones. Cumulative adoption does not indicate that the same interface remained active in every subsequent month: the contemporaneous share of websites with a detected CMP is 15.1% in December 2019, against the cumulative 22.2%. The CMP census is reported as descriptive evidence on the timing of consent-infrastructure adoption; the staggered difference-in-differences estimates in Table W32 use only the Degeling et al. (2019) crawl measures, whose monthly frequency and pre-/post-enforcement crawls define the adoption rates.

*Table W31. GDPR Effect on Weekly Visits by Cookie Consent Banner Adoption Timing*

| Cookie Consent Banner Adoption | N | 3 Months | 6 Months | 9 Months | 12 Months | 18 Months |
|---|---|---|---|---|---|---|
| Displayed a banner before GDPR | 1,428 | -2.31 | -4.50 | -6.28 | -6.81 | -6.69 |
| Adopted a banner after GDPR | 1,000 | -2.34 | -4.81 | -6.46 | -7.05 | -6.95 |
| Never displayed a banner | 881 | -10.10 | -13.33 | -15.98 | -16.91 | -18.90 |
| Ever adopter vs. never gap (pp) | | +7.8 | +8.7 | +9.6 | +10.0 | +12.1 |

Notes. Entries report the mean estimated GDPR effect on weekly visits (in percent) within each group. Websites are grouped by when the crawl of Degeling et al. (2019) first identifies a cookie banner: already present in the pre-GDPR May 2018 crawl ("before GDPR"), first observed in a later 2018 crawl ("after GDPR"), or never observed ("never"). Of the 3,674 websites matched to the Degeling et al. (2019) crawl (Table W29), 363 have no recorded cookie-banner status in any crawl, and 2 lack a website-specific visit-effect estimate; the remaining 3,309 are classifiable here. The ever adopter vs. never adopter gap is significant at every horizon (Welch t-test, all $p < 0.001$; 95% CIs [5.8, 9.7], [6.7, 10.7], [7.6, 11.7], [7.9, 12.1], [9.7, 14.6] pp for 3–18 months) and widens monotonically. Pre- and post-GDPR adopters are statistically indistinguishable in every window (all $p > 0.72$). The N-weighted mean across the three groups is more negative than the aggregate estimates in Table W34 because the samples differ; this table covers the Degeling-matched subset, whereas Table W34 is restricted to websites with a majority-EU audience.

Table W32 reports the corresponding staggered difference-in-differences estimates following Callaway and Sant'Anna (2021), estimated on the Degeling et al. (2019) crawl measures rather than on the CMP census. The treatment date is the first month in which the relevant feature is detected. Whereas Table W31 compares adopters with non-adopters in the cross-section, the estimates exploit within-website variation around the month of adoption, so the two tables address different estimands and need not agree in sign. Privacy policy adoption provides a placebo-style benchmark because a policy rarely interrupts the browsing session in the same way as a cookie consent banner. Cookie consent banner adoption captures the visible interface most directly associated with consent friction.

The table reports the overall average treatment effects, aggregated across adoption cohorts and post-adoption periods, for six outcomes. It also reports, for each specification, the p-value of a joint Wald test that the pre-adoption event-study leads are zero.

*Table W32. First Detected Consent Mechanism Adoption and Website Usage: Staggered Difference-in-Differences*

| Mechanism | Outcome | Overall ATT (%) | p-value | Joint Wald p (pre-adoption leads) |
|---|---|---|---|---|
| Privacy policy | Visits | -0.60 | 0.596 | .48 |
| | Unique Visitors | -0.57 | 0.925 | .84 |
| | Visits per Unique Visitor | -1.17 | 0.240 | .19 |
| | Pages per Visit | +0.85 | 0.619 | .86 |
| | Bounce Rate | +0.01 | 0.980 | .92 |
| | Bouncing Visits | -0.09 | 0.947 | .39 |
| Cookie consent banner | Visits | -1.44 | 0.054 | .001 |
| | Unique Visitors | +3.73 | 0.544 | .90 |
| | Visits per Unique Visitor | -0.88 | 0.198 | .86 |
| | Pages per Visit | -2.18 | 0.100 | .84 |
| | Bounce Rate | +0.64 | 0.028 | .79 |
| | Bouncing Visits | -0.02 | 0.981 | .09 |

Notes. Callaway and Sant'Anna (2021) staggered difference-in-differences estimated on the N = 3,672 matched websites with SimilarWeb traffic data (see Table W29); websites never observed displaying the relevant feature enter as never-treated controls, including the 363 for which no crawl records a cookie consent banner status, which we treat as non-adopters rather than dropping; the overall ATT is aggregated across adoption cohorts and post-treatment periods, using never-adopting and not-yet-adopting websites as controls. The treatment date is the first month in which a privacy policy (respectively, a cookie consent banner) is detected. Privacy policy adoption rarely introduces an interruptive browsing interface and therefore provides a placebo-style comparison; it is not a pure placebo because adoption may coincide with other compliance changes. Joint Wald p reports the p-value of a joint Wald test that the pre-adoption event-study leads are jointly zero. It rejects only for the cookie consent banner visits specification ($\chi^2(6) = 21.6$, $p = .001$), which cautions against a causal reading of that coefficient. All outcomes enter the estimation as $\log(Y + 1)$, so reported effects are percentage changes of $(Y + 1)$; for the count outcomes, this closely approximates the percentage change of the outcome itself, but for the three ratio outcomes, the added one is not negligible. Evaluated at pre-adoption sample means, the bounce-rate estimate of +0.64 corresponds to an increase of roughly 0.9 percentage points in the bounce rate (a 0-1 share; positive values indicate more bouncing), the pages-per-visit estimate of -2.18 to a decline of roughly 2.6 percent in pages per visit, and the visits-per-unique-visitor estimate of -0.88 to a decline of roughly 1.1 percent in visits per unique visitor. Because each outcome is estimated in a separate $\log(Y + 1)$ specification, with cohort composition and the added one differing across outcomes, the visits estimate is not the sum of the unique visitor and visits-per-unique-visitor estimates, and the three should not be read as an exact decomposition. The reported p-values are not adjusted across the six outcomes for each mechanism; we therefore treat the individual estimates as exploratory and emphasize their overall pattern.

Privacy policy adoption is not statistically distinguishable from zero for any of the six outcomes. The joint tests also fail to reject that the pre-adoption coefficients are jointly zero for every outcome (p=.19 to p=.92).

For cookie consent banner adoption, the point estimates indicate a 1.44% decline in visits ($p = .054$), a 2.18% decline in pages per visit ($p = .100$), and a 0.64% proportional increase in the bounce rate ($p = .028$). The estimates for unique visitors, visits per unique visitor, and bouncing visits are imprecise and do not exhibit a consistent negative pattern.

The weekly visits estimate should not be given a causal interpretation because the joint test strongly rejects that its pre-adoption coefficients are zero ($\chi 2(6)=21.6$, p=.001). The pages-per-visit and bounce rate specifications do not reject the joint test of zero pre-adoption leads (p=.84 and p=.79, respectively), but the estimates remain observational and are examined alongside several outcomes. We therefore interpret them as suggestive evidence that cookie consent banner adoption may produce modest changes in selected usage and post-arrival outcomes, rather than as evidence of a general decline across usage outcomes.

The estimates do not constitute a formal one- to two-percent bound on consent friction. The overall ATT averages across all observed post-adoption periods rather than focusing on the adoption week, and its causal interpretation depends on the parallel-trends assumption. Nevertheless, the estimated changes in total visits and pages per visit are modest relative to the sustained aggregate GDPR effect. This conclusion is consistent with Lefrere et al. (2026), who find no detectable usage effect of consent-mechanism adoption among content providers.

Taken together, the adoption data show that visible consent infrastructure expanded sharply around GDPR enforcement. Cookie consent banner prevalence jumped in May 2018, and CMP adoption was detected to accelerate in May and June 2018. The staggered difference-in-differences analysis provides suggestive evidence of modest changes associated with cookie consent banner adoption in selected outcomes but does not reveal broad negative effects across all usage metrics.

Web Appendix I further shows that the aggregate GDPR effect continues to deepen through months 7-9 and remains substantial thereafter. The differing temporal patterns make a short-lived enforcement-period interface effect an incomplete explanation for the sustained usage decline. They do not rule out persistent consent friction, nor do they establish that consent interfaces are exogenous to websites' broader GDPR responses.

## J.2 Enforcement Timing and Google Fine

National data-protection authorities initiated GDPR enforcement actions at different times. Among countries with a publicly identified action during the panel, first actions span September 2018 to September 2019, while several countries have no identified action within the observation window (Table W33).

Table W34 presents a descriptive comparison of the evolution of the estimated GDPR effect for websites whose primary EU-user country initiated enforcement by the end of 2018, versus websites linked to countries that initiated enforcement later or not within the panel. Both groups experience deepening usage declines, including before the later-enforcing countries' first actions, and their effects converge by 18 months.

We do not interpret this comparison causally. Enforcement timing may reflect institutional capacity, complaint activity, prior compliance conditions, or other country characteristics, and relatively few countries remain not yet enforced for long enough to support a stable staggered-adoption design. We therefore use the chronology and group comparison to describe the evolving enforcement environment rather than to identify the effect of enforcement.

*Table W33. First Publicly Recorded GDPR Enforcement-Related Action by Country*

| **Country** | **Date of first recorded action** | **Reported monetary sanction** | **Entity** | **DPA** |
|---|---|---|---|---|
| Austria | 27 Sep 2018 | €300 | (individual) | DSB |
| Czech Republic | 25 Oct 2018 | €388 | (not named in source) | UOOU |
| Germany | 21 Nov 2018 | €20,000 | Knuddels.de | Baden-Württemberg |
| Bulgaria | 04 Dec 2018 | €500 | (not named in source) | KZLD |
| Portugal | 15 Dec 2018 | €400,000 | Barreiro Montijo Hospital | CNPD |
| Hungary | 18 Dec 2018 | €3,200 | (not named in source) | NAIH |
| France | 21 Jan 2019 | €50,000,000 | Google LLC | CNIL |
| Malta | 18 Feb 2019 | €5,000 | Lands Authority | Commissioner |
| Poland | 26 Mar 2019 | €220,000 | (not named in source) | UODO |
| Italy | 17 Apr 2019 | €50,000 | Rousseau platform | Garante |
| Lithuania | 16 May 2019 | €61,500 | MisterTango | VDAI |
| Belgium | 28 May 2019 | €2,000 | (individual) | APD |
| Spain | 11 Jun 2019 | €250,000 | LaLiga | AEPD |
| Denmark | 15 Jun 2019 | — | — | Datatilsynet |
| Norway | 15 Jun 2019 | — | — | Datatilsynet |
| Netherlands | 18 Jun 2019 | €350,000 | Haga Hospital | AP |
| Romania | 27 Jun 2019 | €130,000 | UniCredit Bank | ANSPDCP |
| United Kingdom | 08 Jul 2019 | — | British Airways | ICO |
| Greece | 30 Jul 2019 | €150,000 | PWC Business Solutions | HDPA |
| Sweden | 20 Aug 2019 | €18,630 | School Skelleftea | DPA |
| Latvia | 26 Aug 2019 | €7,000 | (not named in source) | DVI |
| Cyprus | 15 Sep 2019 | — | — | Commissioner |

Notes. For each country, the table reports the date of the first publicly recorded GDPR enforcement action by its national data protection authority (DPA), along with the associated fine, the fined entity, and the DPA. Source: GDPR Enforcement Tracker (enforcementtracker.com), cross-referenced with DPA press releases. A dash (—) indicates that the source reports no fine amount or named entity for the first action (e.g., Denmark, Norway, and Cyprus record early actions without a published fine; the United Kingdom entry reflects the ICO's July 2019 notice of intent to fine British Airways, with the fine imposed in 2020). Dates are approximate where the source gives no exact day. The first recorded action is not necessarily the first economically significant fine. Several EU/EEA countries represented in our sample recorded no enforcement action within the observation window and are omitted from the table.

*Table W34. Evolution of the GDPR Effect on Weekly Visits by Country Enforcement Timing*

| **Post-GDPR window** | **Enforced by end-2018** | | **Not yet enforced by end-2018** | |
|---|---|---|---|---|
| | **Median (%)** | **Mean (%)** | **Median (%)** | **Mean (%)** |
| 3m | -4.46 | -5.65 | -2.34 | -3.64 |
| 6m | -7.20 | -7.86 | -4.27 | -5.76 |
| 9m | -8.17 | -9.21 | -6.25 | -7.58 |
| 12m | -8.75 | -9.25 | -7.18 | -8.11 |
| 18m | -8.71 | -8.75 | -8.15 | -8.53 |

Notes. Websites are grouped by the enforcement timing of their primary EU-user country, the EU country accounting for the largest share of the website's EU traffic (first-action dates as reported in Table W33). "Enforced by end-2018" comprises countries whose data-protection authority took its first recorded enforcement action in 2018; among the eleven EU user markets covered by the traffic panel, these are Austria, Germany, and Hungary (Bulgaria, the Czech Republic, and Portugal also enforced in 2018 but contribute no websites to the sample). "Not yet enforced by end-2018" comprises the remaining eight markets (Denmark, France, Italy, the Netherlands, Poland, Spain, Sweden, and the United Kingdom), all of which took their first action in 2019; no sample market remained unenforced by the end of the observation window. Of the 5,790 treated websites, 5,519 have a primary EU-user country: 1,399 in the 2018-enforcing markets and 4,120 in the 2019-enforcing markets. Entries are website-level synthetic-control-group estimates of the GDPR's effect on weekly visits, aggregated within the group for websites with a majority-EU audience, because the primary EU-user country is informative about regulatory exposure only for websites with a predominantly EU audience. Both groups show deepening declines. Importantly, the later-enforcing group already exhibits substantial declines at the 3- and 6-month horizons, before any of its countries' first recorded actions, and the two groups converge by 18 months.

The most prominent enforcement action during the observation period is the CNIL's €50 million fine against Google on January 21, 2019. The fine falls within months 7-9 after GDPR enforcement, the period in which the non-cumulative mean usage effect reaches its most negative value in Web Appendix I. This temporal coincidence motivates a focused analysis but does not by itself establish that the fine caused the deepening decline.

Using the within-website design of Web Appendix G, we estimate that EU-user traffic declines by an additional 5.1% relative to non-EU-user traffic after the fine across the EU ($p<.001$). We find no incremental France-specific shift relative to the rest of the EU (+1.3%, $p=.81$), and excluding France produces a similar EU-wide estimate (-5.4%, $p<.001$).

The absence of a France-specific change and the similar estimate after excluding France are more consistent with a broader EU-wide shift than with a response confined to the jurisdiction imposing the fine. One possible interpretation is that the highly publicized action increased the salience or perceived credibility of GDPR enforcement across Europe. The analysis does not separately identify that mechanism, however. It examines a single prominent event embedded in an already evolving post-GDPR trajectory, and the post-fine coefficient may capture the continuation of that trajectory or other contemporaneous EU-wide developments.

*Table W35. EU-vs.-Non-EU Traffic Differentials Around the CNIL Google Fine*

| Specification | Estimate (log points) | SE | Effect (%) | p-value |
|---|---|---|---|---|
| All EU countries | -0.052 | 0.014 | -5.1% | <.001 |
| Incremental France-specific differential | 0.012 | 0.052 | 1.3% | .81 |
| Excluding French EU-user instances | -0.056 | 0.015 | -5.4% | <.001 |

Notes: Estimates from the within-website difference-in-differences design of Web Appendix G: each of the 1,104 non-EU-registered websites contributes a treated EU-user instance and a control non-EU-user instance (2,208 website-instances), observed weekly in a fully balanced panel, every instance is observed in all 83 weeks of the estimation window; zero-traffic weeks enter as zeros. The dependent variable is log(weekly visits +1). All models include website-instance and calendar-week fixed effects, cluster standard errors by website, and restrict the sample to the post-GDPR window (May 25, 2018 – December 23, 2019; 83 weeks; N = 183,264 website-instance-weeks), so the GDPR-enforcement response itself is excluded from identification. "All EU countries" codes every EU-user instance as treated from January 21, 2019, the date of the CNIL's €50 million fine against Google, onward. "Incremental France-specific differential" instead codes only the EU-user instances of the 105 websites whose EU market is France as treated, so the common post-fine movement of all instances is absorbed by the week fixed effects and the coefficient measures the French deviation from that common path. "Excluding French EU-user instances" removes those 105 treated instances from the sample while retaining their within-website control instances (N = 174,549). Effects are 100·(exp(β)-1). Each specification includes a single post-fine indicator and does not allow differential pre-event trends. The EU-wide widening, together with the absence of an incremental French response, is consistent with a broader EU-wide shift rather than a response confined to France. Because the model does not allow for differential pre-event trends and examines a single event, it does not separately identify an enforcement-salience effect.

## J.3 Bouncing Behavior and Visit Duration

One behavioral outcome that consent-interface friction could affect is bouncing: users who arrive and leave after viewing a single page. Bouncing is not unique to consent friction and may also reflect page relevance, traffic-source composition, loading performance, or website design.

We compute the GDPR's effect on the number of bouncing visits (Table W36) and on the bounce rate (Table W37) from the same website-level GSC estimates as in the main text; the bounce-rate effect is constructed as ((1 + effect on bouncing visits) / (1 + effect on visits)) − 1.

Panels B and C of Table W36 classify websites using the sign of their estimated visit effect. Because visits determine both subgroup assignments and the denominator of the constructed rate, the subgroup divergence is partly mechanical. We therefore rely primarily on Panel A for substantive interpretation and present Panels B and C as descriptive decompositions.

*Table W36. Effect of GDPR on Bouncing Visits (%)*

| | Post-Treatment Period (months after GDPR) | | | | |
|---|---|---|---|---|---|
| **Statistic** | **3m** | **6m** | **9m** | **12m** | **18m** |
| **Panel A: All treated websites** | | | | | |
| Mean (%) | −4.79 | −7.32 | −9.32 | −9.64 | −9.58 |
| Median (%) | −4.19 | −6.69 | −8.68 | −9.10 | −9.78 |
| Sig. positive (%) | 7.02 | 7.94 | 8.30 | 9.14 | 11.57 |
| Sig. negative (%) | 17.64 | 22.98 | 27.64 | 29.72 | 31.09 |
| Non-significant (%) | 75.34 | 69.08 | 64.06 | 61.14 | 57.34 |
| Share of all treated websites (%) | 100 | 100 | 100 | 100 | 100 |
| **Panel B: Websites gaining visits** | | | | | |
| Mean (%) | +11.09 | +9.39 | +9.67 | +10.65 | +14.83 |
| Median (%) | +7.46 | +5.98 | +6.50 | +7.64 | +10.46 |
| Sig. positive (%) | 15.08 | 17.62 | 19.92 | 22.09 | 27.16 |
| Sig. negative (%) | 2.90 | 5.27 | 7.12 | 8.64 | 8.98 |
| Non-significant (%) | 82.01 | 77.11 | 72.96 | 69.27 | 63.87 |
| Share of treated websites gaining visits (%) | 40.7 | 36.2 | 33.0 | 32.5 | 33.3 |
| **Panel C: Websites losing visits** | | | | | |
| Mean (%) | −15.71 | −16.80 | −18.66 | −19.40 | −21.69 |
| Median (%) | −12.25 | −13.38 | −15.50 | −16.61 | −18.96 |
| Sig. positive (%) | 1.48 | 2.45 | 2.59 | 2.91 | 3.83 |
| Sig. negative (%) | 27.77 | 33.02 | 37.73 | 39.86 | 42.07 |
| Non-significant (%) | 70.75 | 64.52 | 59.68 | 57.22 | 54.10 |
| Share of treated websites losing visits (%) | 59.3 | 63.8 | 67.0 | 67.5 | 66.7 |

Notes: Bouncing visits are the number of weekly visits that end after a single page view, computed as bounce rate × weekly visits from the SimilarWeb data. Effects are estimated at the website-instance level, a website's EU-user and, where present, non-EU-user traffic enter as separate instances, each matched to its own synthetic control, and combined into a website-level effect as the average of the instance-level estimates weighted by the website's pre-GDPR traffic shares; entries express these website-level effects as percentage changes, $100\cdot(\exp(\tau)-1)$, summarized by the cross-website mean and median for each cumulative post-treatment period. "Sig. positive" and "Sig. negative" report the share of websites whose website-level effect is statistically significant at the 5% level (two-sided) with the indicated sign; together with "Non-significant" they sum to 100%. Panels B and C split websites by the sign of the same website's estimated effect on weekly visits in the same window; because websites move between the gaining and losing groups across windows (the gaining share falls from 40.7% at 3 months to 33.3% at 18 months), the panels describe changing compositions rather than fixed groups of websites. The same classification underlies Panels B and C of Table W37, which therefore report identical group shares. The corresponding results for the bounce rate are reported in the following table.

*Table W37. Effect of GDPR on the Bounce Rate (in Percentage Points)*

| | Post-Treatment Period (months after GDPR) | | | | |
|---|---|---|---|---|---|
| **Statistic** | **3m** | **6m** | **9m** | **12m** | **18m** |
| **Panel A: All treated websites** | | | | | |
| Mean (pp) | +0.81 | +0.70 | +0.79 | +1.22 | +2.51 |
| Median (pp) | −0.57 | −0.88 | −1.04 | −0.93 | −0.58 |
| Sig. positive (%) | 3.30 | 5.79 | 7.39 | 8.79 | 11.30 |
| Sig. negative (%) | 3.93 | 7.16 | 9.08 | 10.74 | 12.50 |
| Non-significant (%) | 92.77 | 87.06 | 83.53 | 80.46 | 76.20 |
| Share of all treated websites (%) | 100 | 100 | 100 | 100 | 100 |
| **Panel B: Websites gaining visits** | | | | | |
| Mean (pp) | −2.86 | −3.81 | −4.18 | −4.37 | −3.76 |
| Median (pp) | −3.36 | −4.38 | −4.87 | −5.06 | −4.81 |
| Sig. positive (%) | 2.06 | 3.22 | 3.76 | 4.82 | 7.59 |
| Sig. negative (%) | 4.97 | 9.55 | 11.64 | 13.63 | 15.71 |
| Non-significant (%) | 92.97 | 87.24 | 84.60 | 81.55 | 76.70 |
| Share of treated websites gaining visits (%) | 40.7 | 36.2 | 33.0 | 32.5 | 33.3 |
| **Panel C: Websites losing visits** | | | | | |
| Mean (pp) | +3.38 | +3.29 | +3.27 | +3.95 | +5.67 |
| Median (pp) | +1.36 | +1.19 | +0.80 | +0.96 | +1.36 |
| Sig. positive (%) | 4.16 | 7.27 | 9.20 | 10.74 | 13.17 |
| Sig. negative (%) | 3.21 | 5.78 | 7.81 | 9.33 | 10.88 |
| Non-significant (%) | 92.63 | 86.96 | 83.00 | 79.93 | 75.95 |
| Share of treated websites losing visits (%) | 59.3 | 63.8 | 67.0 | 67.5 | 66.7 |

Notes: The bounce rate is the share of weekly visits that end after a single page view. Its effect is constructed from the website-level effects on bouncing visits and weekly visits as ((1 + effect on bouncing visits) / (1 + effect on visits)) − 1, paralleling the derivation of the usage intensity metrics in Web Appendix B; entries are reported in percentage points: the expression above yields a proportional change in the bounce rate, which we multiply by the website's pre-GDPR mean bounce rate to express in percentage points, and are summarized by the cross-website mean and median. Because the split in Panels B and C is based on the sign of the visit effect and visits enter the bounce-rate denominator, the split is partly mechanical; we report it for consistency with the usage-intensity metrics. The group shares are identical to those in Panels B and C of Table W36, which use the same classification.

Bouncing visits decline by approximately the same order of magnitude as total weekly visits: the median effect becomes more negative from -4.2% at 3 months to -9.8% at 18 months. The aggregate traffic declines, so it does not appear to be driven primarily by a selective disappearance of visits that would otherwise have ended after a single page view. Both total visits and bouncing visits decline.

For the typical website, the constructed bounce-rate effect is modest: the median ranges from -0.6 to -1.0 percentage points across horizons. The mean is positive and rises to +2.5 percentage points at 18 months, indicating a right-skewed distribution in which increases are concentrated among a subset of websites rather than occurring broadly across the median website. The gap between mean and median reflects a small number of websites with large positive constructed rate effects, which arise mechanically where the estimated decline in visits exceeds the decline in bouncing visits.

At 18 months, 76.2% of individual website estimates are not statistically distinguishable from zero. This share also reflects differences in website-level precision and should not be interpreted as proving that those websites experienced no change.

Websites classified as losing visits have positive constructed-rate effects on average, whereas websites classified as gaining visits have negative effects. Because visits determine both the classification and the denominator of the constructed outcome, this divergence is partly mechanical and provides limited independent evidence about the behavioral mechanism.

Visit duration provides a complementary measure of behavior conditional on a visit occurring. The mean duration-per-visit effect ranges from -0.94% to +0.09%, while the median ranges from -1.96% to -0.92% across post-treatment periods (Table W38). The point estimates, therefore, indicate small declines for the typical website rather than literally no change, although their magnitude is modest relative to the decline in total traffic.

These estimates provide little evidence of a large or broad deterioration in time spent per visit among the visits that continue to occur. Duration per visit does not measure longer-term user retention and cannot rule out changes in other dimensions of engagement.

*Table W38. Effect of GDPR on Visit Duration*

| **Post-Treatment Period** | **Mean (%)** | **95% CI (mean)** | **Median (%)** | **95% CI (median)** |
|---|---|---|---|---|
| 3m | -0.83 | [-1.46, -0.16] | -1.96 | [-2.36, -1.39] |
| 6m | -0.48 | [-1.15, +0.13] | -0.95 | [-1.43, -0.46] |
| 9m | -0.57 | [-1.27, +0.12] | -0.92 | [-1.45, -0.32] |
| 12m | -0.94 | [-1.71, -0.19] | -1.27 | [-1.76, -0.66] |
| 18m | +0.09 | [-0.77, +0.91] | -0.93 | [-1.62, -0.29] |

Notes. Duration per visit is the total visit duration divided by the total number of visits. The effect is the proportional website-level synthetic-control-group estimate, ((1 + effect on total duration) / (1 + effect on total visits)) − 1, computed from the same website-level estimates used in the main text. 95% confidence intervals are bootstrap intervals obtained by resampling websites (2,000 replications).

The estimated changes are small relative to the decline in weekly visits. The median is modestly negative at every horizon, whereas the mean confidence interval includes zero at 6, 9, and 18 months. The results, therefore, provide little evidence of a large deterioration in time spent per visit, but they do not establish exact equivalence to zero or measure longer-term user retention.

## J.4 Summary and Conclusion

The evidence in this appendix is more informative about mechanisms that are unlikely to explain the entire effect than about a uniquely identified causal mechanism. Visible consent infrastructure expanded sharply around GDPR enforcement: cookie consent banner prevalence jumped during May 2018, and detected CMP adoption accelerated in May and June 2018. The staggered difference-in-differences estimates suggest modest changes associated with cookie consent banner adoption in selected outcomes but do not indicate broad negative effects across all usage metrics. The weekly visits estimate is particularly difficult to interpret causally because the joint test rejects parallel pre-adoption dynamics.

The aggregate GDPR effect also continues to deepen after the main enforcement-period increase in visible consent infrastructure and remains substantial thereafter. This timing makes a short-lived interface-friction effect an incomplete explanation for the sustained usage decline. Consent interfaces may nevertheless have persistent effects and may constitute one component of the broader response.

The national enforcement chronology similarly does not support a simple account in which website usage begins to decline only after each country's first enforcement action. Websites linked to later-enforcing countries already experience substantial declines before those actions. The EU–non-EU differential also widens after the Google fine without an additional France-specific shift. This pattern is consistent with a broader EU-wide response, potentially reflecting increased enforcement salience, but the single-event analysis cannot distinguish that interpretation from the continuation of the existing GDPR trajectory or other contemporaneous developments.

Finally, the behavioral outcomes point primarily to fewer visits occurring than to a broad deterioration in behavior conditional on a visit. Bouncing visits decline alongside visits, while the median change in the constructed bounce rate and the effects on visit duration remain comparatively small. This combination is consistent with reduced reach, changes in traffic sources, or users reallocating their activity across websites. The available outcomes do not distinguish between these explanations and should not be interpreted as uniquely identifying reduced acquisition.

## WEB APPENDIX REFERENCES